\documentstyle[seceq,preprint,psfig,here,wrapft]{ptptex}

\newcommand{\be}{\begin{equation}}
\newcommand{\ee}{\end{equation}}
\newcommand{\bea}{\begin{eqnarray}}
\newcommand{\eea}{\end{eqnarray}}
\newcommand{\bean}{\begin{eqnarray*}}
\newcommand{\eean}{\end{eqnarray*}}

\newcommand{\ba}{\begin{array}}
\newcommand{\ea}{\end{array}}
\newcommand{\bmat}{\left(\ba}
\newcommand{\emat}{\ea\right)}
\newcommand{\sslash}[1]{\not{\!#1}}
\newcommand{\lslash}[1]{\not{\!\!#1}}

\preprintnumber[3cm]{%<-- [..]: optional width of preprint # column.
HUPD-9819\\ JUPD-9825\\ hep-ph/9812449}
\title{%        %You can use \\ for explicit line-break
Polarized Structure Functions in QCD
}
\author{%       %Use \sc for the family name
Jiro {\sc Kodaira}
and Kazuhiro {\sc Tanaka}$^{*}$
}

\inst{%         %Affiliation, neglected when [addenda] or [errata]
Dept. of Physics, Hiroshima University, Higashi-Hiroshima 739-8526, Japan
\\
$^*$Dept. of Physics, Juntendo University,
Inba-gun, Chiba 270-1606, Japan
}

\recdate{%      %Editorial Office will fill in this.
\today
}

\abst{
Hadron spin physics is now one of the most active fields of physics.
Especially in the last ten years, great progress has been made
both theoretically and experimentally that has
considerably improved our knowledge of the spin structure of nucleons. 
We review the nucleon's polarized structure functions 
from the viewpoint of factorization theorems and the
gauge invariant, nonlocal light-cone operators in QCD.
We discuss a systematic treatment of
the polarized structure functions and the corresponding
parton distribution functions, 
which are relevant to inclusive lepto- and hadro-production.
We give a detailed analysis of these spin-dependent
distribution functions at the twist-2 and twist-3 level,
and present various properties and relations 
satisfied by the parton distributions,
which can be derived directly from QCD.
We emphasize unique features of higher twist distributions,
and the role of the QCD equations of motion to derive their 
sensitivity to the quark-gluon correlation and 
their anomalous dimensions for $Q^{2}$-evolution. 
}

\begin{document}

\maketitle

%-------------------- text -------------------------------
\section{Introduction}

Hadron spin physics is now one of the major fields of particle
and nuclear physics based on quantum chromodynamics (QCD).
The starting point for this field was
the measurement of the polarized nucleon structure function
$g_{1}(x, Q^{2})$ in deep inelastic scattering
by the EMC collaboration.~\cite{emc1}
The data of this EMC experiment indicate that an unexpectedly 
small fraction of the nucleon's spin was carried by its constituents,
quarks, and have caused many physicists to challenge
the so-called \lq\lq spin crisis\rq\rq\ problem in QCD.
After a flood of theoretical papers as well as experimental 
data,~\cite{emc2,emc3,emc4}
our understanding on this problem is much more updated.~\cite{rv}
Through these developments, \lq\lq hadron spin physics\rq\rq\ has 
grown up as a field attracting considerable attention. 
Now our interest has spread to various other processes to explore the 
spin structure of the nucleon.
In particular, in conjunction with 
new projects like the \lq\lq RHIC spin project\rq\rq, \lq\lq polarized
HERA\rq\rq , etc.,  we are now
in a position to obtain more information on the 
deep structure of the nucleon and the dynamics of QCD.
Furthermore, the spin degrees of freedom
in high-energy processes are expected to play an important role 
to verify the Standard Model itself.

The aim of this article is to provide an overview of the 
perturbative methods in QCD used
to study the nucleon's spin structure through deep inelastic processes,
and to summarize the recent theoretical progress on this subject including
the QCD evolutions of polarized structure functions.~\footnote{Those 
who are interested in the present status of phenomenological
studies are recommended to refer to the recent comprehensive review by
Lampe and Reya.~\cite{bodo}}
Although the high-energy (short-distance) behavior of QCD is
described by perturbation theory thanks to  
asymptotic freedom, a general cross section is a complicated combination
of the short- and long-distance interactions,
and is not directly calculable.
For some specific processes, however, the factorization theorems allow us
to separate (factorize) the
short-distance dynamics from the long-distance dynamics
in the cross section in a systematic fashion,
and to derive the predictions for the short-distance dynamics
using perturbation theory.
Many observables are known in
spin related phenomena of high-energy processes, to which
factorization theorems and thus 
perturbative QCD can be successfully applied. 
In this paper, we restrict ourselves to the nucleon's 
polarized structure functions,
which can be measured in high-energy inclusive reactions, like
lepton-nucleon deep inelastic scattering
and the Drell-Yan process in nucleon-nucleon collisions.

In the present context, the proper high-energy limit is
provided by the \lq\lq Bjorken limit\rq\rq. 
In this limit, using the notion of \lq\lq
twist\rq\rq, we can systematically expand a cross section in
inverse powers of a characteristic large momentum scale $Q$,
and extract the leading contributions.
The factorization theorems can be proved (in principle)
order by order in the twist expansion. 
At the leading twist (twist-2) level, the result implied by
the factorization theorems is
realized as the QCD improved parton model.
On the other hand, the higher twist effects describe
the coherent quark-gluon behavior inside the nucleon,
and thus contain information on the correlation of
quarks and gluons beyond the parton model.
The higher twist contributions are suppressed by some powers
of $1/Q$ compared to the leading twist contribution
and are usually hidden by the latter in the cross section.
This makes it very difficult to extract
higher twist effects from experimental studies.
In the case of spin-dependent processes, 
however, we have a good chance to study them; it is known that 
the twist-3 contributions can be measured as leading effects
in certain asymmetries.
Therefore we will discuss in detail
the twist-3 contributions in addition to 
the leading twist effects for the polarized structure functions. 

This paper is organized as follows.
In \S2 we recall the factorization theorems in the above two typical
examples of inclusive processes.
To describe the long-distance part of the factorized cross section,
we introduce the parton distribution functions as the nucleon's
matrix elements of nonlocal light-cone operators in QCD.
We discuss the parton distributions given by the two- and
three-particle operators and classify them with respect to
twist, spin-dependence and chiral properties.
In \S3 we explain the equivalence of the approach in \S2 
to the conventional approaches based on the operator product expansion.
A general feature of the renormalization of gauge invariant operators
will be discussed in \S4 to deal with
the higher twist operators in a covariant approach.
The QCD predictions for the chiral-even structure functions
are summarized in \S\ref{sec:ce}.
The chiral-odd structure functions are discussed in \S\ref{sec:co}.
The final section is devoted to a summary.

\section{Factorization theorems and parton distribution functions}

To describe a variety of high-energy processes in a universal language,
it is desirable to have a definition of parton distribution
functions based on the operators in QCD.
The traditional approach for this purpose relies on 
the operator product expansion (OPE), 
but it can be applied only to a limited
class of processes like deep inelastic lepton-hadron scattering. 
This calls for an approach based on the {\it factorization}
as a generalization of the OPE.
In this section, we explain the factorization theorems, and introduce a 
definition of the parton distribution functions in terms
of the nonlocal light-cone operators in QCD.

\subsection{Factorization theorems for hard processes in QCD}

There are various hard processes that are characterized by the 
large momentum squared $Q^{2}$:\ deep inelastic scattering (DIS),\  
$l + A \rightarrow l' + X$
($A, B, \ldots$ represent hadrons, $l$ a lepton, and  
$X$ a system of hadrons produced through inelastic processes);
the Drell-Yan (DY) processes, 
$A + B \rightarrow l^{+} + l^{-} + X$ ;
jet production, $A + B \rightarrow {\rm jet} + X$ ;
heavy quark production, $A + B \rightarrow {\rm heavy}$ 
${\rm quark} + X,$ etc.
The basis for the QCD analysis of these hard processes 
is provided by the factorization theorems in QCD,\cite{CSS}
which give an extension of the OPE.
It provides a foundation of the ``parton model''
in the Bjorken limit, where $Q^{2}\rightarrow \infty$ 
with the Bjorken variable $x$ fixed,
and also a systematic framework to calculate the QCD
corrections beyond the leading order.
The theorems tell us that 
we can view a high-energy beam of hadrons as if it were a beam of partons
(quarks and gluons), and the hadron reaction is induced by 
hard scattering among individual partons.
Corresponding to this intuitive picture, 
the cross section for the hard processes is given as a product,
or more precisely, convolution of the short- and long-distance parts.
The former contains all the dependence on the large momentum $Q$,
while the latter depends 
essentially on the QCD scale parameter $\Lambda_{\rm QCD}$.
These two parts are divided at a factorization (renormalization) 
scale $\mu$,\footnote{The factorization scale can be different from
the renormalization scale. In this paper, we set them equal for
simplicity.}
and the short-distance (long-distance) part involves the 
momenta larger (smaller) than $\mu$. 
The short-distance part corresponds to the hard scattering cross section for 
the partons with the large momentum $Q$ exchanged,
while the long-distance part corresponds to the 
parton distribution functions in a hadron.
The former is systematically calculable in perturbation theory for 
each process, due to the asymptotic freedom 
of QCD, while the latter is controlled by the 
nonperturbative dynamics of QCD.

Let us recall the factorization formulae for some familiar processes. 
First, consider the DIS, which proceeds via the exchange of a virtual photon
with momentum $q_{\mu}$ between a lepton and a hadron 
($Q^{2} \equiv -q^{2} \ge 0$).~\footnote{More generally,
the vector boson $W^{\pm}$ or $Z^{0}$ can be exchanged.
In this case, $J_{\mu}(z)$ of (\ref{eq:current})
should be replaced by the corresponding charged or neutral current.}
As is well known, the DIS cross section
is given in terms of the leptonic ($L_{\mu\nu}$) and the hadronic
($W_{\mu\nu}$) tensors,~\cite{rv}
\[ k'_0 \frac{d\sigma}{d^3k'} = \frac{1}{k\cdot P}
   \left( \frac{e^2}{4\pi Q^2}\right)^2 L^{\mu\nu}W_{\mu\nu}\ , \]
where $k (k')$ and $P$ denote the momenta of the incident (scattered) 
lepton and hadron ($k-k' = q$). 
All the information of the strong interaction is 
contained in the hadronic tensor:
\begin{eqnarray}
  W_{\mu\nu}
  &=& \frac{1}{2\pi} \sum_{X} \langle P S |
      J_{\mu}(0)|X \rangle \langle X |J_{\nu}(0)| P S \rangle 
     (2\pi )^4 \delta^{(4)} (p_X - P - q ) \nonumber \\
 &=& \frac{1}{2\pi }\int d^{\,4} z e^{- i q\cdot z} \langle P S |
               [J_{\mu}(0)\,,\,J_{\nu}(z)]| P S \rangle \ .\label{ht}
\end{eqnarray}
Here $|PS\rangle$ is the hadron state with momentum $P$ and spin $S$,
$p_{X}$ is the momentum of the hadronic final state $|X \rangle$,
and 
\begin{equation}
  J_{\mu}(z)
  =  \bar{\psi}(z) \{\lambda^{3}/2 + \lambda^{8}/(2\sqrt{3})\} 
  \gamma_{\mu}\psi(z)
  \equiv \bar{\psi}(z) {\cal Q}^{(el)}\gamma_{\mu}\psi(z) \label{eq:current}
\end{equation}
is the hadron's electromagnetic current
composed of the quark field $\psi(z)$
and the flavor matrices $\lambda^{i}$ for the $u$-, $d$- and $s$-quarks
when we consider three flavors.  
For the case of a nucleon target with mass $M$
($P^{2} = M^{2}, P\cdot S = 0$, $S^{2} = - M^{2}$),
the hadronic tensor can be expressed in general by the four structure
functions $F_{1}, F_{2}, g_1$ and $g_2$ as
\bea
 \lefteqn{W_{\mu\nu} =}\nonumber\\ 
      &-&\,\left(\,g_{\mu\nu}-\frac{q_{\mu}q_{\nu}}{q^2}\right)
          F_1(x, Q^{2}) + \left(\,P_{\mu}-\frac{P\cdot q}{q^2}q_{\mu}\right)
              \left(\,P_{\nu}-\frac{P\cdot q}{q^2}q_{\nu}\right)
              \frac{2}{P \cdot q}\,F_2 (x, Q^{2})\nonumber\\
      &+&\,2 i \, \epsilon _{\mu\nu\lambda\sigma}q^{\lambda}
              \left\{S^{\sigma} \frac{1}{P\cdot q}\,g_1(x, Q^{2})
          + (P\cdot q S^{\sigma}-q\cdot S
                   P^{\sigma}) \frac{1}{(P\cdot q)^2}\,g_2(x, Q^{2}) 
           \right\}
         \label{gdef} .
\eea
The symmetric (antisymmetric) part in $\mu\nu$ is relevant
to the unpolarized (polarized) scattering; in (\ref{gdef}),
the decomposition of the spin-independent part
is universal for an arbitrary hadronic target,
while that of the spin-dependent part is specific for the spin
1/2 target.
The structure functions are dimensionless functions 
of two invariants, $Q^{2}$ and the Bjorken variable $x = Q^{2}/(2P\cdot q)$.
Equation~(\ref{ht}) can be expressed by the ``cut diagram'' corresponding to  
the discontinuity of the forward virtual Compton amplitude
between the virtual photon and a hadron. 
In general kinematics, this virtual Compton amplitude
contains all the complicated interactions
between the virtual photon and a hadron, 
including various ``soft interactions'' where soft
momenta are exchanged.
However, a drastic simplification occurs if one 
goes to the Bjorken limit $Q^{2} \rightarrow \infty$ with $x$ fixed:
the amplitude is dominated by the contribution which 
is factorized into the short- and long-distance parts
(see Fig.~1), and other 
complicated contributions are suppressed by the powers of $1/Q$.

%%%%%%%%%%%%%%%%%%%%%%%
\begin{figure}[H]
\begin{center}
\begin{tabular}{cc}
\leavevmode\psfig{file=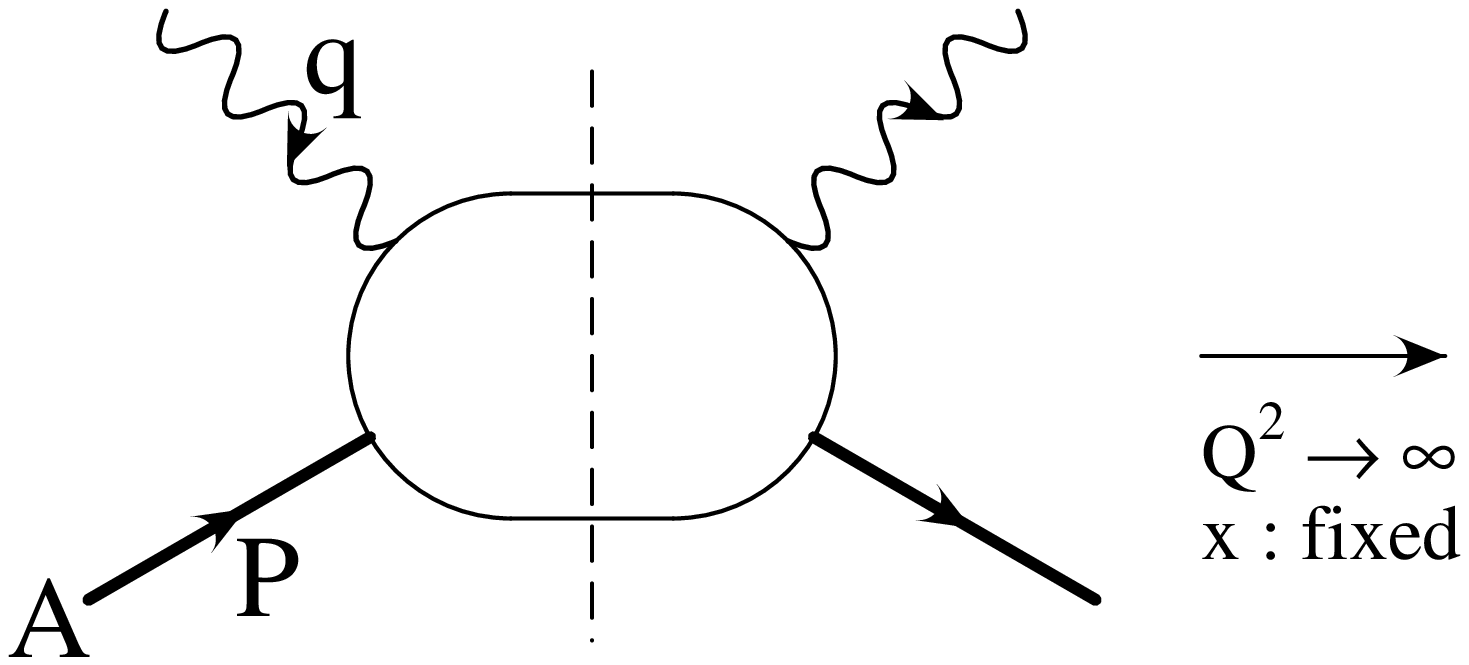,height=3cm} &
\leavevmode\psfig{file=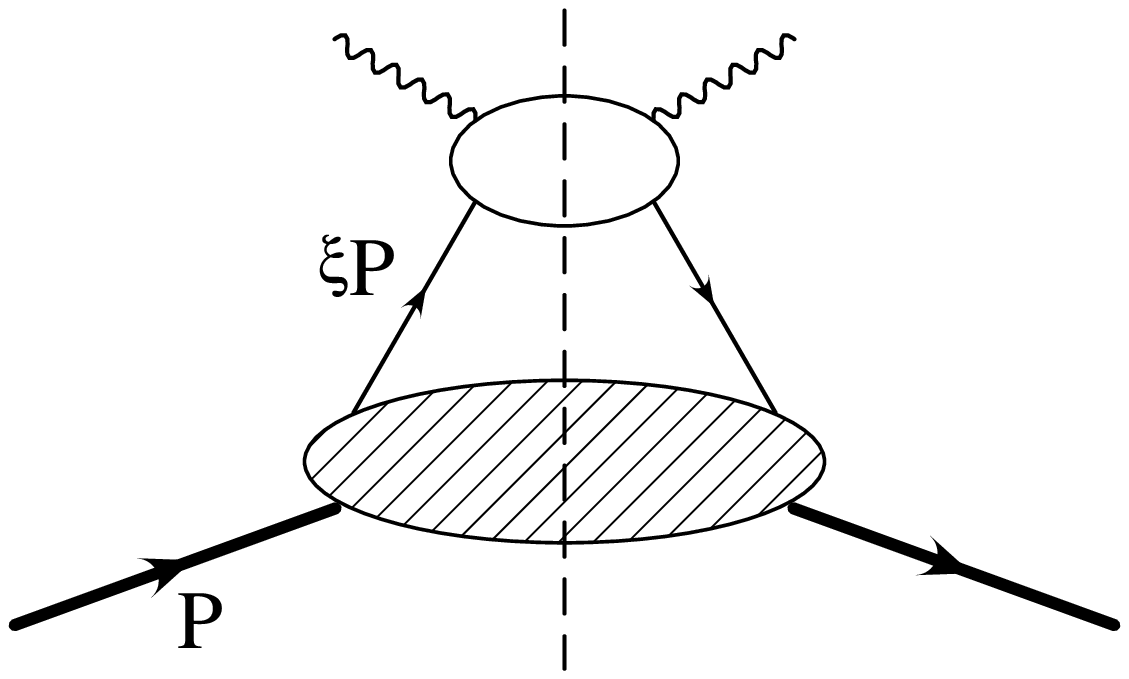,width=5cm} 
\end{tabular}

\caption{The DIS process and the Bjorken limit.}
\end{center}
\end{figure}
%%%%%%%%%%%%%%%%%%%%%%%
\vspace{-0.5cm}

The factorized amplitude corresponds to the process 
in which a parton carrying the momentum $\xi P$
($0 \leq \xi \leq 1$) comes out from the long-distance part, 
followed by the hard impact of the virtual photon,
and then goes back to the long-distance part. 
As a result, the structure functions
$F_{1}(x, Q^{2})$ and $F_{2}(x, Q^{2})$,
which give the unpolarized DIS cross section 
up to kinematical factors, assumes the factorized form
\begin{equation}
  F_{2}(x, Q^{2}) = \sum_{i} 
  \int_{x}^{1} \frac{d\xi}{\xi} \xi 
  f_{i/A}(\xi, \mu^{2}) H_{i}\left(\frac{x}{\xi},
     \frac{Q^{2}}{\mu^{2}}, \alpha_{s}(\mu^{2})\right)\ ,
\label{eq:f2}
\end{equation}
and similarly for $F_{1}(x, Q^{2})$.
The quantity $f_{i/A}(\xi, \mu^{2})$ is the parton distribution 
function corresponding to the long-distance part, and
it is interpreted as the probability density of 
finding a parton of type $i$ ($={\rm gluon},
u, \bar{u}, d, \bar{d}, \cdots$) in a hadron $A$,
carrying a fraction $\xi$ of the hadron's momentum.
The summation of (\ref{eq:f2}) is over all possible types of partons, $i$.
$H_{i}$ denotes the short-distance part;
it corresponds to the hard scattering cross section 
between the virtual photon and a parton $i$,
and is given by a power series in 
$\alpha_{s}=g^{2}/4\pi$ ($g$ is the QCD coupling constant)
with finite coefficients.

Next, we consider the DY process.~\cite{DY70}
The cross section for this process is again written in terms of
the leptonic and hadronic tensor,
\[   d \sigma  = \frac{2}{\sqrt{s (s - 4M^2 )}}
          \left(\frac{e^{2}}{Q^2}\right)^{2} 
      {\cal L}^{\mu\nu} {\cal W}_{\mu \nu}\,
       \frac{d^3 k_1}{(2\pi )^3 \,2 k_1^0} \,
         \frac{d^3 k_2}{(2\pi )^3 \,2 k_2^0}\ ,\]
where $k_i$ is the momentum of the produced lepton,
$Q^2 \equiv q^2 = (k_1 + k_2 )^2$ and $s \equiv (P_{A} + P_{B})^{2}$. 
The hadronic tensor is given by~\cite{rv,DY70}
\begin{equation}
  {\cal W}_{\mu \nu} = \int d^{4}z e^{-iq\cdot z}
     \langle P_{A}S_{A}, P_{B}S_{B}|
   J_{\mu}(z) J_{\nu}(0) |P_{A}S_{A}, P_{B}S_{B} \rangle\ ,
\label{eq:WDY}
\end{equation}
where $|P_{A}S_{A}, P_{B}S_{B} \rangle$ is the 
in-state of two hadrons $A$ and $B$.
$g^{\mu \nu}{\cal W}_{\mu \nu}$ determines
the DY cross section with unpolarized hadron-hadron collision, while
the other components of ${\cal W}_{\mu \nu}$ are relevant 
to polarized collisions.
Equation~(\ref{eq:WDY})
is expressed by the cut of the forward scattering amplitude 
between $A$ and $B$ (see Fig.~2).
In the Bjorken limit $Q^{2} \rightarrow \infty$
with $\tau = x_{A}x_{B} = Q^{2}/s$ fixed, 
the amplitude is dominated by the contribution factorized 
into short- and long-distance parts;
other contributions are suppressed by the powers of $1/Q$.

%%%%%%%%%%%%%%%%%%%%%%%%%%%%%%%%%%%%%%%%%%%%%%
\begin{minipage}{8.5cm}
\begin{figure}[H]
\begin{flushright}
\leavevmode\psfig{file=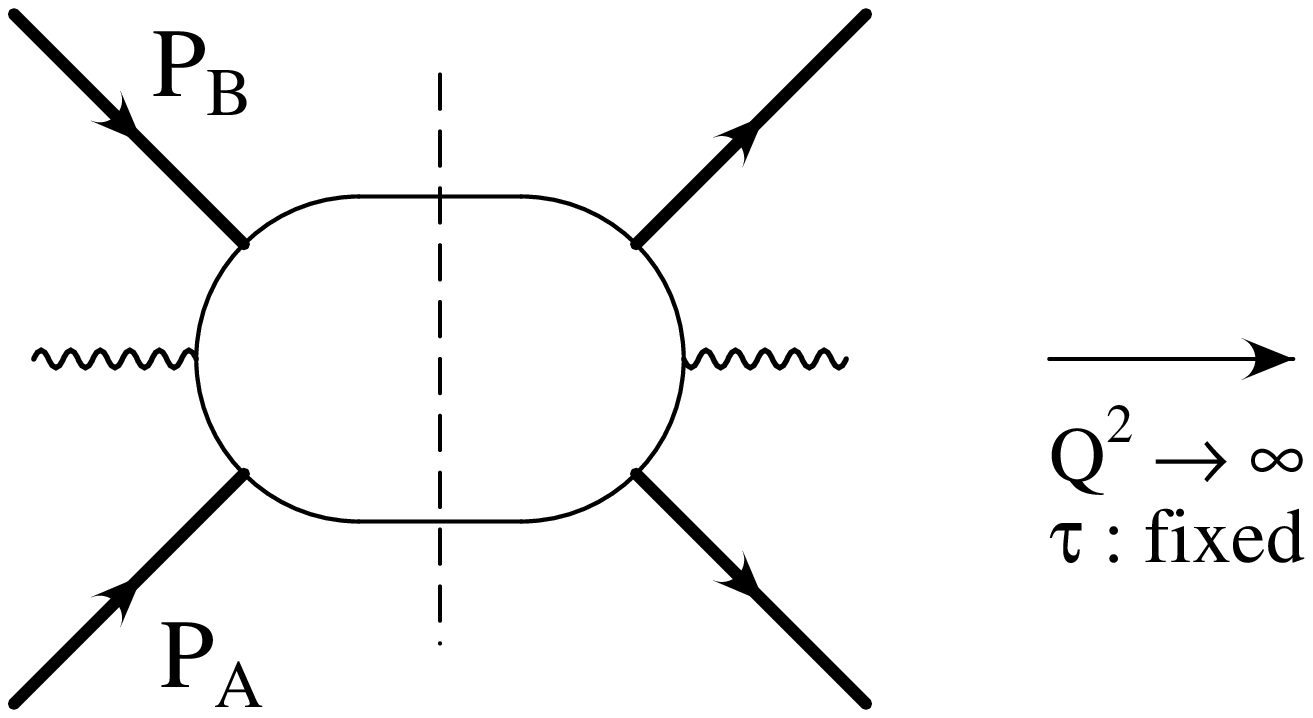,height=3.3cm}
\end{flushright}
\end{figure}
\end{minipage}
\begin{minipage}{6.5cm}
\begin{figure}[H]
\begin{flushleft}
\leavevmode\psfig{file=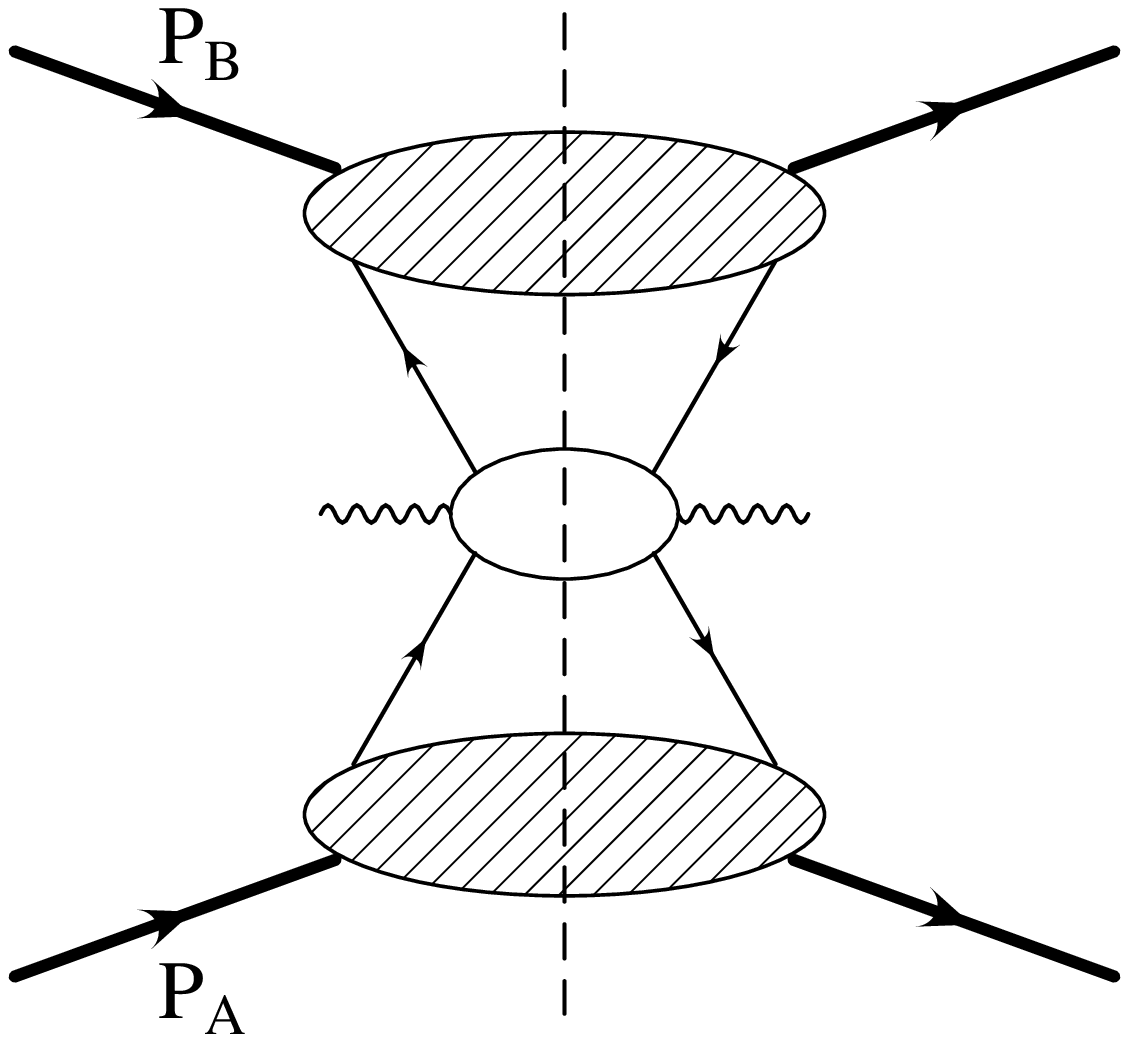,width=4.5cm} 
\end{flushleft}
\end{figure}
\end{minipage}
\begin{center}
Fig. 2. \quad The DY process and the Bjorken limit.
\end{center}
%%%%%%%%%%%%%%%%%%%%%%%%%%%%%
The factorized amplitude corresponds to the process 
in which partons carrying the momenta $\xi_{A} P_{A}$ and $\xi_{B} P_{B}$ 
come out from the lower and the upper long-distance parts, 
followed by the hard scattering between them, and then 
go back to the long-distance parts (see Fig.~2).
Then, the cross section for the unpolarized DY, accurate up to 
corrections suppressed by the powers of $1/Q$, is given by
\begin{equation}
\frac{d^{2}\sigma}{dQ^{2} d y} 
 \sim \sum_{i, j}\int^{1}_{x_{A}}d\xi_{A}\int^{1}_{x_{B}}
    d\xi_{B} f_{i/A}(\xi_{A}, \mu) 
    H_{ij}\left(\frac{x_{A}}{\xi_{A}},\frac{x_{B}}{\xi_{B}},
    Q, \frac{Q}{\mu}, \alpha_{s}(\mu^{2}) \right) f_{j/B}(\xi_{B}, \mu)\ ,
\label{eq:do}
\end{equation}
where $x_A / x_B = e^{2y}$.
Here $H_{ij}$ is the hard-scattering coefficient corresponding to
the short-distance interaction between the 
partons $i$ and $j$.
The long-distance parts appearing in Fig.~2
have exactly the same structure as that appearing in Fig.~1,
and therefore $f_{i/A}$ and $f_{j/B}$ are the same 
as in (\ref{eq:f2}).
This demonstrates the universal nature 
of the long-distance part: it is determined completely if
one specifies the target. 
Though the long-distance part is not calculable
using the perturbation theory, 
it can be determined by experiments of some hard processes.
For example, if one extracts the parton distribution functions
for the nucleon target by DIS experiments, they can be used 
to describe the DY processes involving the nucleon.

The validity of (\ref{eq:f2}) and (\ref{eq:do}) can be
demonstrated by analyzing the Feynman diagrams
for the corresponding amplitude in the Bjorken limit.\cite{CSS}
(For the DIS, the equivalent result 
can be obtained by the OPE.\cite{JC2})
The central point in the proof is to show
that the long-distance contributions,
which appear as infrared-divergent contributions in the perturbative 
diagrams, can be completely factorized into universal hadron matrix
elements (parton distribution functions) or cancel out,
and thus the hard scattering coefficient has purely short-distance
contributions.
Extensions of the factorization theorems to hard processes
involving polarized beams and/or targets
have been discussed by many authors.\cite{rv,CO93}
The corresponding factorization formulae are 
similar to (\ref{eq:f2}) and (\ref{eq:do})
but involve various polarized parton distribution functions.
 
\subsection{Factorization of the DIS in the free field theory}

Once the factorization theorems are proved,
the parton distribution functions
can be studied separately from the short-distance part
and deserve detailed discussions because of their universal nature.
Before going into a systematic study of parton distribution functions,
it is instructive to demonstrate the DIS formulae like (\ref{eq:f2})
in free field theories, that is, neglecting the QCD interaction.
The result provides a guide to
operator definitions of the parton distributions.

Let us first recall some basic points,
which generally hold even in the presence of QCD interactions.
Here and below, we consider the nucleon's structure functions,
and $|PS\rangle$ denotes the nucleon state.
As mentioned above, the relevant quantity for the DIS
through one photon exchange is the hadronic tensor (\ref{ht}).
In the Bjorken limit, the hadronic tensor (\ref{ht}) is governed by
the behavior of the current products near the light-cone.
This {\it light-cone dominance} is easily understood from the following
observation.
We take the Lorentz frame in which the momenta of the nucleon
and the virtual photon are both in the $\hat{\mib{e}}_{z}$ 
direction, without loss of generality.
It is convenient to introduce two auxiliary light-like vectors,
\[  p^{\mu} \equiv \frac{{\cal P}}{\sqrt{2}} ( 1,0,0,1)\ , \quad 
           w^{\mu} \equiv \frac{1}{\sqrt{2} {\cal P}} ( 1, 0,0, -1) \ , \]
with $p^2 = w^2 = 0$ , $p \cdot w = 1$.
We have $P^{\mu} = p^{\mu} + \frac{M^{2}}{2}w^{\mu}$, and 
${\cal P} \equiv P^+$ is a parameter which selects a specific frame
in the direction of $\hat{\mib{e}}_{z}$.~\footnote{
The light-cone coordinates
$a^{\mu} = (a^{+}, a^{-}, \mib{a}_{\perp})$
are related to the usual coordinates as 
$a^{\pm}= (a^{0}\pm a^{3})/\sqrt{2}$, $\mib{a}_{\perp} = (a^{1}, a^{2})$,
so that $a \cdot b = a^{+}b^{-} + a^{-}b^{+} - \mib{a}_{\perp} 
\cdot \mib{b}_{\perp}$.
Note that $p^{\mu}$ ($w^{\mu}$) has only
$p^{+}$ ($w^{-}$) as nonzero components in these coordinates.}
In the Bjorken limit, $Q^{2} = -q^{2} \rightarrow \infty$ with
$x= Q^{2}/(2P\cdot q)$ fixed,
it is easy to derive
\[ \lim_{\rm Bj} q^{\mu} =
      \left( P\cdot q + \frac{1}{2} M^2 x \right) w^{\mu}
        - x p^{\mu} + {\cal O} \left( \frac{1}{Q^2} \right) \ .\]
Therefore, writing the coordinate $z^{\mu}$ as
$z^{\mu} = \eta p^{\mu} + \lambda w^{\mu} + z_{\perp}^{\mu}$, we obtain
$\lim_{\rm Bj}\,  q\cdot z \simeq P\cdot q \eta - \lambda x $.
By the Riemann-Lebesgue theorem, 
only the integration regions
$|q\cdot z | \,\mbox{\raisebox{-0.5ex}{$\stackrel{<}{\sim}$}}\, 1$, namely
$|\eta| \,\mbox{\raisebox{-0.5ex}{$\stackrel{<}{\sim}$}} \,1 / (P\cdot q)$ and 
$|\lambda| \, \mbox{\raisebox{-0.5ex}{$\stackrel{<}{\sim}$}}\, 1 / x$, 
contribute to (\ref{ht}). 
This is the light-cone dominance which implies that only the region
\begin{equation}
 0 \leq z^2 = 2 \eta\lambda - \mib{z}_{\perp}^2 \leq 2 |\eta\lambda|
       \le \frac{{\rm const}}{Q^2} 
\label{eq:lcd}
\end{equation}
is important in the Bjorken limit. 
The lower limit of $z^2$ results from causality.

Now we consider the free field theory.
For simplicity, we neglect the flavor structure of quarks
and also quark mass (we write the current as
$J_{\mu} = \bar{\psi}\gamma_{\mu}\psi$).
The current commutator in (\ref{ht}) can be 
easily worked out by using the free-field anticommutation 
relation:~\cite{BD65}
$\{\psi(0), \bar{\psi}(z)\} =-\sslash{\partial} \Delta(z)$, 
where $\Delta (z ) = - (1/2 \pi ) \varepsilon (z^0 ) \delta (z^2 )$
with $\varepsilon(z^{0}) = z^{0}/|z^{0}|$.
We obtain~\footnote{We keep only terms which contribute to the DIS,
omitting the $c$-number contributions from the nonlocal 
operators ${\cal U}^{\sigma}_{V,A}(0,z)$. Thus
${\cal U}^{\sigma}_{V,A}(0,z)$
should be understood to be normal ordered.}
\be
   [ J_{\mu} (0) \,,\, J_{\nu} (z) ] =
  - \, \left(\partial^{\alpha} \Delta (z )\right) 
    [ S_{\mu \alpha \nu \sigma} 
           {\cal U}_V^{\sigma} ( 0 , z) - i 
        \epsilon_{\mu\alpha\nu\sigma}
           {\cal U}_A^{\sigma} ( 0 , z) ]  \ ,\label{ccinfree}
\ee
with $S_{\mu \alpha \nu \sigma}\equiv  g_{\mu\alpha} g_{\nu\sigma}
- g_{\mu\nu} g_{\alpha\sigma} + g_{\mu\sigma} g_{\nu\alpha}$,
and the \lq\lq nonlocal (bilocal) operators\rq\rq\ defined by
\bea
  {\cal U}_V^{\sigma} (0 , z) &\equiv& 
     \bar{\psi} (0 ) \gamma^{\sigma} \psi (z)
              - \bar{\psi} (z) \gamma^{\sigma} \psi (0 ) \ , \label{bico0}\\
  {\cal U}_A^{\sigma} (0 , z) &\equiv& 
     \bar{\psi} (0) \gamma^{\sigma} \gamma^5 \psi (z)
   + \bar{\psi} (z) \gamma^{\sigma} \gamma^5 \psi (0) \ .\label{bico}
\eea
We take the nucleon matrix element of
(\ref{ccinfree}) and substitute it into (\ref{ht}).
In accord with the general argument (\ref{eq:lcd}),
$\Delta(z)$ of (\ref{ccinfree}) selects
the integration region with $\eta \sim z_{\perp} \sim 0$
in the Bjorken limit.
Therefore, we can expand the matrix element in powers of the deviation 
from the light-cone, $z^{\mu} -\lambda w^{\mu}$, and approximate it as
$\langle P S | {\cal U}_{V,A}^{\sigma} (0 ,z) | P S \rangle \simeq
    \langle P S | {\cal U}_{V,A}^{\sigma} (0 , \lambda w ) 
| P S \rangle$
(the neglected higher-order terms
give contributions suppressed by powers of $1/Q$ after the integration).

Let us consider the contribution of ${\cal U}_V^{\sigma} (0, \lambda w)$
in detail: 
the matrix element corresponding to the first term of (\ref{bico0}) 
can be parameterized as
\begin{equation}
 \langle PS | \bar{\psi} (0 ) \gamma^{\sigma} \psi (\lambda w )| PS \rangle
    = 2 p^{\sigma} \hat{q} (\lambda )+ 2 w^{\sigma} M^2 \hat{f}_4 (\lambda ) 
   \ . \label{eq:decomp}
\end{equation}
The second term has a factor $M^2$ to cancel the
dimensionality of $w^{\mu}$. 
Here, let us use the term \lq\lq twist\rq\rq\ in the following sense:
when a piece of the operator matrix element
contributes to the hard processes
at order $(1/Q)^{t-2}$, such a piece is said to have twist-$t$. 
Actually, this definition of the twist based on the power counting of $1/Q$
has a slight mismatch with the conventional definition
as ``dimension minus spin'' of the relevant operators.~\footnote{
The mismatch gives rise to the 
``Wandzura-Wilczek parts''\cite{WW} for twist-3 contributions
(see \S\S\ref{sec:ce} and \ref{sec:co}).}
However, it is useful to classify 
the operator contribution to the cross sections with respect to
the power dependence in $1/Q$.
In the above example, the first term is the twist-2 contribution,
and the second term is the twist-4 contribution, since the factor $M^2$
should be compensated dimensionally
as $M^{2}/Q^2$ in the physical structure functions.

We insert (\ref{ccinfree}) and (\ref{eq:decomp}) into (\ref{ht}),
and compute the corresponding contribution to $W_{\mu\nu}$
to leading order in $1/Q$.
By introducing the Fourier transformations
\bean
   \Delta(z) &=& \int \frac{d^{4}k}{(2 \pi)^{4}}e^{ik\cdot z} \left[
   2 \pi i
   \varepsilon(k^{0}) \delta(k^{2}) \right] \ , \\
   \hat{q}(\lambda) &=& \int d\xi e^{- i \lambda \xi} q(\xi)
     = \int d\xi e^{-i p \cdot z \xi} q(\xi) \ ,
\eean
it is straightforward to obtain
\begin{equation}
W_{\mu \nu} = 2 S_{\mu \alpha \nu \sigma} p^{\sigma}
          \int d\xi (q + \xi p)^{\alpha}
                \varepsilon(q^{0} + \xi p^{0}) \delta ( (q+\xi p)^{2} )
        \left[ q(\xi) - q(-\xi)\right] + \cdots \ , \label{eq:Wget1}
\end{equation}
where the dots stand for the contributions
from $\hat{f}_{4}(\lambda)$ and 
$i\epsilon_{\mu \alpha \nu \sigma} \langle PS|
{\cal U}_{A}^{\sigma}(0, \lambda w)|PS \rangle$; 
those from $\hat{f}_{4}(\lambda)$
produce the contribution suppressed by $M^{2}/Q^{2}$, as noted above,
and thus are irrelevant here.
On the other hand, those
from $i\epsilon_{\mu \alpha \nu \sigma} \langle PS|
{\cal U}_{A}^{\sigma}(0, \lambda w)|PS \rangle$
are relevant to the antisymmetric part of $W_{\mu \nu}$.
Noting that $(q + \xi p)^{2} = q^{2} + 2 \xi p\cdot q
=  2P\cdot q( \xi - x) + {\cal O}(M^{2}/Q^{2})$, we obtain
\begin{equation}
W_{\mu \nu} = \left(-g_{\mu \nu} + 
         \frac{P_{\mu} (q + xP)_{\nu}}{P\cdot q} 
         + \frac{P_{\nu} (q+xP)_{\mu}}{P\cdot q}
          \right) \int d\xi\delta(\xi - x)
        \left[ q(\xi) - q(-\xi)\right] + \cdots \ . \label{eq:Wget2}
\end{equation}
Comparing this result with the general expression (\ref{gdef}),
one obtains for the unpolarized structure functions,
\begin{equation}
  F_1 (x) = \frac{1}{x}F_{2}(x)
   = \int\frac{d\xi}{\xi}\delta (x/\xi -1) \left[
   q(\xi) - q(-\xi)\right] = q (x) - q(-x)  \ , \label{eq:f1free}
\end{equation}
with
\begin{equation}
q(x) = \int \frac{d\lambda}{4\pi} e^{i\lambda x}
  \langle P S| \bar{\psi} (0)\sslash{w}\, \psi (\lambda w)| P S\rangle \ 
              , \label{eq:f1free2}
\end{equation}
up to corrections of ${\cal O}(M^{2}/Q^{2})$.
We see that $q(\xi)$ corresponds to a parton distribution
$f_{i/A}(\xi, \mu^{2})$ of (\ref{eq:f2}), while
$H_{i}(x/\xi, Q^{2}/\mu^{2}, g(\mu^{2})) \rightarrow \delta(x/\xi-1)$
in the present free case.
As discussed in \S2.3 below,
$q(x)$ and $- q(-x)$ give the quark
and the anti-quark distribution functions, respectively.
Similarly, it is straightforward to see that the 
terms corresponding to the dots in (\ref{eq:Wget2})
give results for the polarized structure functions
of (\ref{gdef}) as~\footnote{The factor 1/2 on the
r.h.s. comes from our definition of $g_i$ in (\ref{gdef}).}
\begin{equation}
g_{1}(x) = \frac{1}{2}\left(\Delta q(x) + \Delta q (-x)\right),
   \quad g_{1}(x) + g_{2}(x) = 
  \frac{1}{2}\left(g_{T}(x) + g_{T}(-x)\right) \ , \label{eq:parton}
\end{equation}
to ${\cal O}(M^{2}/Q^{2})$ accuracy. The distribution functions 
$\Delta q(x)$ and $g_{T}(x)$ are 
defined similarly to (\ref{eq:f1free2}), and their
explicit forms are given in (\ref{eq:axialv}) and (\ref{eq:odg}) below.

Equations (\ref{ccinfree}) and (\ref{eq:Wget1}) demonstrate that the
singular function $\partial^{\alpha} \Delta (z)$ corresponds to
the short-distance part containing all the dependence
on $q^{\mu}$, while the matrix element of the nonlocal operators
corresponds to the long-distance part yielding parton 
distribution functions.
Roughly speaking,
the above results are modified in two respects
in the presence of the QCD interaction.
First, the higher order interactions
produce the logarithmic ($\ln(Q^{2}/\mu^{2})$)
corrections to $\Delta (z)$, and correspondingly
the parton distribution functions acquire a dependence on the
renormalization scale $\mu$.
Second, the coupling of the ``longitudinal'' gluons replaces
$\Delta(z)$ as~\cite{GT71,BB}
\begin{displaymath}
  \Delta (z) \rightarrow \Delta(z)[0, z] \ ,
\end{displaymath}
where
\[    [y,z]= {\rm P} \exp \left(ig\!\!\int_0^1\!\! dt\,
      (y-z)_\mu A^\mu(ty+(1-t)z) \right)\  \]
is the path-ordered gauge phase factor along the
straight line connecting the points $z_{\mu}$ and $y_{\mu}$.
By absorbing this factor,
the nonlocal light-cone operators which define the parton
distribution functions now preserve gauge invariance.

Extension to 3 flavors with the current (\ref{eq:current})
is straightforward.
The quark charge matrix squared,
${{\cal Q}^{(el)}}^{2}= \frac{2}{9} \mbox{\boldmath$1$} +\frac{1}{6} \lambda^{3}
+ \frac{1}{6 \sqrt{3}}\lambda^{8}$,
should be inserted between the quark fields
of (\ref{bico0}) -- (\ref{eq:decomp}) and (\ref{eq:f1free2}).
The long-distance part is, therefore, decomposed
into singlet ({\boldmath$1$}) and non-singlet ($\lambda^{3,8}$) parts. 

\subsection{Classification of the quark distribution functions}
\label{sec:qc}

Now, it is not difficult to extend the above logic in the free case
to the quark distributions in the presence of QCD interaction.~\cite{CS} 
We consider the following quantity involving the nucleon matrix element:
\be
 \int_{-\infty}^{+\infty} \frac{d\lambda}{2\pi} e^{i\lambda x} 
        \langle P S|
       \bar{\psi}(0) [0,\lambda w] \Gamma \psi(\lambda w) |P S \rangle 
    \ . \label{eq:gam}
\ee
Here $\Gamma$ is a generic Dirac matrix, and
the gauge phase factor $[0, \lambda w]$
makes the operators gauge invariant.
Equation (\ref{eq:gam}) defines the distribution function
for a quark with momentum $k^{+}=xP^{+}$
(recall $\lambda x = (xP)\cdot (\lambda w)$).
It is understood that the nonlocal light-cone operator in 
(\ref{eq:gam}) is renormalized at the scale $\mu$.

$\Gamma$ can be any Dirac matrix, depending on which hard process is considered. 
An important observation made by Jaffe and Ji\cite{JJ}
is that one can generate all quark distribution functions 
up to twist-4 by substituting all the possible Dirac 
matrices for $\Gamma$,
although the operator definition (\ref{eq:gam})
is motivated by the factorization at the leading twist.
By decomposing (\ref{eq:gam}) into independent tensor 
structures, one finds nine independent quark distribution functions
associating with each tensor structure:
\begin{eqnarray}
\int { d\lambda  \over 2 \pi } e^{i\lambda x} \langle PS |
 \bar{\psi}(0) [0, \lambda w]\gamma^{\mu} 
  \psi(\lambda w) |PS \rangle 
   &=& 2 \left[ q(x,\mu^2) p^{\mu} 
         + f_{4}(x,\mu^2)M^{2}w^{\mu} \right],\nonumber\\
      \label{eq:vector}\\
\int { d\lambda  \over 2 \pi } e^{i\lambda x} \langle PS |
  \bar{\psi}(0) [0, \lambda w]\gamma^{\mu}\gamma_5 
  \psi(\lambda w) |PS \rangle
 &=& 2 \left[ \Delta q(x,\mu^2) p^{\mu}(S\cdot w)
   + g_{T}(x, \mu^2) S_{\perp}^{\mu} \right. \nonumber \\
  &+& \left. g_{3}(x,\mu^2)M^{2}w^{\mu}(S\cdot w) \right], \label{eq:axialv}\\
\int { d\lambda  \over 2 \pi } e^{i\lambda x} \langle PS |
  \bar{\psi}(0) [0, \lambda w]\sigma^{\mu\nu}i\gamma_5 
  \psi(\lambda w) |PS \rangle 
  &=& 2 \left[ \delta q(x,\mu^2) ( S_{\perp}^{\mu}p^\nu - 
  S_{\perp}^{\nu}p^\mu )/M \right. \nonumber \\
  + h_{L}(x, \mu^2)M(p^\mu w^\nu - p^\nu w^\mu )(S \cdot w)
   &+& \left.h_{3}(x,\mu^2)M
   (S_{\perp}^{\mu} w^\nu -S_{\perp}^{\nu} w^\mu) \right], \nonumber\\
                 \label{eq201}\\
\int { d\lambda  \over 2 \pi } e^{i\lambda x} \langle PS |
  \bar{\psi}(0)[0, \lambda w] \psi(\lambda w) |PS \rangle 
  &=&  2\ M e(x, \mu^{2}), \label{eq:scalar}
\end{eqnarray}
where we have written
$S^{\mu} =  S_{\parallel}^{\mu} + S_{\perp}^{\mu}$ with
$S_{\parallel}^{\mu} = (S \cdot w)\, p^{\mu} + (S \cdot p)\, w^{\mu}$.
The spin vector $S_{\mu}$ can only occur linearly
in the decomposition, because the dependence 
on $S_{\mu}$ is determined by the density matrix
$(1 + \gamma_{5}\rlap/{\mkern-1mu S}/M)/2$
(recall that we normalize $S_{\mu}$ so that $S^{2} = - M^{2}$).

We suppressed the flavor structure in the above definitions.
When the quark fields have a definite flavor
as $\psi \rightarrow \psi_{f}$ ($f = u,d,s$),
the distribution functions on the r.h.s. of
(\ref{eq:vector}) -- (\ref{eq:scalar}) should be
understood as those for the corresponding flavor :
$q^{f}(x, \mu^{2})$, $\Delta q^{f}(x, \mu^{2})$, $\delta q^{f}(x, \mu^{2})$,
$g_{T}^{f}(x, \mu^{2})$, $h_{L}^{f}(x, \mu^{2})$, etc.
Alternatively, we may insert the flavor matrices $\lambda^{i}$
or the unit matrix {\boldmath$1$} in between the quark fields; 
then the distribution functions on the r.h.s. will be
understood as those for the non-singlet or singlet part.

These quark distributions are dimensionless functions
of the Bjorken variable $x$; they depend on the renormalization 
scale $\mu$ as well because the nonlocal light-cone operators on
\begin{wraptable}{l}{\halftext}
\caption{Spin, twist and chiral classification of the
the quark distribution functions.}
\label{tab:1}
\begin{center}
\begin{tabular}{c|ccc}\hline \hline
Twist    & 2 & 3 & 4 \\
    & ${\cal O}(1)$  & ${\cal O}(1/Q)$& ${\cal O}(1/Q^{2})$ \\ \hline
spin ave.& $q(x)$ & $e(x)^{\displaystyle{\star}}$ & $f_{4}(x)$ \\
$S_{\parallel}$ & $\Delta q(x)$ & $h_{L}(x)^{\displaystyle{\star}}$ 
& $g_{3}(x)$ \\
$S_{\perp}$ & $\delta q(x)^{\displaystyle{\star}}$ & $g_{T}(x)$ 
& $h_{3}(x)^{\displaystyle{\star}}$\\
\hline
\end{tabular}
\end{center}
\end{wraptable}
the l.h.s. are renormalized at $\mu$.
Their spin, twist and chiral classifications 
are listed in Table \ref{tab:1}.
The distributions in the first row are spin-independent, 
while those in the second and third rows correspond to the 
longitudinally ($S_{\parallel}$) and transversely ($S_{\perp}$)
polarized nucleons.
Each column refers to the twist.
The distributions marked with ``$\star$'' 
are referred to as chiral-odd, because they correspond to
chirality-violating Dirac matrix structures
$\Gamma = \{\sigma_{\mu \nu} i \gamma_{5},\, 1\}$.~\footnote{
Equation (\ref{eq:gam}) vanishes for $\Gamma = i \gamma_{5}$
due to the time-reversal invariance.}
The other distributions are chiral-even, because of the
chirality-conserving structures
$\Gamma = \{\gamma_{\mu},\, \gamma_{\mu}\gamma_{5}\}$.

In the massless quark limit, chirality
is conserved through the propagation of a quark. This means that
in the DIS one can measure only the chiral-even 
distributions up to tiny quark mass corrections,
because the perturbative quark-gluon 
and quark-photon couplings conserve chirality.
On the other hand, in the DY and certain other  
processes, both chiral-odd and chiral-even distribution
functions can be measured,\cite{RS} because the chiralities of the 
quark lines originating in a single nucleon 
are uncorrelated (see Fig.~2).

One convenient way to understand the twist classification
of distribution functions directly from their definitions
is to go over to the infinite momentum frame
$P^+ \sim Q \rightarrow \infty$, so that
$S \cdot w \sim 1$ and $S_{\perp} \sim M$.
This determines  the power counting in $Q$
of all terms on the r.h.s. of (\ref{eq:vector}) -- (\ref{eq:scalar}).
The first, second, and third terms in
(\ref{eq:axialv}) and (\ref{eq201}) behave as ${\cal O}(Q)$,
${\cal O}(1)$ and ${\cal O}(1/Q)$, respectively, and the terms in
(\ref{eq:vector}) and (\ref{eq:scalar}) are 
of ${\cal O}(Q)$, ${\cal O}(1/Q)$ and of ${\cal O}(1)$.

A conceptually different approach to a similar
twist counting is based on the light-cone quantization
formalism.\cite{JJ,KS} 
In this approach quark fields are decomposed into  ``good''
and ``bad'' components,
so that $\psi = \psi_{+} + \psi_{-}$ with
$\psi_{+} = \frac{1}{2}\gamma^{-}\gamma^{+} \psi$ and
$\psi_{-} = \frac{1}{2}\gamma^{+}\gamma^{-} \psi$, where
$\gamma^{\pm} = (\gamma^{0} \pm \gamma^{3})/\sqrt{2}$.
As discussed in Ref.~\citen{JJ}, a ``bad'' component $\psi_{-}$
introduces one unit of twist.
Therefore, a quark bilinear $\bar{\psi}\Gamma \psi$
contains twist-2 ($\bar{\psi}_{+} \Gamma \psi_{+}$), 
twist-3 ($\bar{\psi}_{+}\Gamma \psi_{-}+ \bar{\psi}_{-}\Gamma \psi_{+}$)
and twist-4 ($\bar{\psi}_{-}\Gamma \psi_{-}$) contributions.
This also explains why a unique distribution function is defined
for each twist and polarization (see Table \ref{tab:1}):
in each of these quark bilinears corresponding to twist-2, -3 and -4,
the total spin of quark pairs can be 0 or 1.
By taking the nucleon matrix element,
the former describes a spin-averaged nucleon state, while
the latter describes a longitudinally and transversely polarized 
state depending on the total helicity.
The physical meaning of this classification is that
a ``good'' component $\psi_{+}$ represents an independent
degree of freedom.
On the other hand, the ``bad'' components are not dynamically independent
and can be reexpressed by a coherent quark-gluon pair.\cite{KS}
Only the twist-2 distribution functions are literally
the ``distributions'', which simply
count the number of independent degrees of freedom
having a definite quantum number (flavor, helicity, 
etc.), and correspond to the parton model.
In particular, by comparing both sides of 
(\ref{eq:vector}) -- (\ref{eq201}), we obtain 
\begin{eqnarray}
q(x, \mu^{2}) &=& 
   \int \frac{d\lambda}{4\pi} 
     e^{i\lambda x} \langle P S|
    \bar{\psi}(0) [0, \lambda w]\rlap/{\mkern-1mu w} 
     \psi(\lambda w) |P S \rangle,
            \label{eq:odf} \\
\Delta q(x, \mu^{2}) &=&
  \int \frac{d\lambda}{4\pi} 
  e^{i\lambda x} \langle P S_{\parallel}|
  \bar{\psi}(0) [0, \lambda w]\rlap/{\mkern-1mu w} \gamma_{5}
  \psi(\lambda w) |P S_{\parallel} \rangle,
  \label{eq:odg} \\
\delta q(x, \mu^{2}) &=& 
  \frac{1}{M}\int \frac{d\lambda}{4\pi} 
 e^{i\lambda x} \langle P S_{\perp}|
 \bar{\psi}(0) [0, \lambda w]\rlap/{\mkern-1mu w} 
 \gamma_{5}\rlap/{\mkern-1mu S}_{\perp}
 \psi(\lambda w) |P S_{\perp} \rangle,
  \label{eq:odh}
\end{eqnarray}
where $|P S_{\parallel} \rangle$ ($|P S_{\perp} \rangle$) denotes
the nucleon's spin being in the helicity
($S_{\parallel} \cdot w =1$) state (in the transverse direction).
Because the bilinear operator 
$\bar{\psi}\rlap/{\mkern-1mu w}\psi =(1/P^{+})\bar{\psi} \gamma^{+}
\psi = (1/P^{+})\psi^{\dagger}_{+} \psi_{+}$ 
corresponds to the number density operator in the light-cone formalism,
$q(x)$ gives the total number density of a quark with the momentum fraction
$x$ of the parent nucleon.
Due to the additional $\gamma_{5}$ between the quark fields,
$\Delta q(x)$ is the helicity distribution as
$\Delta q(x) = q_{\uparrow}(x) - q_{\downarrow}(x)$,
where $q_{\uparrow (\downarrow)}(x)$ is the number density of a quark
with helicity parallel (antiparallel) to the nucleon's spin.
If one recalls that $\Sigma (S) = (1 + \gamma_{5}
\rlap/{\mkern-1mu S}/M)/2$ is the projection operator 
for a Dirac particle with the polarization $S_{\mu}$, one recognizes 
$\delta q(x) = q_{\leftarrow}(x)
- q_{\rightarrow}(x)$, where $q_{\leftarrow (\rightarrow)}(x)$
is defined similarly to $q_{\uparrow (\downarrow)}(x)$,
except that its polarization is defined in reference to
the transverse direction.
We call $\delta q(x)$ the ``transversity distribution'',
following Ref.~\citen{JJ}.
On the other hand, the higher twist distributions are the multiparton 
(quark-gluon) correlation
functions which contain information beyond the parton model.\cite{JJ,JA}
This point is discussed in detail in \S\S\ref{sec:ce} and \ref{sec:co}.

Here it is worth noting some basic properties
of the distributions functions:
(i) By inserting a complete set of the states
$\sum_{n} |n\rangle \langle n |$ between the quark fields
of (\ref{eq:vector})--(\ref{eq:scalar}),
and using the positivity
$P_{n}^{+} = (P_{n}^{0} + P_{n}^{3})/\sqrt{2} >0$ 
for intermediate state $|n\rangle$, 
we obtain for all nine distributions
$\phi=  \{q, \Delta q, \delta q, e, h_{L}, 
g_{T}, f_{4}, g_{3}, h_{3}\}$: 
\begin{equation}
  \phi(x) = 0 \quad , \quad |x| > 1. \label{eq:region}
\end{equation}
(ii) The antiquark distributions $q^{\bar{f}}(x), \Delta q^{\bar{f}}(x)$, 
$h_{L}^{\bar{f}}(x)$, $g_{T}^{\bar{f}}(x)$, etc.,
are defined by (\ref{eq:vector})--(\ref{eq:scalar}) 
for a flavor $f$ with
the nonlocal light-cone operators transformed by the charge conjugation.
We find, for 
$\phi^{f} = \{\Delta q^{f}, g_{T}^{f}, g_{3}^{f}, e^{f}\}$
and 
$\chi^{f} = \{ q^{f}, f_{4}^{f}, \delta q^{f}, h_{L}^{f}, h_{3}^{f}\}$,
\begin{equation}
 \phi^{f}(-x) = \phi^{\bar{f}}(x)\quad , \quad 
 \chi^{f}(-x) = - \chi^{\bar{f}}(x). \label{eq:charge}
\end{equation}
(iii) From the positivity on 
$q^{f}_{\uparrow,\downarrow}(x), q^{f}_{\leftarrow,\rightarrow}(x)$, 
and the formula
$q^{f}(x) = q^{f}_{\uparrow}(x) + q^{f}_{\downarrow}(x)=
q^{f}_{\leftarrow}(x) + q^{f}_{\rightarrow}(x)$,
we obtain the trivial inequalities $| \Delta q^{f}(x)| \leq q^{f}(x)$,
$| \delta q^{f}(x)| \leq q^{f}(x)$.

\subsection{Classification of the gluon distributions}
\label{sec:gldis}

Our analysis can be extended to the gluon distribution functions.
The relevant cut diagram is given by Figs.~1 and 2, with the 
gluon lines connecting the short and long distance parts.
The gauge-invariant definition of the gluon distribution functions
is provided by~\cite{CS,mano,ji}
\begin{eqnarray}
\lefteqn{\frac{2}{x} \int \frac{d\lambda}{2\pi} 
  e^{i\lambda x}
   \langle PS| {\rm tr}\, w_{\alpha} G^{\alpha \mu}(0) [0, \lambda w]
  w_{\beta} G^{\beta \nu}(\lambda w) |PS \rangle} \nonumber \\
& & \qquad\qquad\quad = \, - \frac{1}{2} {\cal G}(x, \mu^{2}) g_{\perp}^{\mu \nu
}
  - \frac{1}{2} \Delta {\cal G}(x, \mu^{2})
  i \epsilon^{\mu \nu \alpha \beta}p_{\alpha}w_{\beta} (S \cdot w)\nonumber\\ 
& & \qquad\qquad\qquad  - \, {\cal G}_{3T}(x, \mu^{2})  
   i \epsilon^{\mu \nu \alpha \beta} S_{\perp \alpha}w_{\beta}
    + {\cal G}_{4} (x, \mu^{2}) M^{2} w^{\mu} w^{\nu},
\label{eq:gphde}
\end{eqnarray}
where 
$G^{\mu \nu}=\partial^{\mu}A^{\nu}-\partial^{\nu}A^{\mu}
-ig [A^{\mu}, A^{\nu}]$ 
is the gluon field strength tensor, ``${\rm tr}$'' refers to the
color matrices, and $g_{\perp}^{\mu \nu} = g^{\mu \nu}
- p^{\mu}w^{\nu}-p^{\nu} w^{\mu}$
is the projector onto the transverse direction.
\begin{wraptable}{l}{\halftext}
\caption{Spin and twist classification of the
the gluon distribution functions.}
\label{tab:2}
\begin{center}
\begin{tabular}{c|ccc}\hline \hline
Twist    & 2 & 3 & 4 \\
    & ${\cal O}(1)$  & ${\cal O}(1/Q)$& ${\cal O}(1/Q^{2})$ \\ \hline
spin ave.& ${\cal G}(x)$ &  & ${\cal G}_{4}(x)$ \\
$S_{\parallel}$ & $\Delta {\cal G}(x)$ & &  \\
$S_{\perp}$ &  & ${\cal G}_{3T}(x)$ & \\ \hline
\end{tabular}
\end{center}
\end{wraptable}
The form of l.h.s. is motivated by the fact that, in the light-cone  
quantization with the $w \cdot A = 0$ gauge,
$w_{\alpha}G^{\alpha \mu} = (1/P^{+})\partial^{+}A^{\mu}$.
The r.h.s. defines four gluon distributions
corresponding to independent tensor structures.

The spin and twist classifications can be inferred
directly from (\ref{eq:gphde}) as in \S\ref{sec:qc},
and are shown in Table~\ref{tab:2}.
The distributions
${\cal G}(x), \Delta{\cal G}(x), {\cal G}_{3T}(x)$ and
${\cal G}_{4}(x)$ mix through renormalization
with the flavor singlet parts of $q(x), \Delta q(x), g_{T}(x)$ and
$f_{4}(x)$ of Table~\ref{tab:1}, respectively,
because of the same twist and spin-dependence
(see \S\S3 and \ref{sec:ce}). 
On the other hand, there exists no gluon distributions
that mix with the chiral-odd quark distributions.

The light-cone quantization formalism again gives some insight
into the twist classification in Table~\ref{tab:2}.
The gluon fields are decomposed into ``good'' and ``bad''
components as $A^{\mu} = A_{\perp}^{\mu} + A_{\parallel}^{\mu}$. 
The ``good'' components $A_{\perp}^{\mu}$ represent the independent
degrees of freedom, which possess helicity $\pm 1$.
The ``bad'' components $A_{\parallel}^{\mu}$
can be re-expressed by a coherent gluon pair which
is coupled to the total spin zero state.
This explains the particular pattern of the spin dependence
for each twist appearing in Table~\ref{tab:2}, and it also implies that
the twist-2 distributions ${\cal G}$ and $\Delta {\cal G}$,
which involve only the ``good'' components,
are literally the ``distributions''.
From (\ref{eq:gphde}), we obtain 
\begin{eqnarray}
{\cal G}(x, \mu^{2}) &=& 
- \frac{2}{x} \int \frac{d\lambda}{2\pi} 
e^{i\lambda x} \langle P S|
{\rm tr}\, w_{\alpha} G^{\alpha \mu}(0) [0, \lambda w]
w^{\beta} G_{\beta \mu}(\lambda w)
|P S \rangle,
      \label{eq:gdt} \\
\Delta {\cal G}(x, \mu^{2}) &=& 
\frac{2 i}{x} \int \frac{d\lambda}{2\pi} 
e^{i\lambda x} \langle P S_{\parallel}|
{\rm tr}\, w_{\alpha} G^{\alpha \mu}(0) [0, \lambda w]
w^{\beta}\widetilde{G}_{\beta \mu}(\lambda w) |P S_{\parallel} \rangle,
\label{eq:gdh} 
\end{eqnarray}
where 
$\widetilde{G}_{\beta \mu} = \frac{1}{2}\epsilon_{\beta \mu \nu \rho}
G^{\nu \rho}$.
It is easy to see that
${\cal G}(x)$ and $\Delta {\cal G}(x)$ represent
the total and helicity distribution of the gluon, respectively.
We introduce the right- and
left-handed circular polarization vectors
$\epsilon^{\mu}_{R}=(0, -1, -i, 0)/\sqrt{2}$ 
and $\epsilon^{\mu}_{L}=(0, 1, -i, 0)/\sqrt{2}$, 
and find $- G^{+\mu}G^{+}_{\mu} = (G^{+ R})^{\dagger}
G^{+ R} + (G^{+ L})^{\dagger}
G^{+ L}$ and $i G^{+\mu}\widetilde{G}^{+}_{\mu} = (G^{+ R})^{\dagger}
G^{+ R} - (G^{+ L})^{\dagger} G^{+ L}$.
Thus ${\cal G}(x) = {\cal G}_{\uparrow}(x)
+ {\cal G}_{\downarrow}(x)$ and 
$\Delta {\cal G}(x) = {\cal G}_{\uparrow}(x)
- {\cal G}_{\downarrow}(x)$, where
${\cal G}_{\uparrow (\downarrow)}(x)$ is the number density
of the gluon with momentum fraction $x$
and helicity parallel (antiparallel) to the nucleon's spin.

Similarly to the quark distributions, we also obtain:
(i) For all four distributions,
$\phi =\{{\cal G}, \Delta {\cal G}, {\cal G}_{3T}, {\cal G}_{4}\}$,
the support property (\ref{eq:region}) holds.
(ii) The charge-conjugation transformation gives,
for $\phi_{\rm even} = \{ \Delta {\cal G}, {\cal G}_{3T} \}$
and $\phi_{\rm odd} = \{ {\cal G}, {\cal G}_{4} \}$,
\begin{equation}
\phi_{\rm even}(-x) = \phi_{\rm even}(x) \quad , \quad 
\phi_{\rm odd}(-x) = - \phi_{\rm odd}(x). \label{eq:chargeg}
\end{equation}
(iii) From the positivity of ${\cal G}_{\uparrow, \downarrow}(x)$,
we obtain ${\cal G}(x) \ge | \Delta {\cal G} (x)|$.

\subsection{The three-particle correlation functions of twist-3}
\label{sec:tp}

Coherent many-particle contents of the nucleon are described by
multiparton distribution functions.
In this paper, we will explicitly deal with the twist-3 quark-gluon
correlation functions, which are defined as
\begin{eqnarray}
\lefteqn{
   \int \frac{d\lambda}{2\pi}\frac{d\zeta}{2\pi}
   \,e^{i\lambda x + i\zeta (x'-x)} 
  \langle PS|\bar{\psi}(0) \gamma^\alpha [0,\zeta w]gG^{\mu\nu}(\zeta w)
     [\zeta w,\lambda w] \psi(\lambda w)|PS\rangle} \nonumber\\
& &\qquad\quad =\, p^\alpha \epsilon^{\mu \nu \xi \rho}p_{\xi} S_{\perp \rho}
      \Psi(x, x', \mu^{2})+\cdots, \label{eq:V3}\\
\lefteqn{\int \frac{d\lambda}{2\pi}\frac{d\zeta}{2\pi}
   \,e^{i\lambda x + i\zeta (x'-x)} 
\langle PS|\bar{\psi}(0) \gamma^\alpha \gamma_5[0,\zeta w]
         g G^{\mu\nu}(\zeta w)[\zeta w,\lambda w]
         \psi(\lambda w)|PS\rangle}\nonumber \\
& &\qquad\quad =\, -ip^\alpha[p^\nu S_{\perp}^{\mu}-p^\mu S_{\perp}^{\nu}]
      \widetilde{\Psi} (x, x', \mu^{2})+\cdots, \label{eq:A3}\\
\lefteqn{\int \frac{d\lambda}{2\pi}\frac{d\zeta}{2\pi}
   \,e^{i\lambda x + i\zeta (x'-x)} 
  \langle PS|\bar{\psi}(0) \sigma^{\alpha\beta} 
    [0,\zeta w] g G^{\mu\nu}(\zeta w)[\zeta w,\lambda w]
         \psi(\lambda w)|PS\rangle} \nonumber \\
& &\qquad\quad =\, i \epsilon^{\alpha \beta \xi \rho}
      p_{\xi} [p^{\mu} g_{\perp \rho}^{\nu} - p^{\nu} g_{\perp \rho}^{\mu}]
          (S \cdot w)  M \widetilde{\Phi}(x, x', \mu^{2}) \nonumber \\
& &\qquad\qquad \, +\, 
    [ p^\alpha p^\nu g_{\perp}^{\beta\mu}
     -p^\beta p^\nu g_{\perp}^{\alpha\mu} 
     -p^\alpha p^\mu g_{\perp}^{\beta\nu}
     +p^\beta p^\mu g_{\perp}^{\alpha\nu}]
     M \Phi (x, x', \mu^{2})+\cdots, \nonumber\\
          \label{eq:T3}
\end{eqnarray}
where the dots stand for Lorentz structures
of the twist higher than 3 
(the Dirac matrices 1 and $i \gamma_{5}$
do not produce any new twist-3 distributions).
Similarly to the quark distribution functions of \S2.3,
the correlation functions can be defined for each quark flavor $f$,
giving $\Psi^{f}(x, x'), \widetilde{\Psi}^{f}(x, x')$, etc.,
or for singlet and non-singlet parts by 
inserting appropriate 
\begin{wraptable}{l}{\halftext}
\caption{Spin and chiral classification of the
quark-gluon correlation functions at twist-~3
(${\cal O}(1/Q)$).}
\label{tab:3}
\begin{center}
\begin{tabular}{c|c}\hline \hline
spin ave.    & $\Phi (x, x')^{\displaystyle{\star}}$\\
$S_{\parallel}$ &$\widetilde{\Phi} (x, x')^{\displaystyle{\star}}$\\  
$S_{\perp}$ &$\Psi (x, x'), \widetilde{\Psi} (x, x')$ \\ \hline
\end{tabular}
\end{center}
\end{wraptable}
flavor matrices between the guark fields. 
These correlation functions describe interferences
between the scattering from a coherent quark-gluon pair
and from a single quark.

The spin and chiral classifications
of the twist-3 quark-gluon correlation functions
are shown in Table~\ref{tab:3}:
those marked with ``$\star$'' are chiral-odd, while
the others are chiral-even. 
By inserting a complete set of intermediate states
between the quark and gluon fields, 
one can obtain the support properties of 
$\Psi, \widetilde{\Psi}, \Phi$ and $\widetilde{\Phi}$,
and their interpretation in the partonic language.
The variables $x, x'$ and $x'-x$ have the physical meaning 
of the momentum fraction carried by the quark, antiquark and gluon
respectively;
positive values correspond to emission of a parton from the nucleon,
and negative ones for absorption of the corresponding antiparton
(see Ref.~\citen{JA}).
The functions $\Psi, \widetilde{\Psi}, \Phi$ and $\widetilde{\Phi}$
vanish unless $|x| < 1, |x'| <1$ and $|x-x'| < 1$.
They also have definite symmetry properties
under the interchange $x \leftrightarrow x'$,
that follow from the parity transformation
combined with the time-reversal transformation. For 
$\Xi_{S} = \{ \Psi, \Phi \}$
and $\Xi_{A} = \{ \widetilde{\Psi}, \widetilde{\Phi} \}$,
\begin{equation}
\Xi_{S}(x, x') = \Xi_{S}(x', x)\quad , \quad 
\Xi_{A}(x, x') = - \Xi_{A}(x', x). \label{eq:symm3}
\end{equation}
Although these correlation functions
assume the rather complicated expressions (\ref{eq:V3})--(\ref{eq:T3}),
they in fact describe the amplitude for finding
a triplet of particles $(q, \bar{q}, G)$
with simple physical configuration in the nucleon: the total spin of
$q \bar{q} G$ can be 0 or 1 so that
it gives a nonzero nucleon matrix element.
The spin-1 state of $q \bar{q}G$ with helicity $\pm 1$ is possible
for the transversely polarized nucleon,
and is described by the chiral-even functions 
$\Psi$ and $\widetilde{\Psi}$.
Because of the chiral-even structure,
$q$ and $\bar{q}$ have opposite helicities,
so that their total helicity is 0
with total spin 0 and 1 for $\Psi$ and $\widetilde{\Psi}$, respectively.
Therefore, the total spin 1 of $q\bar{q}G$ is exactly
that of the transverse gluon in $\Psi$,
while both the $q \bar{q}$-pair and gluon 
have spin 1 in $\widetilde{\Psi}$.
This discussion can be made explicit by considering
the combination (see (\ref{eq:V3}) and  (\ref{eq:A3}))
\begin{eqnarray}
\lefteqn{\frac{1}{2}
  \left( \Psi (x, x') \pm \widetilde{\Psi}(x, x') \right) S_{\perp}^{L}
    =  }\nonumber \\
&& - i \int \frac{d\lambda}{2\pi}\frac{d\zeta}{2\pi}
   \,e^{i\lambda x + i\zeta (x'-x)} 
  \left\langle PS \left| 
  \bar{\psi}(0) \rlap/{\mkern-1mu w} [0,\zeta w]gG^{L\nu}(\zeta w)
  w_{\nu} [\zeta w,\lambda w]\frac{1 \mp \gamma_{5}}{2}
         \psi(\lambda w) \right| PS \right\rangle,\nonumber\\%[-2pt]
    \label{eq:chiralep}
\end{eqnarray}
where $L$ denotes the ``left-handed component'' introduced 
in \S\ref{sec:gldis}. A similar expression with $L\rightarrow R$
and $\gamma_{5} \rightarrow -\gamma_{5}$ can also be obtained.
On the other hand, it is possible to find 
two other relevant configurations of $q\bar{q}G$,
the spin-1 state with helicity 0 and the spin-0 state,
for the longitudinally polarized and unpolarized nucleons.
The corresponding amplitudes are given by the chiral-odd functions
$\widetilde{\Phi}$ and $\Phi$, respectively.
Due to the chiral-odd structure,
$q$ and $\bar{q}$ have the same helicities,
so that the $q\bar{q}$-pair has the spin 1.
In $\Phi$, the spin of the $q\bar{q}$-pair and that of the transverse gluon
are antiparallel, and therefore $q\bar{q}G$ has total spin 0. 
In $\widetilde{\Phi}$, the spin of the gluon is combined with that
of the $q \bar{q}$-pair to give total spin 1 and helicity 0.
These considerations also reveal on a physical basis
that the four functions
in Table~\ref{tab:3} constitute a complete set of the twist-3 
quark-gluon correlation functions.

We note that the quark-gluon correlations
$\Phi (x, x'), \widetilde{\Phi} (x,x')$
and $\Psi (x, x'), \widetilde{\Psi} (x, x')$
have exactly the same spin, twist,
and chiral classification with $e(x), h_{L}(x)$ and 
$g_{T}(x)$, respectively (compare Tables~\ref{tab:1} and \ref{tab:3}).
As discussed in \S\S\ref{sec:ce} and \ref{sec:co} below,
one can actually express the twist-3 quark distributions
$e(x), h_{L}(x)$ and 
$g_{T}(x)$ in terms of a certain integral of the corresponding
twist-3 quark-gluon correlation functions;
only such an integral (over the gluon momentum)
is potentially measurable in inclusive reactions like the DIS and DY.
However, we will show that the description of the QCD evolution 
of $e(x), h_{L}(x)$ and 
$g_{T}(x)$ requires full knowledge of the quark-gluon correlations
$\Phi (x, x'), \widetilde{\Phi} (x,x')$
and $\Psi (x, x'), \widetilde{\Psi} (x, x')$ 
as functions of the two variables $x$ and $x'$.

The treatment in this section can be extended to the case of the twist-3
three-gluon correlation functions, which are relevant to 
the (singlet) quark distribution $g_{T}(x)$ and the gluon distribution
${\cal G}_{3T}(x)$.
Although the extension is straightforward, the actual analysis
is complicated due to participation of three identical particles,
which have color degrees of freedom
and obey the Bose statistics.
We do not go into the details of the treatment of the three-gluon
nonlocal light-cone operators,~\cite{ji,BKL}
but we discuss equivalent results based on the local three-gluon
operators in \S\ref{sec:ce}.

\section{Relation to the local operator approach and renormalization}

In the previous section, we have defined and classified various
parton distribution functions in terms of the
nonlocal light-cone operators.
For the structure functions to which the OPE
can be applied, the above definition, of course, 
leads to the same results as those obtained using the OPE.
We briefly discuss this connection employing 
an example in the DIS.
Based on this connection,
we introduce the renormalization group equations and 
their solutions, which describe the $Q^{2}$-evolution 
of the parton distribution functions.
We also remind the reader of an important issue
regarding the factorization
scheme dependence of the parton distribution functions.

\subsection{Equivalence to OPE and renormalization group equation}

Let us first consider the familiar example of the twist-2 term of the 
unpolarized structure function $F_{2}(x, Q^{2})$.
To reveal the connection between  
the nonlocal operator approach and the traditional OPE,
it is convenient to work in the moment space defined by
\begin{equation}
 M_n (Q^2 ) \equiv \int_{0}^{1} dx x^{n-2} F_2 (x , Q^2 ) \ .
\label{eq:mom0}
\end{equation}
The factorization formula (\ref{eq:f2}) with $i = (f, \bar{f}, G)$ 
and $A=N ({\rm nucleon})$ gives us [ substituting
$f_{f/N}(\xi, \mu^{2}) = q^{f}(\xi, \mu^{2}),
f_{\bar{f}/N}(\xi, \mu^{2}) = q^{\bar{f}}(\xi, \mu^{2})$ and 
$f_{G/N}(\xi, \mu^{2}) = {\cal G}(\xi, \mu^{2})$
introduced in \S\S\ref{sec:qc} and \ref{sec:gldis}
(see also (\ref{eq:f1free}))]
\begin{eqnarray}
  M_n(Q^2) &=& \sum_{f} C_n^{f} 
     \left(Q^2 / \mu^2\,,\,
      \alpha_{s}(\mu^{2})
         \right) v_n^{f} ( \mu^2) +  C_n^G \left(
    Q^2 / \mu^2\,,\,\alpha_{s}(\mu^{2})\right) v_n^G ( \mu^2)  
\nonumber \\
&=&\frac{2}{9}C_n^{S} \left(Q^2 / \mu^2\,,\,
\alpha_{s}(\mu^{2})
         \right) v_n^q ( \mu^2) +  C_n^G \left(
    Q^2 / \mu^2\,,\,\alpha_{s}(\mu^{2})\right) v_n^G ( \mu^2) 
\nonumber \\
  && + \,\frac{1}{6}C_n^{NS} \left(Q^2 / \mu^2\,,\,
\alpha_{s}(\mu^{2})
         \right) \left(v_n^{(3)} ( \mu^2) 
   + \frac{1}{3} v_n^{(8)} ( \mu^2) \right)\ ,
\label{fmoment}
\end{eqnarray}
where we have defined
\begin{eqnarray}
 v_{n}^{f}(\mu^{2}) &=& \int_{0}^{1}dx x^{n-1}\left[ q^{f}(x, \mu^{2})
       + q^{\bar{f}}(x, \mu^{2}) \right] \,  , \, 
 v_n^G (\mu^{2}) = \int_{0}^{1}dx x^{n-1} {\cal G}(x, \mu^{2})\ ,
   \nonumber\\
               \label{eq:anqg}\\
        & &       C_{n}^{f,G}(Q^{2}/\mu^{2}, \alpha_{s}(\mu^{2})) = 
    \int_{0}^{1}dx x^{n-2}H_{f,G}\left(x, Q^{2}/\mu^{2}, 
       \alpha_{s}(\mu^{2})\right).
         \label{eq:cnqg}
\end{eqnarray}
Here $f=u,d,s$ for three flavors, and
we have used $H_{f}=H_{\bar{f}}$,
because the DIS is a charge-conjugation even process.
Note that $H_{f}\left(\xi, Q^{2}/\mu^{2}, \alpha_{s}(\mu^{2})\right) 
= ({{\cal Q}^{(el)}}^{2})_{ff}
\left[ \delta(\xi -1) + {\cal O}(\alpha_{s})\right]$ and 
$H_{G}\left(\xi, Q^{2}/\mu^{2}, 
\alpha_{s}(\mu^{2})\right)  = {\cal O}(\alpha_{s})$,
so that $C_{n}^{f} = 
({{\cal Q}^{(el)}}^{2})_{ff} \left(1 + {\cal O}(\alpha_{s})\right)$,
and $C_{n}^{G} = {\cal O}(\alpha_{s})$. 
The second expression of (\ref{fmoment}) corresponds to
the singlet ($v_{n}^q (\mu^{2}), v_n^G (\mu^{2})$) and the nonsinglet
($v_{n}^{(3)}(\mu^{2}), v_{n}^{(8)}(\mu^{2})$) decomposition
(see the last paragraph of \S2.2):
$v_{n}^q = v_{n}^{u} + v_{n}^{d} + v_{n}^{s}$,
$v_{n}^{(3)} = v_{n}^{u} - v_{n}^{d}$,
$v_{n}^{(8)} = v_{n}^{u} + v_{n}^{d} -2 v_{n}^{s}$,
and $C_{n}^{S}$ and $C_{n}^{NS}$ are defined by
factoring out the flavor structure so that
$C_{n}^{S , NS} =  1 + {\cal O}(\alpha_{s})$.~\footnote{
In general, $C_{n}^{S}\neq C_{n}^{NS}$.
The coefficient functions corresponding to 
$v_{n}^{(3)}$ and $v_{n}^{(8)}$
coincide up to the flavor structure because of the flavor independence
of the perturbative quark-gluon coupling.}

In the traditional OPE approach, 
(\ref{fmoment}) is known as the moment sum rule.~\cite{rv}
The $C_n^j$ are called Wison's coefficient functions
(short distance parts), and the $v_n^j$ are given by the nucleon
matrix elements of the local composite operators.
In the present case, the relevant twist-2 local operators are
\bea
 O_{V,\psi}^{\mu_1 \cdots \mu_n}  &=&  i^{n-1} \bar{\psi}\gamma^{\{\mu_1}
          D^{\mu_2}\cdots D^{\mu_n \}} \psi - {\rm traces}\ , 
\label{eq:o}\\
 O_{V,G}^{\mu_1 \cdots \mu_n}  &=&  i^{n-2} G_{c_{1}}^{\alpha \{\mu_1}
          D_{c_{1}c_{2}}^{\mu_2}\cdots 
 D_{c_{n-1}c_{n}}^{\mu_{n-1}} G_{c_{n}}^{\mu_n\}}{}_{\alpha}
                   - {\rm traces}  \ ,\label{eq:og}
\eea
where $D_{\mu}=\partial_{\mu} - igA_{\mu}$
is the covariant derivative, and
$\{ \, \}$ denotes the symmetrization over all Lorentz indices. 
Here ``$- {\rm traces}$'' represents for the subtraction of the trace terms
to make the operators traceless, which will be suppressed
in the following. 
$D_{\mu}$ of (\ref{eq:og}) is in the adjoint representation,
and the color indices $c_{i}$ are explicitly shown.
In the OPE, the $v_n$ are defined by
\be
 \langle PS| O_{V,\psi}^{\mu_1 \cdots \mu_n}| PS \rangle  =  2\, v_n
           P^{\mu_1}\cdots P^{\mu_n}\quad 
 {\rm or}\quad v_n = \frac{1}{2}
         w_{\mu_1} \cdots w_{\mu_{n}}
    \langle PS | O_{V,\psi}^{\mu_1 \cdots \mu_n}| PS \rangle
         \ , \label{adef}
\ee
where we have used  $P \cdot w = 1$.  
For $O_{V,\psi}^{\mu_1 \cdots \mu_n}$,
corresponding to the flavor structure,
$v_n^j$ becomes $v_{n}^{f}, v_{n}^{q}$ and $v_{n}^{(3),(8)}$,
respectively. Similarly, $v_n^G$ is given by
\begin{equation}
 \langle PS| O_{V,G}^{\mu_1 \cdots \mu_n}| PS \rangle  =  
             2\, v_n^G  P^{\mu_1}\cdots P^{\mu_n}\ .
\label{eq:calvdef}
\end{equation}

Now let us calculate $v_{n}^{f}$ of (\ref{eq:anqg})
using the quark distribution function (\ref{eq:odf})
based on the nonlocal operator approach.
Using the basic property of the distribution functions
(\ref{eq:region}) and (\ref{eq:charge}),
we obtain (suppressing the flavor index),
\bea
v_{n} &=& \int_{- \infty}^{+ \infty} dx x^{n-1} q (x)\nonumber\\
    &=& \frac{1}{4\pi} \int d \lambda 
           \left\{ \left( - i\frac{\partial}{\partial
              \lambda} \right)^{n-1} 2\pi \delta (\lambda)\right\}
        \langle P S| \bar{\psi} (0)\rlap/{\mkern-1mu w} 
     [0, \lambda w] \psi (\lambda w)| P S\rangle \nonumber \\
       &=& \frac{1}{2}\,
              \langle PS | \bar{\psi} (0) \rlap/{\mkern-1mu w}
     \left( i w\cdot D \right)^{n-1} \psi (0) | P S\rangle \ ,
\label{eq:moment}
\eea
where $n={\rm even}$ \, in the DIS.~\footnote{
By extending $v_{n}^{f}$ of (\ref{eq:anqg})
as $v_{n}^{f}= \int_{0}^{1}dx x^{n-1} [ q^{f}(x, \mu^{2})
+(-1)^{n} q^{\bar{f}}(x, \mu^{2}) ]$,
the result (\ref{eq:moment}) holds for odd $n$ as well.}
Comparing this result with (\ref{eq:o}) and (\ref{adef}),
one realizes that our definition of the 
quark distribution function, (\ref{eq:odf}),
gives results equivalent to the OPE. It is easy to see that 
$v_n^G$ of (\ref{eq:anqg}), combined with ({\ref{eq:gdt}),
also gives the matrix element of the local composite operator
(\ref{eq:og}), and is equivalent to the OPE results.
For other structure functions, one can show that the situation is the same.
In general, the moments of the parton distribution functions 
are given by the matrix elements of the corresponding
gauge-invariant, local composite operators (see \S\S5 and
6).\footnote{An exception is the first moment of $\Delta {\cal G}(x)$.
This will be discussed in \S5.}

The bare nonlocal light-cone operators 
contain ultraviolet divergences
due to loop corrections in QCD perturbation theory.
The renormalization of the ultraviolet divergences
induces the dependence of the parton distribution functions
on the renormalization scale $\mu$,
which is governed by the renormalization group (RG) equation
for the corresponding nonlocal operators.
In partonic language, this is the
Dokshitser-Gribov-Lipatov-Altarelli-Parisi (DGLAP)
equation,~\cite{AP} and reads 
\begin{eqnarray}
\lefteqn{\mu\frac{d}{d\mu} q^{\tilde{f}}(x, \mu^{2})}\nonumber\\
  &=&\frac{\alpha_{s}(\mu^{2})}{\pi}
\int_{x}^{1}\frac{dy}{y}\left[ \sum_{\tilde{f}'} 
P_{\tilde{f}\tilde{f}'}\left(\frac{x}{y}, 
\alpha_{s}(\mu^{2})\right)
q^{\tilde{f}'}(y, \mu^{2}) 
+ P_{\tilde{f}G}\left(\frac{x}{y}, \alpha_{s}(\mu^{2})\right)
         {\cal G}(y, \mu^{2})\right],\nonumber \\
\lefteqn{\mu\frac{d}{d\mu} {\cal G}(x, \mu^{2})}\nonumber\\
  &=&\frac{\alpha_{s}(\mu^{2})}{\pi}
\int_{x}^{1}\frac{dy}{y}\left[ 
\sum_{\tilde{f}'}P_{G\tilde{f}'}\left(\frac{x}{y}, 
\alpha_{s}(\mu^{2})\right)
q^{\tilde{f}'}(y, \mu^{2}) 
+ P_{GG}\left(\frac{x}{y}, \alpha_{s}(\mu^{2})\right)
  {\cal G}(y, \mu^{2})\right],\nonumber\\[-1pt]
\label{eq:dglap}
\end{eqnarray}
where the indices $\tilde{f}$ and $\tilde{f}'$ 
run over quarks and antiquarks
of all flavors.
The $P_{ij}$ with $i, j = G, u, \bar{u}, d, \bar{d}, \cdots$ 
are the DGLAP-kernels (splitting functions)
given by perturbative series in $\alpha_{s}(\mu^{2})$.
In the dimensional regularization in $4-2\varepsilon$ dimensions,
these kernels show up as the pole residue in $\varepsilon$
for the loop corrections to the corresponding parton distribution.
Physically, $P_{ij}$ gives the probability density of
parton decay process $j \rightarrow i + ({\rm anything})$.
Thus the equations in (\ref{eq:dglap})
have the simple physical interpretation as the master equations
of the probability density for partons.
Although they can be obtained straightforwardly by renormalizing 
the corresponding nonlocal light-cone operators,~\cite{CS,BB}
the actual procedure is rather complicated.
For our purposes, it is more convenient to discuss the equivalent
results based on the local operator language,
utilizing the one-to-one correspondence between the nonlocal and
local operator approaches established above.

Taking the moment of (\ref{eq:dglap}),
we go over to the RG equations for the corresponding local composite
operators.
It is convenient to write down these equations for
singlet and nonsinglet combinations.
For the singlet channel, we obtain ($a = q, G$)
\begin{equation}
\mu \frac{d}{d\mu}\, v_{n}^{a}(\mu^{2})
     + \sum_{b=q,G} \left[ \widehat{\gamma}^{V}_{n}(\alpha_{s}(\mu^{2})) 
\right]_{ab}
           v_{n}^{b}(\mu^{2})= 0\ .
\label{eq:RGeq}
\end{equation}
Here, $\widehat{\gamma}^{V}_{n}$ 
is given by (an appropriate linear combination of) the $n$-th moment
of the DGLAP-kernel
$\int_{0}^{1} dx x^{n-1} (\alpha_{s}/\pi) P_{ab}(x, \alpha_{s})$,
e.g.,
\begin{equation}
\left[ \widehat{\gamma}^{V}_{n}(\alpha_{s}(\mu^{2})) \right]_{GG}
  = - \int_{0}^{1}dx x^{n-1} \frac{\alpha_{s}(\mu^{2})}{\pi}
  P_{GG}\left(x, \alpha_{s}(\mu^{2})\right).
\label{eq:momAPK}
\end{equation}
In the local operator language,
$\widehat{\gamma}^{V}_{n}$ is the anomalous dimension (matrix),
which describes the mixing 
between the singlet part of the quark operator (\ref{eq:o})
and the gluon operator (\ref{eq:og})
under renormalization.
Equation (\ref{eq:RGeq}) can be derived directly
from the definition of the renormalized composite operators:
\[ 
\bmat{c}
 O_{V,S}^{\mu_1 \cdots \mu_n} \cr
 O_{V,G}^{\mu_1 \cdots \mu_n} \cr
 \emat_0
=
\bmat{cc}
   (\widehat{Z}_{n} )_{qq} & (\widehat{Z}_{n} )_{qG}\\
   (\widehat{Z}_{n} )_{Gq} & (\widehat{Z}_{n} )_{GG}
\emat
\bmat{c}
 O_{V,S}^{\mu_1 \cdots \mu_n} \cr
 O_{V,G}^{\mu_1 \cdots \mu_n} \cr
 \emat_R \, ,
\]
where $O_{V,S}^{\mu_1 \cdots \mu_n}$
is the singlet part of the quark operator (\ref{eq:o}), 
the suffix $R$ $(0)$ denotes renormalized (bare) quantities,
and $(\widehat{Z}_{n} )_{ab}$ represents the corresponding
renormalization constants.
From the fact that the bare operators do not depend on $\mu$,
we immediately obtain (\ref{eq:RGeq}) with
\[  \left[ \widehat{\gamma}^{V}_{n} ( \alpha_{s} (\mu^2 ))\right]_{ab} 
   \equiv \sum_{c=q,G} 
\,({\widehat{Z}_{n}}^{-1} )_{ac} \, \mu \frac{d}{d \mu} \, 
(\widehat{Z}_{n} )_{cb} \ .\]
Similarly, we obtain the RG equation for local nonsinglet
operators as
\begin{equation}
\mu \frac{d}{d\mu} v_{n}^{(k)}(\mu^{2}) 
+ \gamma_{n}^{V}(\alpha_{s}(\mu^{2}))v_{n}^{(k)}(\mu^{2}) =0 \ ,
\label{eq:RGNS}
\end{equation}
where $k=3,8$.
The anomalous dimension $\gamma_{n}^{V}$ for the nonsinglet
part is also related to the $n$-th moment of the DGLAP-kernels,
and it is defined through the corresponding renormalization constant as
\begin{displaymath}
\gamma_{n}^{V}(\alpha_{s}(\mu^{2})) = \frac{\mu}{Z_{n}}
\frac{d}{d\mu} Z_{n}, \;\;\;\;\;\;\;
\left(O_{V,NS}^{\mu_1 \cdots \mu_n}\right)_{0}
= Z_{n} \left(O_{V,NS}^{\mu_1 \cdots \mu_n}\right)_{R} \ ,
\end{displaymath}
where $O_{V,NS}^{\mu_1 \cdots \mu_n}$ is the nonsinglet part of (\ref{eq:o}). 
Note that the RG equations (\ref{eq:RGeq}) and (\ref{eq:RGNS}) 
decouple for each $n$
because the Lorentz spin $n$ is a trivial invariant under the evolution.

The coefficient functions in (\ref{fmoment}), 
$C_{n}^{j} (Q^{2}/\mu^{2})$, obey the RG equations ``conjugate'' to
(\ref{eq:RGeq}) and (\ref{eq:RGNS}), respectively,
so that $M_{n}(Q^{2})$ is independent of (arbitrary) scale $\mu$.
For phenomenological applications,
it is convenient to take $\mu^{2} = Q^{2}$ in (\ref{fmoment}).
Then the $Q^{2}$-dependence of $M_{n}(Q^{2})$ is 
dominated by the moment of the parton distribution functions 
(nucleon matrix element of the local operators)
renormalized at $Q^2$,~\cite{kubk} which
is given by solving (\ref{eq:RGeq}) and (\ref{eq:RGNS}) as
\begin{eqnarray}
  v_n^a (Q^2 ) &=& 
          \sum_{b=q,G}\left[ 
               {\rm T}_{( Q )} \exp\left\{-  \int_{Q_0}^{Q}
                \frac{d \kappa}{\kappa} \widehat{\gamma}^{V}_{n}
      (\alpha_{s}(\kappa^{2} )) \right\}\right]_{ab} v_n^b (Q_0^2 ) \ ,
         \label{eq:RGsol}\\
  v_{n}^{(k)}(Q^{2}) &=& \exp\left\{-  \int_{Q_0}^{Q}
                \frac{d \kappa}{\kappa} {\gamma}^{V}_{n}
      (\alpha_{s}(\kappa^{2} )) \right\} v_n^{(k)} (Q_0^2 ) \ ,
\label{eq:RGsol2}
\end{eqnarray}
where ${\rm T}_{(Q)}$ denotes the ``time ordering''
with respect to $Q$, and $Q_0$ is the starting scale of the
perturbative evolution.
On the other hand, the coefficient functions 
$C_{n}^{j}\left(1, \alpha_{s}(Q^{2})\right)$ 
depend weakly on $Q^{2}$ only through $\alpha_{s}(Q^{2})$.

The coefficient functions 
$C_{n}^{j}\left(1, \alpha_{s}(Q^{2})\right)$ of (\ref{fmoment})
as well as the anomalous dimensions 
$\widehat{\gamma}_{n}^{V}(\alpha_{s}(\mu^{2}))$ and
$\gamma_{n}^{V}(\alpha_{s}(\mu^{2}))$ of (\ref{eq:RGsol}) and 
(\ref{eq:RGsol2}) are calculable using perturbation theory.
The results can be expressed as power series in $\alpha_{s}$.
We have
\begin{equation}
C_{n}^{j}(1, \alpha_{s}) = C_{n(0)}^{j} 
+ \frac{\alpha_{s}}{4\pi} C_{n(1)}^{j}
+ \cdots \ ,
\label{eq:Pc}
\end{equation}
where $C_{n(0)}^{S,NS} = 1$, $C_{n(0)}^{G} = 0$
(see the discussion below (\ref{eq:cnqg})) and
\begin{eqnarray}
\widehat{\gamma}^{V}_{n}(\alpha_{s}) &=& \frac{\alpha_{s}}{4 \pi} 
\widehat{\gamma}^{V}_{n(0)} +\left(\frac{\alpha_{s}}{4 \pi}\right)^{2}
\widehat{\gamma}^{V}_{n(1)} + \cdots \ , \nonumber \\
\gamma^{V}_{n}(\alpha_{s}) &=& \frac{\alpha_{s}}{4 \pi} 
\gamma^{V}_{n(0)} +\left(\frac{\alpha_{s}}{4 \pi}\right)^{2}
\gamma^{V}_{n(1)} + \cdots \ .
\label{eq:Pg}
\end{eqnarray}
The running of $\alpha_{s}(\mu^{2})$ is driven  
by the $\beta$ function
\begin{equation}
\mu \frac{d}{d\mu} g \equiv \beta (g) 
= - \beta_0 \frac{g^3}{16 \pi^{2}} - \beta_1 \frac{g^5}{(16\pi^{2})^{2}}
+ \cdots\ ,
\label{eq:beta}
\end{equation}
where $\beta_{0} = 11 - 2 N_{f}/3$ and $\beta_{1} = 102 - 38N_{f}/3$,
with $N_{f}$ the number of flavors.
When we substitute only the leading term
of (\ref{eq:Pc})--(\ref{eq:beta}) into (\ref{fmoment}),
(\ref{eq:RGsol}) and (\ref{eq:RGsol2}),
we obtain the leading order (LO) prediction for the $Q^{2}$-evolution
of $M_{n}(Q^{2})$ in the RG improved perturbation theory.
By including also the next-to-leading order terms
(one-loop term of the coefficient functions (\ref{eq:Pc}),
and the two-loop term of the anomalous dimensions (\ref{eq:Pg})
and the $\beta$ function (\ref{eq:beta})),
we obtain the next-to-leading order (NLO) prediction for $M_{n}(Q^{2})$.
In the NLO approximation, (\ref{eq:beta}) is solved to give
the running coupling constant,
\begin{equation}
  \alpha_{s}(Q^2 ) =\frac{g^2 (Q^2 )}{4\pi} 
     = \frac{4 \pi}{\beta_0 \ln (Q^2 / \Lambda_{\rm QCD}^2 )}
    \left[1 - \frac{\beta_1 \ln (\ln (Q^2 / \Lambda_{\rm QCD}^2))}
      {\beta_0^2 \ln (Q^2 / \Lambda_{\rm QCD}^2 )}  \right] \ ,
\label{eq:2loop}
\end{equation}
where $\Lambda_{\rm QCD}$ is the QCD scale parameter 
at two loops~\footnote{$\Lambda_{\rm QCD}$ is renormalization
scheme dependent (see \S3.2 below).}
and in the corresponding results of (\ref{eq:RGsol}) and (\ref{eq:RGsol2}), 
the leading ($[\alpha_{s}\ln(Q^{2}/Q_{0}^{2})]^{n}$)
and the next-to-leading ($\alpha_{s}[\alpha_{s}\ln(Q^{2}/Q_{0}^{2})]^{n}$)
logarithms are correctly summed to all orders.
For the nonsinglet part (\ref{eq:RGsol2}),
the integration over $\kappa$ is straightforward
at the NLO level and gives
\begin{equation}
v_{n}^{(k)} (Q^{2}) = 
L^{\gamma_{n(0)}^{V}/2\beta_{0}}
\left[ 1 + \frac{\alpha_{s}(Q^{2}) - \alpha_{s}(Q_{0}^{2})}{4\pi}
\frac{\beta_{1}}{\beta_{0}}\left( \frac{\gamma_{n(1)}^{V}}{2 \beta_{1}}
- \frac{\gamma_{n(0)}^{V}}{2 \beta_{0}} \right) \right]
v_{n}^{(k)} (Q_{0}^{2}) \ ,
\label{eq:NLONS}
\end{equation}
where $L \equiv \alpha_{s}(Q^{2})/\alpha_{s}(Q_{0}^{2})$.
If we set $\beta_{1} \rightarrow 0$ and $\gamma_{n(1)}^{V} \rightarrow 0$,
(\ref{eq:2loop}) and (\ref{eq:NLONS}) reduce to the LO results. 
The NLO result for the singlet part (\ref{eq:RGsol})
is somewhat complicated due to the mixing,
but it is not difficult to show that, at LO, (\ref{eq:RGsol}) gives 
\begin{equation}
v_{n}^{a}(Q^{2}) 
= \sum_{b=q,G}
\left[ L^{\widehat{\gamma}^{V}_{n(0)}/2\beta_{0}} \right]_{ab}
v_{n}^{b}(Q_{0}^{2}) \ .
\label{eq:LOS}
\end{equation}
We note that, for $M_{n}(Q^{2})$ of (\ref{eq:mom0}),
all quantities necessary for the prediction of the NLO 
$Q^{2}$-evolution are known.

The correspondence between the nonlocal light-cone and local operators
like (\ref{eq:moment}) holds for other twist-2 distribution functions,
the helicity distributions, 
$\Delta q(x, \mu^{2})$ and $\Delta{\cal G}(x, \mu^{2})$,
and the transversity distribution, $\delta q(x, \mu^{2})$,
with appropriate substitutions.
The scale-dependence of these distributions
is also governed by an evolution equation
similar to (\ref{eq:RGsol}), (\ref{eq:RGsol2}), (\ref{eq:NLONS}) and
(\ref{eq:LOS}) with the anomalous dimensions for the corresponding
local operators (see \S\S5 and 6). 

A similar correspondence between the nonlocal light-cone operators
and the local composite operators is taken over by the
three-parton correlation functions.
Manipulations similar to (\ref{eq:moment})
give, for $\Psi (x, x')$ of (\ref{eq:V3}),
\begin{eqnarray}
\lefteqn{S_{\perp}^{\mu}\int_{-1}^{1}dx dx' x^{k-1}x'^{l-1} \Psi (x, x')}
          \nonumber \\
    & & \qquad = \,\langle PS_{\perp}|
        \bar{\psi}(0) \rlap/{\mkern-1mu w}
       (iw\cdot D)^{l-1} g \widetilde{G}^{\mu \nu}(0)w_{\nu}
        (iw \cdot D)^{k-1} \psi(0)|PS_{\perp}\rangle,
\label{eq:V3L}
\end{eqnarray}
with $k,l = 1, 2, \ldots$.
Similar results can be obtained for 
$\widetilde{\Psi}(x, x'), \Phi (x, x')$
and $\widetilde{\Phi} (x, x')$ of (\ref{eq:A3}) and (\ref{eq:T3}). 
Therefore, the double moment of the three-parton correlation
functions correspond to the gauge-invariant local
three-particle operators.
In particular, (\ref{eq:V3L}) and the corresponding
expression for $\widetilde{\Psi} (x, x')$ generate
a set of three-particle operators,
which coincide with those obtained using the OPE
for the transverse spin structure function $g_{2}(x, Q^{2})$ 
(see \S\ref{sec:ce}).

The scale-dependence of the three-particle correlation functions
is governed by the RG equation
for the corresponding three-particle light-cone operators.
It gives a generalization of the DGLAP equation into the three-body case,
which is schematically given by
\begin{equation}
\mu \frac{d}{d \mu} \Xi(x, x', \mu^{2})
= \frac{\alpha_{s}(\mu^{2})}{\pi}\int_{x}^{1}dy\int_{x'}^{1}dy'
{\cal P}(x,x'; y, y'; \alpha_{s}(\mu^{2})) \Xi (y, y', \mu^{2}),
\label{eq;DGLAP3}
\end{equation}
for the flavor nonsinglet channel,
up to terms proportional to the quark mass.
Here $\Xi (x, x')$ generically denotes $\Phi(x, x')$,
$\widetilde{\Phi}(x, x')$ and 
some linear combination of \{$\Psi(x, x')$, $\widetilde{\Psi}(x, x')$\}, 
and ${\cal P}$ is the corresponding kernel obtained by renormalizing the 
ultraviolet divergence of the relevant quark-gluon nonlocal operator.
As in the twist-2 case, it is convenient to work in the moment space,
based on (\ref{eq:V3L}). The moments with different sums $k+l$
correspond to different spins, and thus they do not mix with each other. 
Introducing $\Omega_{nl} (\mu^{2})\equiv \int dx dx' 
x^{n-l-2}x'^{l-1} \Xi (x, x', \mu^{2})$,
we obtain ($l,j = 1, 2, \ldots, n-2$)
\begin{equation}
\mu \frac{d}{d \mu}\Omega_{nl}(\mu^{2})
+  \sum_{j} \left[\Gamma_{n}(\alpha_{s}(\mu^{2}))\right]_{lj}
\Omega_{nj}(\mu^{2})=0.
\label{eq:RG3}
\end{equation}
Here
$\Gamma_{n}\left(\alpha_{s}(\mu^{2})\right)$ is related to
the double moment of the kernel ${\cal P}$.
In the local operator language,
it is the anomalous dimension matrix describing the renormalization
mixing between the relevant quark-gluon local operators
\begin{equation}
 \Omega_{nl} \sim \langle PS|
     \bar{\psi} \rlap/{\mkern-1mu w}\Lambda
      (iw\cdot D)^{l-1} gG^{\nu \rho}w_{\rho}
          (iw \cdot D)^{n-l-2}\psi|PS\rangle \ ,
\label{eq:Omega}
\end{equation}
with $\Lambda$ some Dirac matrix structure.
The suffix $l$ here labels many independent local operators having the same spin
$n$, and the operators with different $l$ for the same $n$ are 
allowed to mix with each other.
Note that the number of independent operators increases with spin $n$.
We emphasize that this sophisticated mixing
originates from the fact that the three-particle correlation functions
depend on the two variables $x$ and $x'$, and that it is characteristic
of higher twist distributions.

The solution of (\ref{eq:RG3}) is obtained from (\ref{eq:RGsol}) and 
(\ref{eq:LOS}) with the formal replacement 
$v_{n}^{a} \rightarrow \Omega_{nl}$ and 
$\widehat{\gamma}_{n}^{V} \rightarrow \Gamma_{n}$.
The mixing matrix $\Gamma_{n}$ can be obtained
by renormalizing the corresponding local composite operators
of (\ref{eq:Omega}) following the standard procedure. 
In the course of the renormalization of higher twist operators, however,
we encounter some novel phenomena which we discuss in \S4.

\subsection{Scheme dependence of distribution functions}

Once the anomalous dimensions (\ref{eq:Pg})
(or the DGLAP kernels of (\ref{eq:dglap})) are known, 
one can predict the $Q^2$-dependence of the parton distribution
functions with (\ref{eq:RGsol}) and (\ref{eq:RGsol2}).
However, the observable cross sections are the convolution
of parton distribution functions and the 
hard scattering parts (short-distance parts)
(see (\ref{fmoment}), (\ref{eq:f2}) and (\ref{eq:do})).
Here an arbitrariness comes into play,
beyond the LO, in defining (separating)
the parton distribution functions and the hard parts.
This arbitrariness is called the ``factorization scheme dependence'', and is
inherent in the factorization procedure of mass singularities
(collinear divergences).~\cite{CSS}
In the local operator language, this corresponds to an arbitrariness
in defining the renormalized
local operators.~\footnote{The finite terms of the
renormalization constants are in general arbitrary,
and are fixed by specifying a scheme.}
Both the parton distribution functions and the hard parts depend on
the scheme which one adopts, and only the convolution of them 
that corresponds to an observable quantity is scheme independent.

The scheme dependence can be stated in the following way
in the example discussed in \S3.1.
Let us concentrate on the singlet part given by the second line
of (\ref{fmoment}).
Taking a different scheme is equivalent to the following
change (we take $\mu^{2} = Q^{2}$): 
\bean
   v_n^a (Q^2 ) &\to& v^{\prime\,a}_n (Q^2 ) =
     \sum_{b=q,G} U^{ab}_n (\alpha_{s}(Q^2) ) v_n^b (Q^2 ) \ ,\\
   \tilde{C}_n^a ( 1 , \alpha_{s} (Q^2 )) &\to&  
  \tilde{C}^{\prime\,a}_n ( 1 , \alpha_{s} (Q^2 ))
       = \sum_{c=q,G} \tilde{C}_n^c ( 1 , \alpha_{s} (Q^2 ))
        \left( U_n^{-1}(\alpha_{s}(Q^2) ) \right)^{ca} \ .
\eean
Here, $U$ is an arbitrary matrix whose components are functions of 
$\alpha_{s}(Q^2)$, and we set
$\tilde{C}_{n}^{q}= C_{n}^{S}$ and $\tilde{C}_{n}^{G} = C_{n}^{G}$.
Clearly, the moment $M_n (Q^2)$ does not change as a result of this 
replacement, but
the scale dependence of $v^{\prime\,a}_n$ is now controlled by the
RG equation (\ref{eq:RGeq}) with a \lq\lq new\rq\rq\ 
anomalous dimension,
\[ (\widehat{\gamma}^{V}_n )_{ab} \to 
       (\widehat{\gamma}^{V \prime}_n )_{ab}
      = \left[ \beta ({g})
     \left( \frac{\partial}{\partial {g}} U_n \right)
      U_n^{-1} + U \widehat{\gamma}^{V}_n U_n^{-1} \right]_{ab} \ .\]
Thus, the matrix elements (parton distribution functions), 
as well as the hard parts, can be redefined {\it simultaneously}
from one scheme to another.
In view of this, it is important to use the same scheme
for the parton distribution functions as the 
corresponding hard parts when one predicts an observable quantity.
From the phenomenological viewpoint,
we must specify the scheme to be adopted
when we derive the parton distribution functions from
the experimental data.

It is worth noting here that
the parameter $\Lambda_{\rm QCD}$ of (\ref{eq:2loop}),
and equivalently the QCD coupling constant, is
also a \lq\lq scheme dependent\rq\rq\  quantity.
This dependence comes from the renormalization (not factorization)
scheme in which the coupling constant is defined.
In contrast to the factorization scheme dependence which
cancels between the hard parts and the distribution functions,
the renormalization scheme dependence of the coupling constant
remains and is unavoidable in the perturbation theory.
The perturbative QCD prediction
with $\Lambda_{\rm QCD}$ fixed unambiguously
(up to neglected higher order corrections)
requires at least NLO accuracy.

\section{Renormalization mixing for gauge invariant operators}

Before proceeding to the detailed study of each polarized structure 
function and its QCD evolution, we discuss the renormalization of
the gauge invariant local, composite operators 
from a general point of view.
We demonstrate that the renormalization mixing
among composite operators obeys a particular pattern in QCD
(and more generally in non-Abelian gauge theories).
This knowledge will be useful to derive sophisticated 
mixing among the three-particle operators relevant to
the $Q^{2}$-evolution of twist-3 structure functions.

A local composite operator ${\cal O}$ is renormalized by determining 
the corresponding counterterms ${\cal C}$
such that arbitrary Green's functions with insertion 
of ${\cal O} + {\cal C}$,
\begin{equation}
   \left\langle 0 \left| {\rm T} \prod_{i=1}^n \phi_{\alpha_{i}} (x_i )
                 ( {\cal O}\,(y) + {\cal C} (y) )
     \right| 0 \right\rangle \ ,
\label{eq:GF}
\end{equation}
become finite.
Here $\phi_{\alpha_{i}} (x_{i})$ generically denote the (renormalized) 
elementary fields in the Lagrangian, 
$\phi_{\alpha_{i}} = \{\psi, \bar{\psi}, A_{\mu}, \cdots\}$;
the subscript $\alpha_{i}$ indicating various fields 
as well as Lorentz and internal symmetry indices
is suppressed for simplicity in the following. 
We employ the minimal subtraction
(MS) scheme in the dimensional regularization in $4-2\varepsilon$
dimensions for the discussion in this section.
Then ${\cal C}$ is given by the sum 
of a power series in $1/\varepsilon$ with appropriate
composite operators as coefficients, and these operators
mix with ${\cal O}$ under renormalization.

As is well known, operators that have
common canonical dimension and quantum numbers
in general mix with each other.~\cite{JC2}
A particular case of this pattern of ``operator mixing''
is represented by those operators which are invariant under some transformation,
and this case can be easily understood by the Ward-Takahashi identities:
Suppose that we have a transformation $\delta_t$ that
leaves the action $S$ invariant, and 
that the operator ${\cal O}$ is invariant
under the same transformation as $\delta_t {\cal O} = 0$.
A Ward-Takahashi identity for (\ref{eq:GF}) reads
\be
  0 = \sum_{k=1}^n \left\langle 0 \left| {\rm T} \left(\delta_t \phi (x_k )\right)
      \prod_{i \neq k}^n \phi (x_i )
          ( {\cal O}\,(y) + {\cal C} (y) ) \right| 0 \right\rangle 
      + \left\langle 0 \left| {\rm T} \prod_{i=1}^n \phi (x_i )
               \delta_t {\cal C} (y) \right| 0 \right\rangle \label{compowt} \ .
\ee
When $\delta_{t} \phi$ is again an elementary field, e.g.
$\delta_t \phi (x_k )\propto \phi (x_k )$,
as in ordinary global transformations, 
the first term of (\ref{compowt}) is finite
by our construction of the counterterms.
Therefore, the second term is also finite; but, it is finite only if
$\delta_t {\cal C} (y) = 0$, since ${\cal C}$ 
is given by power series in $1/\varepsilon$.
This means that only those operators, which 
are invariant under $\delta_{t}$ participate in ${\cal C}$ 
and thus mix with ${\cal O}$ under renormalization.

To explore the operator mixing in the renormalization of gauge invariant
operators, which we denote as ${\cal O}_{i}$ ($i = 1,2, \cdots$), however, 
the above simple argument should be modified in two points
(we employ a covariant gauge).
First, due to the gauge fixing, 
the action for quantum theory is invariant under
the BRST transformation, not under the gauge transformation.
Therefore, $\delta_{t}$ should represent the BRST transformation
$\delta_{\rm BRST}$ in the present context.
Note that the gauge invariant operators are 
BRST invariant, but the converse is not true. 
The BRST invariant operators involving the Faddeev-Popov ghosts
are not gauge invariant, and such operators
are now allowed to mix with gauge invariant operators
under renormalization.
These gauge noninvariant operators include those which vanish by
the equations of motion (see (\ref{eq:EOM}) below).
We call the rest of gauge noninvariant operators, which do not vanish by
the equations of motion, the ``alien operators'' denoted by
${\cal B}_{j}$ ($j=1,2, \cdots$).
From the nilpotency of the BRST transformation, $\delta^{2}_{\rm BRST}=0$, 
the basis of the alien operators can be chosen so that
they are all BRST exact;~\cite{JC2} i.e.,
they can be written as ${\cal B}_{i} = \delta_{\rm BRST} \hat{\cal B}_{i}$ 
with $\hat{\cal B}_{i}$ being some local operator.

The second modification comes from the fact that the BRST
transformation of an elementary field
$\delta_{\rm BRST} \phi$ is a composite operator.
Therefore, it is possible that the first term of (\ref{compowt})
ceases to be finite.
If so, we must have $\delta_{\rm BRST} {\cal C} \neq 0$ in order to
satisfy (\ref{compowt}).
Fortunately, these BRST variant counterterms turn out to be particular
operators, as can be understood from the following observation: Because 
the BRST transformed composite operators $\delta_{\rm BRST} \phi$
are themselves finite operators,
the possible divergence in the first term of (\ref{compowt})
occurs only when $x_k = y$, and this divergence 
should be canceled by the second term.
Now we argue that such cancellation can be realized 
when ${\cal C}(y)$ is given by 
the composite operators ${\cal E}_{i}(y)$ ($i = 1,2, \cdots$),
which are proportional to the equations of motion:
\begin{equation}
{\cal E}_{i}(y) = {\cal F}_{i}\left(\phi(y)\right)\frac{\delta S}
                       {\delta \phi (y)} \ . \label{eq:EOM}
\end{equation}
Here ${\cal F}_{i}$ is some function of $\phi$, 
and from this point we call $\{{\cal E}_{i}\}$
the equations-of-motion (EOM) operators.
It is easy to see that 
a Green's function containing the operators $\{ {\cal E}_{i}\}$ reads 
\be
 \left\langle 0 \left|{\rm T} {\cal E}_{j}(x) \prod_{i=1}^n \phi (x_i )
         \right| 0 \right\rangle 
    =  \,i \sum_k {\delta}^{(4)} (x - x_k )
        \left\langle 0 \left| {\rm T} 
         {\cal F}_{j} (\phi (x_k )) \prod_{i \neq k} \phi (x_i )
               \right| 0 \right\rangle \label{eomconst}\ ,
\ee
by integrating by parts in the functional integral
over $\phi$.~\footnote{The contact term due to 
$\delta {\cal F}_{i}(\phi(y))/\delta \phi(y)
\propto \delta^{(4)}(0)$ vanishes in $4-2\varepsilon$ dimensions.}
This indicates that $\delta_{\rm BRST} {\cal C}(y)$ can generate 
the divergent terms for $x_{k} = y$ in the second term of (\ref{compowt})
when ${\cal C}$ contains the ${\cal E}_{j}$ with
$\delta_{\rm BRST}{\cal E}_{j} \neq 0$. 
Therefore, we conclude that the BRST variant EOM operators,
as well as the BRST invariant EOM operators, now mix with $\{{\cal O}_{i}\}$. 

Summarizing, under the renormalization of gauge invariant
operators $\{{\cal O}_{i}\}$, the BRST invariant alien operators 
$\{{\cal B}_{i}\}$ and the EOM operators $\{{\cal E}_{i}\}$
mix.~\footnote{No other types of operators mix. 
For a complete proof, see, e.g., Refs.~\citen{JC2} and ~\citen{JL76}.}
Here $\{ {\cal E}_{i}\}$ involves both BRST invariant and variant EOM operators.
The renormalization matrix describing the entire mixing 
among these three kinds of operators obeys a characteristic
triangular pattern, and it is expressed schematically as 
\begin{equation}
\bmat{c}
 {\cal O} \cr
 {\cal B} \cr
 {\cal E} \cr
\emat_0
=
\bmat{ccc}
   Z_{\cal O O} & Z_{{\cal O B}} & Z_{{\cal O E}}\\
        0   & Z_{\cal B B} & Z_{\cal B E}\\
        0   &    0        &      Z_{\cal E E}
\emat
\bmat{c}
 {\cal O}\\
  {\cal B}\\
  {\cal E}
\emat_R \, .
\label{eq:REMAT}
\end{equation}
The reason for this particular form is simple: 
Both the bare and the renormalized EOM operators should vanish
when we {\it use} the equations of motion, and thus 
$Z_{\cal EO} = Z_{\cal EB} = 0$.
On the other hand, for the alien operators,
a Ward-Takahashi identity tells us
\bean
  \left\langle 0 \left| {\rm T} {\cal B}_{j} (y) \prod_{i=1}^n 
           \phi (x_i ) \right| 0 \right\rangle 
        &=& \left\langle 0 \left| {\rm T} \left(\delta_{\rm BRST} 
             \hat{{\cal B}}_{j}(y) \right) 
             \prod_{i=1}^n \phi (x_i ) \right| 0 \right\rangle \\
        &=& - \sum_{k=1}^n \left\langle 0 \left| {\rm T} \hat{{\cal B}}_{j} (y) 
           \left(\delta_{\rm BRST} \phi (x_k )\right)
      \prod_{i \neq k}^n \phi (x_i ) \right| 0 \right\rangle \ .
\eean
Here, suppose that we determine the counterterms 
${\cal C}(\hat{{\cal B}}_{j})$ for $\hat{{\cal B}}_{j}$ 
so that the second line becomes finite at $y \neq x_i$
($i= 1, \cdots, n$).
Then, from the first line, we see that 
$ \delta_{\rm BRST} ( \hat{{\cal B}}_{j} + {\cal C}(\hat{{\cal B}}_{j}))$
$=$ ${\cal B}_{j} + \delta_{\rm BRST} {\cal C}(\hat{{\cal B}}_{j})$
is a finite operator, and thus 
$\delta_{\rm BRST} {\cal C}(\hat{{\cal B}}_{j})$
gives the counterterms for ${\cal B}_{j}$,
except for the points $y=x_i$ where $\{{\cal E}_{j}\}$ participate.
Therefore, $Z_{\cal BO}=0$ in the third row of the above matrix.

One important point for actual applications is that
the physical (on mass-shell) matrix elements of both $\{ {\cal B}_{i}\}$ 
and $\{ {\cal E}_{i}\}$
vanish, and these operators do not contribute to the final results.
The proof is straightforward.~\cite{JC2,P80} 
Using the LSZ reduction formula,~\cite{BD65}
a matrix element of ${\cal E}_{i}$ between ``physical'' 
states can be written as~\footnote{The disconnected, 
forward-scattering term $\langle 0| {\cal E}_{i}(y) |0\rangle$
vanishes in $4-2\varepsilon$ dimensions.} 
\bea
 \lefteqn{\langle P|{\cal E}_{i}(y)|P' \rangle}\nonumber\\
    &=&  \lim_{\rm on-shell} \int 
            d^{4}zd^{4}z'e^{iP\cdot z}
            e^{-iP'\cdot z'}\left(P^2 -M^2 \right)
            \left(P^{\prime 2} - M^2 \right)
          \langle 0|{\rm T} {\cal K}^{\dagger}(z)
                 {\cal E}_{i}(y) {\cal K} (z') |0 \rangle\ .\nonumber\\
                \label{eq:LSZ}
\eea
Here ${\cal K}$ denotes an interpolating field
to create a hadron state $|P \rangle$ with mass $M$,
or, at the level of pure perturbative calculations,
${\cal K}$ denotes an elementary field $\phi$ 
to create a state with an on-shell quark or an on-shell (transverse) gluon.   
With the help of (\ref{eomconst}), the Green's function on the 
r.h.s. becomes
\[  \langle 0|{\rm T} {\cal K}^{\dagger}(z) 
{\cal E}_{i}(y) {\cal K}(z') |0 \rangle
    \sim \delta^{(4)} (y-z) \tau(z,z') + \delta^{(4)} (y-z') 
\tilde{\tau} (z,z') \ ,\]
where $\tau$ and $\tilde{\tau}$ are the two-point functions.
Based on their spectral representation,~\cite{BD65}
$\tau$ and $\tilde{\tau}$ 
cannot have a double pole which cancels the on-shell
factors of (\ref{eq:LSZ}). Thus we obtain
\begin{equation}
  \langle P | {\cal E}_{i}(y) |P'\rangle = 0 \ . \label{eq:EOMV}
\end{equation}
Similarly to the situation with (\ref{eq:LSZ}),
an on-shell matrix element of ${\cal B}_{i}$
is again proportional to the Green's function 
$\langle 0|{\rm T} {\cal B}_{i} {\cal K}^{\dagger}{\cal K} |0 \rangle$.
This can be rewritten using a Ward-Takahashi identity:
\begin{equation}
 \langle 0|{\rm T} {\cal B}_{i} {\cal K}^{\dagger}{\cal K}
         |0 \rangle =
         \left\langle 0 \left| {\rm T} \left(\delta_{\rm BRST} 
             \hat{\cal B}_{i}\right)  
           {\cal K}^{\dagger}{\cal K} \right| 0 \right\rangle 
     = - \left\langle 0 \left| {\rm T} \hat{\cal B}_{i} \delta_{\rm BRST} 
         \left({\cal K}^{\dagger}{\cal K}\right) \right| 0 \right\rangle \ .
\label{eq:WTB}
\end{equation}
When ${\cal K}$ is a gauge invariant interpolating field, 
(\ref{eq:WTB}) vanishes.
On the other hand, ${\cal K}$ can be an elementary field $\phi$,
which gives a BRST transformed composite operator $\delta_{\rm BRST}\phi$
on the r.h.s. of (\ref{eq:WTB}).
Because $\delta_{\rm BRST}\phi$ involves a ghost field, 
it cannot produce (perturbatively) physical particle poles
to cancel the on-shell factors of (\ref{eq:LSZ}).
Therefore, we obtain generally
\begin{equation}
\langle P | {\cal B}_{i}(y) |P'\rangle = 0 \ .
\label{eq:BRSTV}
\end{equation}
  
How is the above general argument relevant to the structure functions?
At the lowest twist level, the above complicated operator 
mixing does not come into play,
because there exists neither an EOM nor BRST invariant alien
operator of twist-2. 
(The EOM as well as the alien operators always have smaller
spin by at least one unit than the possible highest spin operators
of the same dimension (see \S\S5 and 6).)
Therefore, only gauge invariant
fermionic and gluonic operators mix with each other under renormalization,
as in \S3.1.

At the higher twist ($ \geq 3$) level, both EOM and BRST invariant alien
operators participate in the renormalization mixing 
with the gauge invariant operators.
This is a characteristic feature of the higher twist operators.~\cite{P80}
Because of (\ref{eq:EOMV}) and (\ref{eq:BRSTV}), however,
only the gauge invariant operators contribute 
to the matrix elements corresponding to the moments of
the structure functions.
Namely, only $Z_{\cal O O}$ in the renormalization matrix (\ref{eq:REMAT})
is of physical importance.
From this one can obtain the anomalous dimensions necessary for the 
prediction of the $Q^{2}$-evolution:
\[  \gamma_{\cal OO} 
= \mu \frac{d}{d\mu} \ln \left(Z_{\cal O O}\right)\ .\]
This would suggest that, by treating only the on-shell matrix
elements of composite operators to work out renormalization, 
we would obtain enough information
without considering the EOM and BRST invariant alien operators.
However, the calculation of the on-shell
matrix elements in terms of purely perturbative Feynman diagrams
in massless theory introduces another complexity.
To these matrix elements, not only the one-particle-irreducible (1PI)
but also the one-particle-reducible (1PR) diagrams contribute.
It often happens, especially for the latter class
of diagrams, that the infrared singularity coming from
the collinear configuration cannot be regulated,~\cite{efp}
so that the perturbative calculation becomes subtle
and potentially dangerous.~\cite{JCOL2}
To avoid this risky situation, one of the best ways is 
to evaluate the (off-shell) Green's functions like (\ref{eq:GF}).
In this case, 
renormalization can be performed by evaluating the 1PI diagrams only,
and off-shell external momenta provide an infrared cutoff. 
The calculation is quite straightforward with the manifest Lorentz 
covariance being maintained.
One price is that we must take into account mixing of 
the EOM and BRST invariant alien operators;
but, with the help of the general results discussed above,
we can assess unambiguously the necessary renormalization constants.  
We will discuss explicit examples for the twist-3
operators in \S\S5 and 6.

\section{Chiral-even structure functions}
\label{sec:ce}

In this section, we discuss the chiral-even polarized 
structure functions of twist-2 and -3,
which can be measured in the DIS.
The QCD evolution is considered with an emphasis on
the Bjorken, Ellis-Jaffe sum rules
and the twist-3 contribution to $g_2(x, Q^{2})$.

\subsection{Twist expansion}

The chiral-even polarized quark and gluon distribution functions are given
by (\ref{eq:axialv}) and (\ref{eq:gphde}), up to twist-4. 
These distribution functions have been related to the 
structure functions $g_{1}(x, Q^{2})$ and $g_{2}(x, Q^{2})$
by (\ref{eq:parton}) in the free field theory 
(for a single flavor case); the results can actually 
be made correct at the LO level
by incorporating the LO $Q^{2}$-dependence into the
distribution functions on the r.h.s.
As mentioned in \S2.2, the definition of the twist 
based on power counting in $1/Q$ is not
exactly the same as that based on the representation
of the Lorentz group (``${\rm twist} = {\rm dimension} - {\rm spin}$''
of the relevant operators).
As a result, $g_{2}$ ($g_{T}$) receives contributions
from both the twist-3 and twist-2 operators.

Let us explain how to identify the twist-2 and twist-3 operators~\cite{JJ2}
for $g_{2}$ ($g_{T}$).
(A similar argument is possible for the twist-3 gluon
distribution ${\cal G}_{3T}$ of (\ref{eq:gphde}).)
Taking the moment of (\ref{eq:axialv}) with respect to $x$ 
and applying manipulations similar to (\ref{eq:moment}),
\begin{eqnarray}
  \frac{1}{2}w_{\mu_{1}} \cdots w_{\mu_{n-1}} \langle PS |
  R^{\sigma \{ \mu _1 \cdots \mu_{n-1}\} } |PS \rangle
     &=&  p^{\sigma}(S\cdot w)
\int_{-1}^{1} dx x^{n-1} \Delta q(x,\mu^2) \nonumber \\
 && + \, S_{\perp}^{\sigma} \int_{-1}^{1} dx x^{n-1} g_{T}(x, \mu^2)  \ ,
\label{eq:momg}
\end{eqnarray}
where we have kept only terms through twist-3, we get 
the local composite operator 
\be
  R^{\sigma \{ \mu _1 \cdots \mu_{n-1}\} }  =  i^{n-1} \bar{\psi}
  \gamma^{\sigma} \gamma_5 D^{\{{\mu_1}} \cdots D^{{\mu_{n-1}}\}} \psi
                  \label{mixedop}\ .
\ee
Here the subtraction of the trace terms
$g_{\mu_{i}\mu_{j}} R^{\sigma \{ \mu _1 \cdots \mu_{n-1}\} }$ and
$g_{\sigma \mu_{i}} R^{\sigma \{ \mu _1 \cdots \mu_{n-1}\} }$, 
which generate 
${\rm twist} \ge 4$ terms, should be understood, and
the Lorentz indices $\mu_i$ are symmetrized (compare with (\ref{eq:o})).
However, the index $\sigma$ on the $\gamma$ matrix is not symmetrized
nor antisymmetrized.
In order for the operator (\ref{mixedop})
to have a definite twist (spin),
it must be decomposed into operators which each
has a definite symmetry with respect to the Lorentz indices:
\begin{eqnarray}
 R^{\sigma \{ \mu _1 \cdots \mu_{n-1}\} }
     &\equiv& R^{\{ \sigma \mu _1 \cdots \mu_{n-1}\} }
         + R^{[ \sigma \{ \mu _1 ] \cdots \mu_{n-1}\} } \nonumber \\
      &=&  R^{\{ \sigma \mu _1 \cdots \mu_{n-1}\} }
         + \frac{1}{n}\, \left[
            (n -1) R^{\sigma \{ \mu _1 \cdots \mu_{n-1}\} }
           - \sum_{l=1}^{n-1} 
            R^{\mu_l \{ \sigma \mu _1 \cdots \mu_{n-1}\} }
         \right] \ .\nonumber\\
\label{eq:53}
\end{eqnarray}
Here $R^{\{ \sigma \mu _1 \cdots \mu_{n-1}\} }$
and $R^{[ \sigma \{ \mu _1 ] \cdots \mu_{n-1}\} }$
are the twist-2 and twist-3 operators because they have
spin $n$ and $n-1$, respectively.
The nucleon matrix element of these operators can be parameterized as,
according to their definite symmetry,
\bea
   \langle P S |  R^{\{ \sigma \mu _1 \cdots \mu_{n-1}\} } | P S
     \rangle &=& 2 a_n S^{\{\sigma} P^{\mu _1}\cdots P^{\mu_{n-1}\}}
       \nonumber \ ,\\
   \langle P S |  R^{[ \sigma \{ \mu _1 ] \cdots \mu_{n-1}\} } | P S
     \rangle &=& 2 \frac{n - 1}{n}
         d_n \left( S^{\sigma} P^{\mu_1} - S^{\mu_1} P^{\sigma} \right)
           P^{\mu_2} \cdots P^{\mu_{n-1}}\ . \label{eq:dn}
\eea
Substituting these expressions into (\ref{eq:momg})
and combining the result with the relation (\ref{eq:parton}),
we obtain ($n= {\rm odd}$ in the DIS)
\bea
  \int_0^1 dx x^{n-1} g_1 (x) &=& 
\frac{1}{2}\int_{-1}^{1} dx x^{n-1} \Delta q(x)
= \frac{1}{2} a_n \ , \label{eq:momg1f} \\
   \int_0^1 dx x^{n-1} \left( g_1 (x) \right.&+& \left. 
g_2 (x) \right) = 
\frac{1}{2} \int_{-1}^{1}dx x^{n-1} g_{T}(x)
=
\frac{1}{2} 
     \left( \frac{1}{n} a_n + \frac{n-1}{n} d_n \right)  \ .\nonumber\\
         \label{eq:momg2f}
\eea
From the above expression, one can write $g_2$ as ($x \ge 0$)
\[   g_2 (x) \equiv g_2^{WW} (x) + \bar{g}_2 (x)
   = - g_1 (x) + \int_x^1 \frac{dy}{y} g_1 (y) + \bar{g}_2 (x)\ .\]
It is $\bar{g}_2 (x)$ which receives a contribution from genuine
twist-3 operators,
\begin{equation}
 \int_0^1 dx x^{n-1} \bar{g}_2 (x) = \frac{n-1}{2 n} d_n \ ,
\label{eq:genuine}
\end{equation}
while the contribution from the twist-2 operators, $g_2^{WW} (x)$,
is called the Wandzura-Wilczek part.~\cite{WW}

\subsection{The structure function $g_{1}(x, Q^{2})$}

We are now in a position to discuss the spin structure function $g_1$. 
Let us recall the moment sum rules
for $g_1$ based on the local composite operators.~\cite{HM,SAR}
The following twist-$2$ operators, corresponding to the moments
of the nonlocal operators (\ref{eq:odg}) and (\ref{eq:gdh}), 
are relevant here (see also (\ref{mixedop}) and (\ref{eq:53})):
\bea
R_{A,\psi}^{\sigma\mu_1 \cdots \mu_{n-1}} & = & i^{n-1} \bar{\psi}
         \gamma^{\{\sigma} \gamma _5 D^{\mu_1}\cdots D^{\mu_{n-1}\}} \psi
                  \ , \label{g1op0}\\
R_{A,G}^{\sigma\mu_1 \cdots \mu_{n-1}} & = & i^{n-1} 
       G^{\{\sigma}{}_{\alpha} D^{\mu_1}
       \cdots D^{\mu_{n-2}} \widetilde{G}^{\mu_{n-1}\} \alpha} \ .
\label{g1op}
\eea
In these expressions, we employ
notation similar to that in (\ref{eq:o}) and (\ref{eq:og}), but
the color indices are suppressed in (\ref{g1op}).
We define the matrix element of these 
composite operators renormalized at $\mu^2$ by
\begin{eqnarray}
\langle P S| R_{A,\psi}^{\sigma\mu_1 \cdots \mu_{n-1}}(\mu^{2})
|P S \rangle
   &=&  2\, a_n (\mu^2 ) S^{\{\sigma} P^{\mu_1}\cdots P^{\mu_{n-1}\}}\ , 
\label{eq:Djn}\\
\langle P S| R_{A,G}^{\sigma\mu_1 \cdots \mu_{n-1}}
(\mu^{2})
|P S \rangle
   &=&  2\, a_n^G (\mu^2 ) 
S^{\{\sigma} P^{\mu_1}\cdots P^{\mu_{n-1}\}}\ , 
\label{eq:DjnG}
\end{eqnarray}
where $a_{n}$ can be defined for a definite flavor structure
(see the discussion below (\ref{adef})).
The moment sum rules for $g_{1}(x, Q^{2})$ read
(we set $\mu^{2} = Q^{2}$)
\begin{eqnarray}
 I_n (Q^2 ) &\equiv& \int_0^1 dx x^{n-1} g_1 (x,Q^2) \nonumber\\
   &=& \frac{1}{2}\, 
   \left[\frac{2}{9}E_{n}^{S}(1, \alpha_{s}(Q^{2}))a_{n}^{q}(Q^{2})
   + E_{n}^{G}(1, \alpha_{s}(Q^{2})) a_n^G (Q^{2})\right. \nonumber \\
   && + \,\left.\frac{1}{6}E_{n}^{NS}(1, \alpha_{s}(Q^{2}))
   \left( a_{n}^{(3)} (Q^{2})+ \frac{1}{3} a_{n}^{(8)}(Q^{2}) \right)\right]\ ,
\label{eq:momsrg}
\end{eqnarray}
where the first (second) line on the r.h.s. gives
the flavor singlet (nonsinglet) part,
and we denote the relevant hard scattering coefficients
as $E_{n}^{j}$, in analogy to the coefficient functions of (\ref{fmoment}).
Note that (\ref{eq:momsrg}) reduces to (\ref{eq:momg1f})
in the LO for a single flavor case. 

The calculation of the hard scattering coefficients at
NLO was performed many years ago for both the
non-singlet~\cite{KO,KO1} and singlet~\cite{KO2} parts.
The corresponding one- and two-loop~\cite{MvNV} anomalous 
dimensions (DGLAP kernels) have been also calculated. 
In the following, we restrict our
discussion to only the first ($n=1$) moment,
which is related to particularly interesting sum rules:
the Bjorken sum rule~\cite{B}(the nonsinglet part) and 
the Ellis-Jaffe sum rule~\cite{EJ} (both the singlet and nonsinglet parts).
Equation (\ref{eq:momsrg}) reads for $n=1$ ($p$: proton, $n$: neutron)
\[  I_1^{p,n} (Q^2 ) = \frac{1}{12}\,  E_{1}^{NS} 
(1 , \alpha_{s}(Q^2))
   \left[ \pm \, a_1^{(3)}(Q^2 ) + \frac{1}{3} 
   a_1^{(8)}(Q^2 ) \right]
      + \frac{1}{9} E_{1}^{S} (1 , \alpha_{s}(Q^2)) 
    \Delta \Sigma (Q^2 )\ ,\]
with
\[ a_1^{(3)}= a^{u}_1 - a^{d}_1 \ ,\ 
    a_1^{(8)}= a^{u}_1 + a^{d}_1
       - 2 a^{s}_1 \ ,\ 
   \Delta \Sigma \equiv a^{u}_1 + a^{d}_1
       + a^{s}_1 = a^{q}_1  \ .\]
Note that the gauge invariant gluon operator (\ref{g1op})
(the corresponding coefficient function) does not exist for $n=1$ 
in a gauge invariant renormalization 
scheme, which we assume here and in the following.
Using $E_{1}^{NS}(1,\alpha_{s}) = E_{1}^{S} (1,\alpha_{s}) 
= 1 - \alpha_s / \pi$ to the one-loop accuracy,
the NLO expression for $I_1^{p,n}$ becomes
\begin{equation}
  I_1^{p,n} (Q^2 ) = 
   \left[ \pm \frac{1}{12}\, a_1^{(3)}(Q^2 ) + \frac{1}{36} 
     a_1^{(8)}(Q^{2})
      + \frac{1}{9}  \Delta \Sigma (Q^2 ) \right]\, 
           \left( 1 - \frac{\alpha_s(Q^{2})}{\pi} \right) \ .
\label{eq:EJ}
\end{equation}
We stress that all systematic QCD analyses of experimental data
have been performed at this NLO accuracy.~\footnote{
For some higher order calculations, see Ref.~\citen{ltv}.}

The fermion bilinear operators which contribute to (\ref{eq:EJ})
are the axial vector currents $\bar{\psi}\gamma^{\sigma}\gamma_{5}\psi$. 
The nonsinglet axial vector current is conserved in (massless) QCD,
so that $a_1^{(3)}$ and $a_1^{(8)}$ of (\ref{eq:EJ}) are
independent of the scale $Q^{2}$
(the corresponding anomalous dimension vanishes
to all orders in perturbation theory).
Therefore, these values can be fixed using information
obtained at low energy.
Since the Bjorken sum rule is 
for the difference $I_{1}^{p} - I_{1}^{n}$ and thus
receives a contribution from only the
nonsinglet channel, we are led to a definite QCD prediction for this sum rule
in the sense that the relevant (nonperturbative) matrix element 
$a_1^{(3)}$ is known. We have $a_1^{(3)} =  \frac{G_A}{G_V} \cong  1.26$ from
the experimental value of the neutron $\beta$-decay.
At the NLO of the QCD correction, the Bjorken sum rule reads
\[ I_1^p(Q^{2}) - I_1^n(Q^{2}) = \frac{1}{6}\frac{G_A}{G_V}
         \left[ 1 - \frac{\alpha_s (Q^{2})}{\pi} \right]\ .\]
This sum rule has been well established by detailed
studies~\cite{emc1,emc2,emc3,emc4} of the experimental data.
On the other hand, the situation for
the flavor singlet channel is quite different than that for
the nonsinglet case.
At the LO approximation, the singlet axial vector current
is conserved, so that $\Delta \Sigma$ of (\ref{eq:EJ})
is also independent of the scale $Q^2$.
However, there is no available information at low energy to fix this.
If we use the information $a_1^{(8)} \cong  0.58$,
which is obtained from the experimental value of
the hyperon $\beta$-decay with the flavor SU(3) symmetry relation,
and \lq\lq assume\rq\rq\  that we can neglect the strange quark
contribution $a_1^s = 0$, we are led to the Ellis-Jaffe sum rule,
\[  I_1^{p,n} (Q^2 ) = 
   \left[ \pm \frac{1}{12}\, a_1^{(3)} + \frac{5}{36} 
     a_1^{(8)}\right] \ .\]
The experimental value for this sum rule is much smaller than
the above prediction. 
This suggests the importance of the contributions
from the strange quark and/or the higher order QCD corrections.
When one goes beyond LO,     
the singlet axial vector current is no longer conserved, due to the
Adler-Bell-Jackiw anomaly.~\cite{abj}
The corresponding anomalous dimension $\widehat{\gamma}_{1}^{A}$ 
starts at the two-loop order~\cite{KO2} (compare with (\ref{eq:Pg})): 
\[   \left[\widehat{\gamma}^{A}_1 (\alpha_{s})\right]_{qq} = 
     12 C_{2}(R)N_{f} \left(\frac{\alpha_{s}}{4\pi}\right)^{2}+ \cdots \ .\]
Here $C_2 (R) = 4/3$ for $N_{c}=3$, and $N_{f} (=3)$ is 
the number of flavors.
This makes $\Delta \Sigma(Q^{2})$ of (\ref{eq:EJ}) 
scale dependent.~\cite{KO2,j}
Because there is no operator mixing in this moment,
the corresponding NLO $Q^{2}$-evolution is obtained by
(\ref{eq:NLONS}) with $\gamma_{n(0)}^{V} \rightarrow 0$,
$\gamma_{n(1)}^{V} \rightarrow 12 C_{2}(R) N_{f}$: 
\[   \Delta\Sigma (Q^2) = 
  \left( 1 + \frac{\alpha_s(Q^{2}) }{\pi} \frac{6N_{f}}{33-2N_{f}} 
      \right) \Delta\Sigma (\infty) \ .\]
The scale dependence is weak because it is purely a two-loop effect.

Concerning the disagreement between the experimental data and
the Ellis-Jaffe sum rule, and the so-called
\lq\lq spin crisis\rq\rq\ problem,
namely $\Delta \Sigma \simeq 0.2$
from the analysis of the DIS data, which corresponds to the total
spin carried by the quarks in the naive parton model,
the role of the polarized gluon has received much attention. 
As noted above, the gauge invariant local operator 
which would contribute to (\ref{eq:EJ}) does not exist.
On the other hand, the polarized gluon distribution 
$\Delta {\cal G}(x)$ has been defined in (\ref{eq:gdh}) based
on a gauge invariant nonlocal operator, and their higher moments 
certainly give the local operator (\ref{g1op}) 
as $\int dx x^{n-1} \Delta {\cal G} (x) = a_n^G$
for $n\ge 3$ (see (\ref{eq:DjnG})).
Where does the first moment of $\Delta {\cal G}(x)$ go?
To resolve this puzzle,
let us consider the first moment of 
$\Delta {\cal G}(x)$ in detail.~\cite{mano}
In this case, the situation is somehow subtle, due to
the presence of $x$ in the denominator of (\ref{eq:gdh}).
As usual, this type of singularity in the distribution functions 
should be understood as a principal
value distribution (in the mathematical sense). 
From the charge conjugation property (\ref{eq:chargeg}), we obtain
\bean
   \lefteqn{\int_{0}^{1} dx \Delta {\cal G}(x)}\\
  &=&   \frac{i}{2 \pi} \int_{-\infty}^{+\infty} dx \,{\rm P}
         \left( \frac{1}{x} \right) \int d\lambda  
              e^{i \lambda x} \langle P S_{\parallel}|
                   {\rm tr}\, w_{\alpha} G^{\alpha \mu}(0)
         [ 0 , \lambda w ]w^{\beta} \widetilde{G}_{\beta \mu}(\lambda w )
                     |P S_{\parallel} \rangle \ ,
\eean
where ${\rm P} (1/x)$ denotes the Cauchy principal part.
Since $\int dx {\rm P} (1/x) e^{i \lambda x} =
i\pi \varepsilon (\lambda )$, we obtain for the r.h.s.,
\[  \frac{- 1}{2} \int d\lambda \,\varepsilon (\lambda )   
          \, \langle P S_{\parallel}|
                   {\rm tr}\, w_{\alpha} G^{\alpha \mu}(0)
         [ 0 , \lambda w ]w^{\beta} \widetilde{G}_{\beta \mu}(\lambda w )
                     |P S_{\parallel} \rangle \ .\]
This result shows that, in general, the first moment of 
$\Delta {\cal G}(x)$ cannot be expressed by a local operator.
However, if one chooses the light-cone gauge, $w \cdot A = 0$,
we obtain $[0, \lambda w] \rightarrow 1$ and
$w_{\nu} G^{\nu \mu}(\lambda w) 
= (w \cdot \partial ) A^{\mu}(\lambda w) =
(d / d\lambda ) A^{\mu}(\lambda w)$, so that,
\begin{eqnarray}
 \int_{0}^{1}dx \Delta {\cal G}(x) &=& 
         \frac{1}{2} \int d\lambda \varepsilon (\lambda )   
          \langle P S_{\parallel}|
         {\rm tr}\, \frac{d}{d \lambda } A^{\mu} (\lambda w )
           w^{\beta} \widetilde{G}_{\beta \mu}(0) |P S_{\parallel} \rangle 
\nonumber \\
      &=& - \, \langle P S_{\parallel}|
         {\rm tr}\,  A^{\mu} (0)
             w^{\beta} \widetilde{G}_{\beta \mu}(0) |P S_{\parallel} \rangle 
\ , \label{eq:CS}
\end{eqnarray}
provided that the matrix element 
$\langle P S_{\parallel}|{\rm tr}\, A^{\mu} (\lambda w)
w^{\beta} \widetilde{G}_{\beta \mu}(0) |P S_{\parallel} \rangle$ 
vanishes when $\lambda \to \pm \infty$.
Here, ${\rm tr}\,A^{\mu} (0) w^{\beta} \widetilde{G}_{\beta \mu}(0)
= w^{\beta}K_{\beta}/2$
with
\[ K_{\beta} \equiv 2 {\rm tr}\,
  A^{\mu} (0) \left( \widetilde{G}_{\beta \mu}(0)
  - \frac{1}{3}\,\epsilon_{\beta \mu \lambda \sigma}
     A^{\lambda}(0)A^{\sigma}(0) \right) \]
being the Chern-Simons current.
This argument indicates that the first moment of $\Delta {\cal G}(x)$
can be expressed by a \lq\lq local operator\rq\rq\ in 
a specific (light-cone) gauge.
Once the first moment of $\Delta {\cal G}(x)$
is introduced as a finite quantity,
it is in principle possible to {\it define} a scheme~\cite{BQ90} where
$\Delta \Sigma$ of (\ref{eq:EJ}) is replaced by some
mixture of $a_{1}^{u} + a_{1}^{d} + a_{1}^{s}$
and (\ref{eq:CS}), that is $a_{1}^{q} \neq \Delta \Sigma
\equiv a_{1}^{u} + a_{1}^{d} + a_{1}^{s}$.
Among possible schemes of this type, the so-called ``AB
scheme''~\cite{arccm} is often used in the
literature.~\footnote{For the phenomenological studies on the $g_1$ structure
function and the spin crisis problem, see, e.g., Ref.~\citen{bodo}.}
However, it should be stressed that any scheme is acceptable
on general grounds of QCD, as explained in \S4.

\subsection{The structure function $g_{2}(x, Q^{2})$}

There have been many works on the transverse spin 
structure function $g_{2}(x,Q^{2})$.~\cite{SV,BKL,Rat,DMu,JiChou}
The main subject in this section regards the derivation of the $Q^2$-evolution
of its twist-3 part.
In contrast to the traditional covariant approach based on the
local composite operators, many authors have employed other approaches
to investigate it.~\cite{BB,BKL,DMu}
These works are based on the renormalization of the nonlocal operators
in the light-like axial gauge, or in the background field gauge.
We develop a framework based on the local composite operators
in a covariant gauge.~\cite{KOD1,KOD2,KOD3,KNTTY97}
According to the discussion in \S4, this approach is 
convenient and straightforward.

To identify the relevant twist-3 operators and a relation among them,
one convenient method is to utilize the exact operator
identities among the gauge invariant nonlocal operators.
We first recall that we have already obtained a twist-3 local operator
$R^{[ \sigma \{ \mu _1 ] \cdots \mu_{n-1}\} }$
of (\ref{eq:53}), which we denote as 
$R_{n,F}^{\sigma\mu_{1}\cdots \mu_{n-1}}$:
\bea
  R_{n,F}^{\sigma\mu_{1}\cdots \mu_{n-1}} &=&
         \frac{i^{n-1}}{n} \Bigl[ (n-1) \overline{\psi}
       \gamma^{\sigma}\gamma_5 D^{\{\mu_1} \cdots D^{\mu_{n-1}\}}
          \psi \nonumber\\
   & & \quad - \sum_{l=1}^{n-1} \overline{\psi} 
       \gamma^{\mu_l }\gamma_5 D^{\{\sigma} D^{\mu_1} \cdots D^{\mu_{l-1}}
            D^{\mu_{l+1}} \cdots D^{\mu_{n-1}\}}
             \psi \Bigr] \ . \label{quark}
\eea
As mentioned in \S2, there also exist other gauge invariant operators,
the number of which increases with spin
(moment of the structure function), and they mix with each other. 
To infer these additional operators,
we note the following operator identity, which can be
obtained by explicit differentiation
(for the derivation, see Refs.~\citen{BB} and ~\citen{BB98}):
\bea
  \lefteqn{ z_{\mu} \left(
     \frac{\partial}{\partial z_{\mu}}
        \bar{\psi}(0) \gamma^{\sigma} \gamma_5 [0, z] \psi( z )
  -  \frac{\partial}{\partial z_{\sigma}}
        \bar{\psi}(0) \gamma^{\mu} \gamma_5 
       [0, z] \psi( z ) \right)}\nonumber\\
  &=& \int^1_0 dt \bar{\psi}(0) [0, tz] \sslash{z} \left\{
          i \gamma_5 \left( t - \frac{1}{2} \right) g G^{\sigma\rho}(tz)
           z_{\rho} - \frac{1}{2} 
g \tilde{G}^{\sigma\rho}(tz) z_{\rho} \right\} 
     [tz , z] \psi ( z ) \nonumber\\
   & & + \,2  m_{q} \bar{\psi}(0) \gamma_5 \sigma^{\sigma\rho} z_{\rho}
       [ 0 , z] \psi (z )\nonumber\\
   & & + \, \left[ \bar{\psi}(0)
           \gamma_5 \sigma^{\sigma\rho} z_{\rho} [ 0 , z]
        ( i \lslash{D} - m_{q} ) \psi (z)  - \bar{\psi}(0)
          ( i \overleftarrow{\lslash{D}} + m_{q}) 
          \gamma_5 \sigma^{\sigma\rho} z_{\rho} [ 0 ,z] \psi ( z )
                \right] \nonumber \ .\\
\label{eq:identity}
\eea
This holds for each quark flavor.
Here $m_{q}$ represents the quark mass generically.
This identity is exact through twist-3.
We have neglected the twist-4 contributions and the total derivatives which are
irrelevant for the forward matrix elements.  

Taylor expanding (\ref{eq:identity}) around $z_{\mu} = 0$, 
it is easy to see that the l.h.s. generates 
operators (\ref{quark}) with $n= 2, 3, \cdots$.
Similarly, we find the following twist-3 operators from the r.h.s.
(flavor matrices and subtraction of trace terms suppressed):
\bea
  R_{n,l}^{\sigma \mu_{1} \cdots \mu_{n-1}} &=& 
     \frac{1}{2n}\left(V_{l}-V_{n-1-l} + U_{l} + U_{n-1-l} \right)
     \;\;\; (l = 1, \cdots, n-2) \ , \label{eq:rl}\\
  R_{n,m}^{\sigma \mu_{1} \cdots \mu_{n-1}}&=& - i^{n}
         \,{\cal S}\,m_{q}
          \overline{\psi}\gamma^{\sigma}\gamma_5
           D^{\mu_{1}}\cdots D^{\mu_{n-2}}\gamma^{\mu_{n-1}}
          \psi \ , \label{eq:rm}\\
  R_{n,E}^{\sigma \mu_{1} \cdots \mu_{n-1}} &=& i^{n-2}\frac{n-1}{2n}
   {\cal S} \left[ \bar{\psi}(i \lslash{D} - m_{q} ) 
   \gamma^{\sigma} \gamma_5 D^{\mu_{1}} \cdots D^{\mu_{n-2}}
        \gamma^{\mu_{n-1}} \psi \right.\nonumber \\
     & & \qquad\quad + \,\left.\bar{\psi} \gamma^{\sigma} \gamma_5
    D^{\mu_{1}} \cdots D^{\mu_{n-2}}\gamma^{\mu_{n-1}}(i\lslash{D}
      - m_{q} ) \psi \right] \ , \label{eq211}
\eea
where
\bean
   V_{l}&=& - i^{n} g {\cal S} \bar{\psi} 
       D^{\mu_1} \cdots G^{\sigma \mu_l} \cdots
       D^{\mu_{n-2}}\gamma^{\mu_{n-1}}\gamma_{5} \psi \ ,\\
   U_{l}&=& - i^{n-1} g {\cal S}\bar{\psi}
      D^{\mu_1} \cdots \tilde{G}^{\sigma \mu_l} \cdots
      D^{\mu_{n-2}}\gamma^{\mu_{n-1}} \psi \ .
\eean
Here, ${\cal S}$ denotes the symmetrization over $\mu_i$.
The $R_{n,l}$ correspond to the first term of (\ref{eq:identity});
it can be shown that this term is related to the integral
of a quark-gluon correlation function 
$\Xi(x, x') = \Psi(x, x') + \Psi(x', x) + \widetilde{\Psi}(x, x')
- \widetilde{\Psi}(x', x)$ (see (\ref{eq:V3}) and (\ref{eq:A3})),
and thus the $R_{n,l}$ are given by the double moment
of $\Xi(x, x')$ in analogy to (\ref{eq:V3L}).~\footnote{
It is possible to derive a relation 
like (\ref{eq:solhL}) and (\ref{eq:sole}) below
among $g_{T}(x)$, $\Xi(x, x')$, $m_{q}\delta q(x)$
and $\Delta q(x)$ from (\ref{eq:identity}).}
The $R_{n,E}$ are the gauge invariant EOM operators,
and the $R_{n,m}$ represent the quark mass effects,
which cause the mixing of the chiral-odd operators.
In the flavor nonsinglet case, neither the pure gluonic operators nor the
BRST invariant alien operators is involved, and
(\ref{quark}) and (\ref{eq:rl})--(\ref{eq211})
form a (over)complete set of the twist-3 operators
(except for the BRST noninvariant EOM operators, as discussed below).

From the above Taylor expansion of (\ref{eq:identity}),
we recognize that the local operators 
(\ref{quark}) and (\ref{eq:rl})--(\ref{eq211})
satisfy the relation~\footnote{
This operator identity can also be obtained 
by applying the relation $[D_{\mu}, D_{\nu}] = -ig G_{\mu\nu}$ 
to (\ref{quark}).~\cite{JJ2,SV}}
\be
 R_{n,F}^{\sigma\mu_{1}\cdots \mu_{n-1}} = 
         \frac{n-1}{n} \,R_{n,m}^{\sigma\mu_{1}\cdots \mu_{n-1}}
             + \sum_{l=1}^{n-2} (n-1-l)
                 R_{n,l}^{\sigma\mu_{1}\cdots \mu_{n-1}} +
             R_{n,E}^{\sigma\mu_{1}\cdots \mu_{n-1}} \ . \label{operel}
\ee
This equation implies that these operators are not all independent.
Therefore we have freedom in the choice of the independent operator
basis to work out renormalization. 
We stress here that we can take any basis, and the final results
do not depend on the choice.~\cite{KOD1}
A convenient choice of the independent operators is 
(\ref{eq:rl}), (\ref{eq:rm}) and (\ref{eq211}).
For the $n$-th moment, $n$ gauge-invariant operators
participate in the renormalization for the nonsinglet channel.

The RG equation for the matrix elements of the local composite operators 
read (compare with (\ref{eq:RG3}))
\be
\mu \frac{d}{d\mu}
 f_{nk} (\mu^2 ) + \sum_j \left[ \Gamma^{A}_n  (\alpha_{s}(\mu^2 )) 
\right]_{kj}\, 
   f_{nj} (\mu^2 )= 0 \ , \quad  (k,j = 1, \cdots, n-2, m) \label{rgeqg2}
\ee
where
\bean
  \langle P S | R_{n,l}^{\sigma \mu_{1} \cdots \mu_{n-1}} (\mu^2 )
              | P S  \rangle &=& 2 \, f_{nl} (\mu^2 )\,
              ( S^{\sigma}P^{\mu_1} - S^{\mu_1}P^{\sigma})
                    P^{\mu_2} \cdots P^{\mu_{n-1}}\ ,\\
  \langle P S | R_{n,m}^{\sigma \mu_{1} \cdots \mu_{n-1}} (\mu^2 )
              | P S  \rangle &=& 2 \, f_{nm} (\mu^2 )\,
              ( S^{\sigma}P^{\mu_1} - S^{\mu_1}P^{\sigma})
                    P^{\mu_2} \cdots P^{\mu_{n-1}} \ .
\eean
Equation (\ref{rgeqg2}) at the LO is solved to give,
similarly to (\ref{eq:LOS}),
\begin{equation}  
f_{nk} (Q^2 ) = \sum_j \left[ L^{X_n / \beta_0} \right]_{kj}
      f_{nj} (\mu^2 ) \ ,
\label{eq:solfnk}
\end{equation}
where we defined the mixing matrix $X_{n}$ as
$  \Gamma^{A}_n (\alpha_{s}) = (\alpha_s/2 \pi) X_n\ .$
     
The anomalous dimension (matrix) $\Gamma^{A}_n (\alpha_{s}(\mu^2 ))$ 
can be obtained by calculating the renormalization constants for
the $n$ operators ((\ref{eq:rl})--(\ref{eq211}))
according to the general argument in \S4.
The calculation proceeds in the standard way.
We multiply the operators by the light-like vector $w^{\mu_i}$
to symmetrize the Lorentz indices and eliminate the trace terms.
We calculate the quark-quark-gluon off-shell three-point Green's functions
with the insertion of the relevant operators.
As discussed in \S4, the gauge (BRST) noninvariant EOM operators
are also required to complete the renormalization.
These operators are obtained~\cite{KOD1,KOD5}
by replacing some of the covariant
derivatives $D_{\mu_i}$ by the ordinary derivatives $\partial_{\mu_i}$
or $A_{\mu_i}$ in (\ref{eq211}). 
This causes the actual calculation to become very complicated
because there are $(2^{n - 2}-1)$ operators of this kind.
This technical problem can be overcome by introducing
the vector $\Omega_{\mu}$, which satisfies $w^{\mu}\Omega_{\mu} =
0$,~\cite{KT} and by contracting the Lorentz index of the external gluon 
with this vector.
This has two merits: First, the tree vertices of the
gauge invariant and noninvariant EOM operators coincide.
Thus essentially only one EOM operator is involved in the mixing.
Second, the structure of the vertices for the twist-3 operators
are extremely simplified, and the computation becomes more tractable.
An explicit one loop calculation appears in
Refs.~\citen{KOD1} and \citen{KOD2}.
The results are consistent with the general argument in \S4,
namely the renormalization constants take the following form:
\[ \bmat{c}
 R_{n,l}\\
 R_{n,m}\\
 R_{n,E}
\emat_0
  =  \bmat{ccc}
     ({\cal Z}_n )_{lj} & ({\cal Z}_n )_{lm} & ({\cal Z}_n )_{lE}\\
     0 & ({\cal Z}_n )_{mm} & 0 \\
     0 & 0 & ({\cal Z}_n )_{EE}
     \emat
   \bmat{c}
 R_{n,j}\\
 R_{n,m}\\
 R_{n,E}
\emat_R \ .
\]
\setcounter{footnote}{0}
The necessary anomalous dimensions are unambiguously determined
from $({\cal Z}_n )_{lj}$ , $({\cal Z}_n )_{lm}$ and $({\cal Z}_n )_{mm}$.
For details, we refer the reader to 
Refs.~\citen{KOD1} and \citen{KOD2}.~\footnote{For 
the description of other processes like DY, $\Gamma_{n}^{A}$
for even $n$, as well as for odd $n$, are required
(see (\ref{eq:ALT})), and these anomalous dimensions have 
recently been worked out in connection with renormalization
of light-cone wave functions.~\cite{KNT98}}

Now let us consider the moment sum rule for $g_2(x, Q^{2})$.
We define the matrix elements of composite operator (\ref{quark})
by (\ref{eq:dn}) with 
$R^{[ \sigma \{ \mu _1 ] \cdots \mu_{n-1}\} } \rightarrow
R_{n,F}^{\sigma\mu_{1}\cdots \mu_{n-1}}(\mu^{2})$
and $d_{n} \rightarrow d_{nF}(\mu^{2})$.
The moment sum rule for the genuine
twist-3 part of $g_2(x, Q^{2})$ reads
(we consider a single flavor case for simplicity)
\bea
    \lefteqn{\int_0^1 dx x^{n-1} \bar{g}_2 (x,Q^2)
   =  \frac{n-1}{2n} d_{nF} (Q^2 ) \bar{E}_n^F ( 1 , \alpha_{s}(Q^2))}\nonumber\\
     &+& \, {1\over 2} \left[ f_{nm}(Q^2) \bar{E}_n^m ( 1 , \alpha_{s}(Q^2))
         + \sum_{l=1}^{n-2} f_{nl} (Q^2) \bar{E}_n^l ( 1 , \alpha_{s}(Q^2)) \right],
                          \label{g2sumrule}
\eea
where $\bar{E}_n^k ( 1 , \alpha_{s}(Q^2))$ 
is the hard scattering part corresponding to each operator.
Note that, at LO, (\ref{g2sumrule}) is consistent 
with (\ref{eq:genuine}) (see below).
From the constraint (\ref{operel}), the matrix elements satisfy
\[   {{n-1}\over n}d_{nF} (Q^2 )={{n-1}\over n}
          f_{nm} (Q^2 ) + \sum_{l=1}^{n-2}(n-l-1) f_{nl}(Q^2 ) \ .\]
It is to be noted that the explicit form of the $Q^2$-evolution
of each term in (\ref{g2sumrule}) depends on the choice of operator
basis. The anomalous dimensions as well as the hard parts
$\bar{E}_n^{k} (1 , \alpha_{s} (Q^2))$ take different forms,
depending on the basis. 
Let us demonstrate this point at LO:
in the basis of independent operators, which includes $R_{n,F}$,
it turns out that the hard parts become
\[  \bar{E}_n^{F} (1,0) = 1 \ , \quad
    \bar{E}_n^{m} (1,0) = \bar{E}_n^{l} (1,0) = 0 \ ,\]
and the result coincides with (\ref{eq:genuine}).
On the other hand, if we eliminate $R_{n,F}$, we have
\[  \bar{E}_n^{m} (1,0)= {{n-1}\over n} \ ,
   \quad \bar{E}_n^{l} (1,0) = n-l-1 \ .\]
However, the moment (\ref{g2sumrule}) itself remains the same, as
should be the case.

Now the present discussion reveals that the $Q^{2}$-evolution
of the moment (\ref{g2sumrule}) obeys a quite sophisticated 
renomalization mixing due to (\ref{eq:solfnk}).
Even if we neglect the quark mass operator (\ref{eq:rm}),
$(n-2)$ independent terms contribute to
the evolution of the $n$-th moment, and mix with each other.
As a result, the comparison of this $Q^{2}$-evolution 
with experimental data is very difficult;
the numerical diagonalization
of the $(n-2)\times (n-2)$ mixing matrix, as well as estimates of 
the $(n-2)$ independent nonperturbative parameters.
However, we note that in the $N_{c} \rightarrow \infty$ limit,
this complicated renormalization mixing
is drastically simplified: Each component of
the mixing matrix of (\ref{eq:solfnk}), 
$[X_{n}]_{kl}$ $(k,l = 1, \cdots, n-2)$,
involves two Casimir operators, 
$C_{2}(R) = \frac{N_{c}^{2}-1}{2N_{c}}$ and $C_{2}(A) = N_{c}$.
In the large $N_{c}$ limit, that is, 
with the replacement $C_{2}(R) \rightarrow \frac{N_{c}}{2}$,
the following exact relation holds:~\cite{ABH}
\be
   \sum_{k=1}^{n-2}(n-k-1) \left[X_{n}\right]_{kl}
   = (n-l-1)\varpi_{n} \ ,\label{eq:g2LN}
\ee
where $\varpi_{n}$ is the {\it smallest} eigenvalue
of $X_{n}$ in this limit and is given by
\be
 \varpi_{n}= 2N_{c} \left( \sum_{j=1}^{n-1}\frac{1}{j}
               - \frac{1}{4} + \frac{1}{2n} \right) \ .\label{eq:varpi}
\ee
As a result, the moment obeys simple $Q^{2}$-evolution
without mixing,
\be
 \int_{0}^{1}dx x^{n-1} \bar{g}_{2}(x, Q^{2})
  = L^{\varpi_{n}/\beta_{0}} \int_{0}^{1}dx x^{n-1} \bar{g}_{2}(x, Q_{0}^{2}) \ .
\label{eq:g2LNsol}
\ee
Therefore, in this limit,
the twist-3 nonsinglet structure function $\bar{g}_{2}(x, Q^{2})$
obeys the $Q^{2}$-evolution of the twist-2 nonsinglet type,
and involves one single nonperturbative parameter.
We discuss further implications of this $N_{c} \rightarrow \infty$
simplification in \S6.2, in connection with similar simplifications
of the chiral-odd twist-3 distribution functions.

In the flavor singlet case, the situation becomes much more complicated.
We have operators that are made of two or three gluon fields, 
as well as the BRST-invariant alien operators, in addition to the above set of 
operators bilinear in quark fields (with the flavor matrix being removed):
\bea
    T_{n,G}^{ \sigma \mu_{1} \cdots \mu_{n-1} }
       & = &
      i^{ n-1 } \mbox{$ \cal{S} \mit $} \mbox{ $ \cal{A} \mit $ }
      \mbox{$ \cal{S} \mit $} \left[ \widetilde{G}^{ \nu \mu_{1} }
     D^{ \mu_{2} } \cdots D^{ \mu_{n-1} } G_{\nu}{}^{\sigma }
       \right] , \label{eqn:1} \\
    T_{n,l}^{ \sigma \mu_{1} \cdots \mu_{n-1} }
       & = &
      i^{n-2} g \mbox{ $ \cal{ S } \mit $ } \left[
     {G}^{\nu \mu_{1}} D^{ \mu_{2} } \cdots \widetilde{G}^{ \sigma \mu_{l}}
      \cdots D^{ \mu_{n-2} } G_{ \nu }{}^{\mu_{n-1} } \right]
          \label{eqn:2}\\
       & & 
\qquad\qquad\qquad\qquad\qquad \left( l=2,\ldots,\frac{n-1}{2}
                   \right), \nonumber\\
   T_{n,B}^{ \sigma \mu_{1} \cdots \mu_{n-1} }
      & = &
     i^{n-1} {\cal S} \{ \widetilde{G}^{ \sigma \mu_{1} } D^{ \mu_{2} }
     \cdots D^{ \mu_{n-2} } \}^{a} \left\{
      - \frac{1}{ \alpha } \partial^{ \mu_{n-1} }(  \partial^{ \nu }
      A^{a}_{ \nu } ) \right. \nonumber\\
        & & \qquad\qquad\qquad\qquad\qquad + \left. 
      g f^{abc}(  \partial^{ \mu_{n-1}} \overline{\chi}^{ b } ) \chi ^{c}
     \right\}  \label{eqn:3}\\
   T_{n,E}^{ \sigma \mu_{1} \cdots \mu_{n-1} }
      & = &
       i^{n-1} {\cal S}\{ \widetilde{G}^{ \sigma \mu_{1} } D^{ \mu_{2} }
        \cdots D^{ \mu_{n-2} } \}^{a} \{
       \left ( D^{ \nu }G_{ \nu}{}^{ \mu_{n-1} } \right )^{a} 
                                   \nonumber \\
     & &  + \ g \overline{ \psi } t^{a} \gamma^{ \mu_{n-1} } \psi
         + \frac{1}{ \alpha } \partial^{ \mu_{n-1} }(  \partial ^{ \nu }
            A^{a}_{ \nu }) -
        g f^{abc}(  \partial^{ \mu_{n-1}} \overline{ \chi } ^{ b } ) \chi ^{c}
              \}  . \label{eqn:4}
\eea
Here $\cal{A}\mit$ antisymmetrizes $\sigma$ and $\mu_{1}$.
The color structure of these operators
should be understood similarly to (\ref{eq:og}).
$\chi$ and $\overline{\chi}$ are the ghost fields,
$t^{a}$ is the color matrix in the fundamental representation,
$[t^a , t^b ] = i f^{abc} t^c $, and $\alpha$ is the gauge parameter.
$T_{n,l}$ is trilinear in the gluon field strength $G_{ \mu \nu }$ and 
the dual tensor, and thus represents the effect of the three
gluon correlations. It satisfies the symmetry relation
$T_{n,l}^{ \sigma \mu_{1} \cdots \mu_{n-1} }
= T_{n,n-l}^{ \sigma \mu_{1} \cdots \mu_{n-1} }$
due to the Bose statistics of the gluon.
$T_{n,E}$ is the gluon EOM operator which is BRST invariant.
$T_{n,B}$ is the BRST invariant alien
operator, which is the BRST variation of the operator,
\[ i^{n-1} {\cal S}\{ \widetilde{G}^{ \sigma \mu_{1} } D^{ \mu _{2} }
  \cdots D^{ \mu_{n-2} } \}^{a} \partial^{\mu_{n-1}}\overline{\chi}^{a}
                        \ .\]

The gluon bilinear operator (\ref{eqn:1})
is related to the trilinear ones (\ref{eqn:2}) by~\cite{KOD3,KNTTY97}
\bea
  T_{n,G}^{ \sigma \mu_{1} \cdots \mu_{n-1}}  &=&
    \sum^{ \frac{n-1}{2} }_{l = 2 } \left[ 
     \frac{( l - 2 )C^{n -2}_{l}}{2(n-1)} -
           \frac{( n - l - 2 )C^{n-2}_{n-l}}{2(n-1)}\right. \nonumber \\
      & & \qquad\qquad  -\ \left. \frac{n(n-2l)C^{n-2}_{l-1}}{2(n-2)(n-1)}
              + (-1)^{l+1} \right] \ 
         (-1)^{l} \,T_{n,l}^{ \sigma \mu_{1} \cdots \mu_{n-1}}\nonumber \\
      &+&  \frac{1}{n-1} \left\{
         T_{n,E}^{ \sigma \mu_{1} \cdots \mu_{n-1}}
          + T_{n,B}^{ \sigma \mu_{1} \cdots \mu_{n-1}}
          + \sum^{n-2}_{l = 1} n C^{n - 3}_{l-1} (-1)^{l+1} 
           R^{ \sigma \mu_{1} \cdots \mu_{n-1}}_{n,l}\right\},
                          \nonumber\\
\label{eq:oi}
\eea
where $C_{r}^{n}=n!/[ r!(n - r)! ]$. To derive this relation,
we have used $[D_{\mu}, D_{\nu}] = -ig G_{\mu \nu}$ and 
\[ D_{\sigma}G_{\nu \alpha} + D_{\nu} G_{\alpha \sigma}
           + D_{\alpha} G_{\sigma \nu} = 0, \quad
           D_{\sigma}\widetilde{G}_{\nu \alpha} + D_{\nu}
            \widetilde{G}_{\alpha \sigma}
            + D_{\alpha} \widetilde{G}_{\sigma \nu} =
           \epsilon_{\nu \alpha \sigma \rho} D_{\lambda} G^{\lambda \rho}, \]
where the first identity is the usual Bianchi identity
and the second one is a consequence of the relation
\[ g_{\mu \nu} \epsilon_{\alpha \beta \gamma \delta}
 = g_{\mu \alpha} \epsilon_{\nu \beta \gamma \delta}
 + g_{\mu \beta} \epsilon_{\alpha \nu \gamma \delta}
 + g_{\mu \gamma} \epsilon_{\alpha \beta \nu \delta}
 + g_{\mu \delta} \epsilon_{\alpha \beta \gamma \nu} \ .\]
As a result of (\ref{eq:oi}),~\footnote{When one takes
the physical matrix element of (\ref{eq:oi}), the EOM and the
BRST invariant alien operators drop out ((\ref{eq:EOMV}) and
(\ref{eq:BRSTV})), and (\ref{eq:oi}) becomes the
identity suggested in Ref.~\citen{BKL}.} 
we can conveniently choose a set of independent operators as
(\ref{eqn:2}) -- (\ref{eqn:4}). For the $n$-th moment,
these $\frac{n + 1}{2}$ independent operators
mix with each other and with the $n$ gauge-invariant operators
bilinear in the quark fields discussed above under
the renormalization.~\footnote{Up to the gauge (BRST)
noninvariant EOM operators.} 
An explicit calculation is given for the $n=3$ case in
Ref.~\citen{KOD3}, and the result is again consistent with the general
argument in \S4.

Finally, let us comment on the Burkhardt-Cottingham sum rule~\cite{bc}
which corresponds to the first moment of $g_2(x, Q^{2})$.
Due to the lack of local operators for $n=1$ moment, 
the operator analysis suggests the sum rule
\[  \int_0^1 d x g_2 (x , Q^2 ) = 0\ .\]
The validity of this sum rule has been a long-standing
problem.~\cite{JJ2}
However, it was shown that this sum rule is satisfied in 
QCD perturbation theory.~\cite{KOD4}

\section{Chiral-odd structure functions}
\label{sec:co}

This section is devoted to discussion of
chiral-odd distribution functions of twist-2 and -3.
As mentioned in \S\ref{sec:qc},
these distributions are inaccessible to totally inclusive
DIS because of their chiral property,
and thus there exist no experimental data for them. 
These ``new'' distributions are expected to open
a new window to explore the nucleon spin structure,
and several experiments designed to access them have been proposed.
For example, the twist-2 distribution
$\delta q(x, \mu^{2})$ of (\ref{eq201}) and (\ref{eq:odh})
can be measured through the polarized DY process with
two transversely polarized proton beams,
where the LO double transverse-spin asymmetry
$A_{TT}$ reads~\cite{RS}
\begin{equation}
A_{TT} = a_{TT} \frac{
\sum_{f}
({{\cal Q}^{(el)}}^{2})_{ff}\left[
\delta q^{f}(x_{1}, Q^{2}) \delta q^{\bar{f}}(x_{2}, Q^{2})
+ (f \leftrightarrow \bar{f} ) \right]}
{\sum_{f}
({{\cal Q}^{(el)}}^{2})_{ff} \left[
q^{f}(x_{1}, Q^{2})  
q^{\bar{f}}(x_{2}, Q^{2}) + (f \leftrightarrow \bar{f} )
\right]}\ ,
\label{eq:ATT}
\end{equation}
up to ${\cal O}(M^{2}/Q^{2})$ corrections.
Here $f = u, d, s, \cdots$, 
$a_{TT}$ is the partonic asymmetry 
corresponding to the short-distance part,
and we set $\mu^{2} = Q^{2}$ with $Q$ the dilepton mass.
The variables
$x_{1}$ and $x_{2}$ refer to the parton's light-cone momentum fraction
in the two nucleons labeled as ``1'' and ``2'', respectively.
Other processes to measure 
$\delta q(x, Q^{2})$ include the direct photon production,
two-jet production, and heavy-quark production in transversely
polarized nucleon-nucleon collisions,~\cite{XJi92,JS96}
and also semi-inclusive DIS
observing a certain hadronic final state 
with one~\cite{JJ93} or two pions~\cite{JJT97} 
or a polarized $\Lambda$~\cite{JAFFE96}
in the current fragmentation region.

On the other hand, the twist-3 distribution functions
$h_{L}(x, \mu^{2})$ and $e(x, \mu^{2})$
of (\ref{eq201}) and (\ref{eq:scalar})
usually constitute small corrections to the
leading twist-2 term, so it is difficult to extract
them experimentally.
For instance, $e(x, \mu^{2})$ contributes as 
an ${\cal O}(M^{2}/Q^{2})$ correction
to (\ref{eq:do}) for the DY process
between two unpolarized nucleons.~\cite{JJ}
However, $h_{L}(x, \mu^{2})$ is somewhat immune
to this difficulty,
like the chiral-even twist-3 distribution $g_{T}(x, \mu^{2})$
discussed in the last section.
$h_{L}(x, \mu^{2})$ reveals itself as
a leading contribution to the polarized DY process
between the longitudinally and transversely polarized proton beams.
The corresponding LO double spin asymmetry $A_{LT}$ is given by~\cite{JJ}
\bea
\lefteqn{A_{LT} =}\nonumber\\
&& a_{LT} \frac{\sum_{f}
({{\cal Q}^{(el)}}^{2})_{ff} \left[
\Delta q^{f}(x_{1}, Q^{2}) x_{2} g_{T}^{\bar{f}}(x_{2}, Q^{2})
+ x_{1} h_{L}^{f}(x_{1}, Q^{2}) \delta q^{\bar{f}}(x_{2}, Q^{2}) 
+ (f \leftrightarrow \bar{f} ) \right]}
{\sum_{f}
({{\cal Q}^{(el)}}^{2})_{ff}
\left[ q^{f}(x_{1}, Q^{2})  q^{\bar{f}}(x_{2}, Q^{2})
+ (f \leftrightarrow \bar{f} ) \right]}\ ,\nonumber\\
\label{eq:ALT}
\eea
up to corrections of twist-4 and higher.
Here the nucleon ``1''
(``2'') is longitudinally (transversely) polarized,
and $a_{LT}$ is the partonic asymmetry.
Actually, $a_{LT}$ is of ${\cal O}(M/Q)$, 
while $a_{TT} = {\cal O}\left( (M/Q)^{0}\right)$.
Thus $A_{LT}$ is smaller by a factor of $M/Q$ compared to $A_{TT}$.
This is similar to the situation with the chiral-even
structure function $g_{2}(x, Q^{2})$
compared to $g_{1}(x, Q^{2})$.

These chiral-odd distribution functions
will be measured at BNL, DESY, CERN, etc.,
in the near future.~\cite{BUNCE}
In particular, RHIC at BNL will provide us with the first data
of $\delta q(x, Q^{2})$ and $h_{L}(x, Q^{2})$
through $A_{TT}$ and $A_{LT}$.
In view of this, it is essential to work out a detailed study
of these distribution functions based on QCD,
including $Q^{2}$-evolution.~\cite{KOI98}

\subsection{The transversity distribution $\delta q(x, \mu^{2})$}

The transversity distribution $\delta q (x, \mu^{2})$
constitutes a complete set of the nucleon's 
twist-2 quark distribtion
functions together with the spin-independent distribution
$q(x, \mu^{2})$ and the helicity distribution $\Delta q(x, \mu^{2})$
(see (\ref{eq:odf})--(\ref{eq:odh}) and Table~\ref{tab:1}).
As demonstrated in \S\ref{sec:qc},
$\delta q(x, \mu^{2})$ has a probabilistic 
interpretation as 
$\delta q(x, \mu^{2}) = q_{\leftarrow}(x, \mu^{2}) 
- q_{\rightarrow}(x, \mu^{2})$
corresponding to a simple parton model.
For nonrelativistic quarks,
where boosts and spatial rotations commute,
one would obtain $\delta q(x, \mu^{2}) = \Delta q (x, \mu^{2})$
(compare (\ref{eq:odg}) and (\ref{eq:odh})).
However, it is well known that the quarks inside a nucleon 
cannot be nonrelativistic,
so that $\delta q(x, \mu^{2})$ and $\Delta q(x, \mu^{2})$
should contain different and complementary information
with regard to the nucleon spin structure.

In the local operator formalism, the transversity distribution
is described by the matrix elements of a set of twist-2 composite
operators, which are obtained by following the procedure 
outlined in \S3.1. Taking the moment of (\ref{eq201})
and using (\ref{eq:region}), we obtain
\begin{equation}
\int_{-1}^{1} dx x^{n-1} \delta q (x, \mu^{2})
= \int_{-\infty}^{+\infty} dx x^{n-1} \delta q (x, \mu^{2})
= t_{n}(\mu^{2})\ ,
\label{eq:momh1}
\end{equation}
where $n = 1, 2, 3, \ldots$, and
\begin{equation}
\langle PS |O_{T}^{\nu \mu_{1} \cdots \mu_{n}}(\mu^{2})
|PS \rangle = \frac{2}{M} t_{n}(\mu^{2}) {\cal S}
\left( S^{\nu}P^{\mu_{1}}- S^{\mu_{1}}P^{\nu}\right)
P^{\mu_{2}} \cdots P^{\mu_{n}}\ ,
\label{eq:tnq}
\end{equation}
with $O_{T}^{\nu \mu_{1} \cdots \mu_{n}}(\mu^{2})$
the twist-2 local operator 
renormalized at $\mu^{2}$:~\footnote{
Compared with the notation of (\ref{eq:o}) and (\ref{g1op0}),
we suppress the subscript ``$\psi$'' in (\ref{eq:OT}),
because the gluonic operator 
$O_{T,G}^{\nu \mu_{1} \cdots \mu_{n}}$
does not exist in the present case 
(see the discussion below (\ref{eq:RGeqT})).}
\begin{equation}
O_{T}^{\nu \mu_{1} \cdots \mu_{n}}(\mu^{2})= \left. i^{n-1}
{\cal S} \bar{\psi}\sigma^{\nu \mu_{1}} i\gamma_{5}
D^{\mu_{2}} \cdots D^{\mu_{n}} \psi \right|_{\mu^{2}}\ .
\label{eq:OT}
\end{equation}
Note that 
the subtraction of all the trace terms is implied,
and $t_{n}$ can be defined for a definite flavor structure
as $t_{n}^{f}$ ($f=u, d, s$) or $t_{n}^{j}$ ($j=q, (3), (8)$)
(see the discussion below (\ref{adef}))
on the r.h.s. of (\ref{eq:tnq}) and (\ref{eq:OT}).
In particular, the first ($n=1$) moment $t_{1}$ 
is called the ``tensor charge''. 
For a quark of flavor $f$ and in the rest frame of the nucleon, 
(\ref{eq:momh1})--(\ref{eq:OT}) give 
\begin{equation}
\int_{0}^{1} dx \left( \delta q^{f} (x, \mu^{2}) 
- \delta q^{\bar{f}} (x, \mu^{2})
\right) = t^{f}_{1}(\mu^{2})\ , \;\;\;\;\;\;\;
\langle PS| \bar{\psi}_{f}\sigma_{j} \psi_{f} |PS \rangle
= 2 t^{f}_{1}(\mu^{2}) S_{j}\ ,
\label{eq:t1q}
\end{equation}
where we have used (\ref{eq:charge})
and set $\sigma_{j} \equiv \sigma_{j0}i\gamma_{5}$. 
It is instructive to compare (\ref{eq:t1q}) with the corresponding 
expression for the axial charge $a_{1}$ 
(see (\ref{eq:momg1f}) and (\ref{eq:Djn})):
\begin{equation}
\int_{0}^{1} dx \left( \Delta q^{f}(x, \mu^{2})
+\Delta q^{\bar{f}} (x, \mu^{2})
\right) = a^{f}_{1}(\mu^{2})\ , \;\;\;
\langle PS| \psi_{f}^{\dagger} \sigma_{j} \psi_{f} |PS \rangle
= 2 a^{f}_{1}(\mu^{2}) S_{j}\ .
\label{eq:Delta1q}
\end{equation}
Here, the relevant operator corresponds to 
the spin operator for a Dirac particle,
and differs by $\gamma^{0}$ from that of (\ref{eq:t1q}).
Therefore, $t_{1}$ is certainly different from $a_{1}$
for relativistic quarks.
Comparison between (\ref{eq:t1q}) and (\ref{eq:Delta1q})
also reveals that $t_{1}$ probes the contribution
from the valence quarks only,
while $a_{1}$ includes the helicity of the sea quarks.
This reflects the difference between the charge-conjugation properties,
expressed by (\ref{eq:charge}), for $\delta q(x, \mu^{2})$ 
and $\Delta q(x, \mu^{2})$.
This discussion suggests that measurements of $t_{1}$,
combined with those of $a_{1}$,
will provide us with deeper knowledge
about the effects of relativistic and sea-quarks
on the nucleon spin structure.

The $\mu^{2}$-dependence of $\delta q(x, \mu^{2})$
is governed by the RG equation
for the moments (\ref{eq:momh1})
as (compare with (\ref{eq:RGNS}))
\begin{equation}
\mu \frac{d}{d\mu} t_{n}(\mu^{2}) 
+ \gamma_{n}^{T}(\alpha_{s}(\mu^{2})) t_{n}(\mu^{2}) = 0\ ,
\label{eq:RGeqT}
\end{equation}
where $\gamma_{n}^{T}\left(\alpha_{s}(\mu^{2})\right)$ 
is the anomalous 
dimension of the operator $O^{\nu \mu_{1} \cdots \mu_{n}}_{T}$
of (\ref{eq:OT}),
and is related to the corresponding DGLAP-kernel. 
As discussed in \S\ref{sec:gldis},
there is no gluon analogue for $\delta q(x, \mu^{2})$,
due to the chiral-odd nature (compare Tables \ref{tab:1} and \ref{tab:2}).
Therefore, $\delta q(x, \mu^{2})$
does not mix with gluons under renormalization,
in contrast to the chiral-even distributions discussed in \S5;
the evolution of each moment is governed homogeneously
by a single anomalous dimension $\gamma_{n}^{T}(\alpha_{s}(\mu^{2}))$, 
as (\ref{eq:RGeqT}), both for 
the nonsinglet ($t^{(3)}, t^{(8)}$) and singlet ($t^{q}$) channels.
We expand $\gamma_{n}^{T}(\alpha_{s})$
similarly to (\ref{eq:Pg}) with one- and two-loop coefficients
$\gamma_{n(0)}^{T}$ and $\gamma_{n(1)}^{T}$.  
The quantity $\gamma_{n(0)}^{T}$ was calculated~\cite{KT,AM} to be  
\begin{equation}
\gamma_{n(0)}^{T} = 2 C_{2}(R)
\left( 1 + 4 \sum_{j=2}^{n} \frac{1}{j}\right)\ ,
\label{eq:gamT0}
\end{equation}
and the LO $Q^{2}$-evolution of $t_{n}(Q^{2})$ 
is given by (\ref{eq:NLONS})
with $\gamma_{n(0)}^{V} \rightarrow \gamma_{n(0)}^{T}$,
$\gamma_{n(1)}^{V} \rightarrow 0$.
In order to demonstrate unusual evolution properties
due to (\ref{eq:gamT0}), we compare it 
with the well known result
for the one-loop contribution to
$\left[ \widehat{\gamma}_{n}^{V} \right]_{qq}$ of (\ref{eq:RGeq}),
\begin{equation}
\gamma_{n(0)}^{V} = \left[ \widehat{\gamma}_{n(0)}^{V} \right]_{qq}
= 2 C_{2}(R) \left(1 - \frac{2}{n(n+1)} 
+ 4 \sum_{j=2}^{n}\frac{1}{j}\right)\ .
\label{eq:gam0}
\end{equation}
$\gamma_{n(0)}^{V}$ wholly governs the LO evolution
of the nonsinglet part of $q(x, Q^{2})$ as well as 
of $\Delta q(x, Q^{2})$; i.e. 
$\gamma_{n(0)}^{V} = \gamma_{n(0)}^{A}$.~\cite{AP}
We see that $\gamma_{n(0)}^{T} > \gamma_{n(0)}^{V}\ge 0$
holds for all $n$. This indicates that $\delta q(x, Q^{2})$
displays a stronger $Q^{2}$-dependence than the chiral-even
distributions $q(x, Q^{2})$ and $\Delta q (x, Q^{2})$.
In particular, $\gamma_{1(0)}^{T} >0$ while $\gamma_{1(0)}^{V} = 0$.
Therefore, the tensor charge $t_{1}$ 
is a scale-dependent quantity, 
unlike the vector and axial charges;
its absolute value decreases with increasing energy scale as
\begin{equation}
t_{1}(Q^{2}) = 
L^{C_{2}(R)/\beta_{0}} t_{1}(Q^{2}_{0})\ .
\label{eq:t1qQ2}
\end{equation}
This reflects the fact that the local operator 
$\bar{\psi}\sigma_{\nu \mu_{1}} i \gamma_{5} \psi$ relevant 
to the tensor charge is not a conserved current.
[ On the other hand, as discussed in \S5.2,
$\gamma_{1}^{V}(\alpha_{s}(\mu^{2})) 
= \gamma_{1}^{A}(\alpha_{s}(\mu^{2})) = 0$
to all orders, because of conservation of vector (nonsinglet
axial vector) current.]
We also note that the difference between $\gamma_{n(0)}^{T}$
and $\gamma_{n(0)}^{V}$ is larger for smaller $n$.
This suggests that the evolution of $\delta q(x, Q^{2})$
has a rather different behavior in the small-$x$ region
compared with that of $q(x, Q^{2})$ and $\Delta q(x, Q^{2})$,
because the small-$n$ region in the moment space
corresponds to the small-$x$ region in the Bjorken-$x$ space.

Recently, the two-loop anomalous dimension $\gamma_{n(1)}^{T}$
has been calculated~\cite{HKK97,VO98,KM97} in the $\overline{\rm MS}$ 
scheme.
It was shown that all the evolution properties
of $\delta q(x, Q^{2})$ discussed above are 
preserved and become even more pronounced
by including the NLO effects.~\cite{HKK97b}
It is also worth noting the following point
to demonstrate the importance of the NLO corrections:
the rightmost singularity of the anomalous dimension 
on the real axis in the complex $n$ plane
is known to determine the small-$x$
behavior of the corresponding parton distribution function
within the DGLAP evolution
(see, e.g., the second reference of Ref.~\citen{CSS}).
The rightmost singularity of $\gamma_{n(0)}^{T}$
is located at $n=-1$,
because $\sum_{j=1}^{n}1/j = \psi(n+1) + \gamma_{E}$,
where $\psi(z)$ is the di-gamma function, and 
$\gamma_{E}$ is the Euler constant;
this gives $\delta q(x, Q^{2}) \sim x$ (ignoring logarithms)
for $x \rightarrow 0$.
On the other hand, the corresponding singularity
of $\gamma_{n(1)}^{T}$ is at $n=0$.~\cite{HKK97,VO98}
Therefore, only after including the NLO effects,
the DGLAP asymptotic behavior gives $\delta q(x, \mu^{2}) \sim {\rm const}$,
and becomes
consistent with the Regge asymptotics discussed in Ref.~\citen{KMSS97}.

Now, the NLO evolution of all the twist-2
distribution functions of the nucleon is available 
both for the nonsinglet and singlet parts.
It is possible to perform the QCD analysis for physical quantities
and to make a systematic comparison between those distribution functions
at the NLO level.
As emphasized in \S3.2, this NLO evolution must be
combined with the NLO short-distance parts calculated
with the same scheme as the former, to compute an observable quantity.
For the present chiral-odd case, the corresponding 
NLO correction to $a_{TT}$ of (\ref{eq:ATT}) appears in 
Refs.~\citen{VO98} and ~\citen{KA96}, and applied to give
a prediction of $A_{TT}$.~\cite{MSSV98}

Theoretical estimates of the $x$-dependence of $\delta q(x, \mu^{2})$,
or equivalently, the matrix elements (\ref{eq:tnq}),
require elaborate calculations based on nonperturbative methods.
In this connection, it is worth noting that
$\delta q(x, \mu^{2})$ obeys some inequalities:
the first one is $|\delta q^{f}(x, \mu^{2})| \le q^{f}(x, \mu^{2})$,
due to the positivity of the probability density,
as discussed below (\ref{eq:charge}) in \S\ref{sec:qc}.
The second one was proposed by Soffer,~\cite{SO95}
based on the positivity properties of (diagonalized)
helicity amplitude for the forward quark-nucleon scattering,
\begin{equation}
2 | \delta q^{f} (x, \mu^{2}) | \le q^{f}(x, \mu^{2}) 
+ \Delta q^{f} (x, \mu^{2})\ .
\label{eq:SO}
\end{equation}
Both inequalities hold for each flavor of quark and antiquark.
These inequalities could provide useful constraints
on $\delta q(x, \mu^{2})$, combined with the known information 
of $q(x, \mu^{2})$, $\Delta q(x, \mu^{2})$,
and/or with reliable assumptions.~\footnote{Soffer's inequality
(\ref{eq:SO}) cannot be converted into that
among the matrix elements $t_{n}, a_{n}$, and $v_{n}$
of (\ref{eq:tnq}), (\ref{eq:Djn}) and (\ref{adef})
corresponding to the moments of the distributions
because of the mismatch in the charge-conjugation property 
(\ref{eq:charge}) among $\delta q(x, \mu^{2}), \Delta q (x, \mu^{2})$
and $q(x, \mu^{2})$. 
In particular, it is useless in deriving a model-independent constraint
on the tensor charge $t_{1}$.}
We note that there have been several discussions
in which it has been pointed out that 
the inequality (\ref{eq:SO}) might suffer from higher order
corrections beyond the LO.~\cite{GJJ95,KCM96,BST98}

There have been various theoretical estimates of
the tensor charge $t_{1}$ based on lattice QCD,~\cite{ADHK97}
QCD sum rules,~\cite{HJ96} nucleon models,~\cite{IS97} etc.
There also exist calculations of the $x$-dependence of 
$\delta q(x, \mu^{2})$ based on QCD sum rules,~\cite{IK95}
and nucleon models.~\cite{JJ,GRW98,WK98}
We do not go into the details of these results here, but simply note
that the results suggest that $\delta q (x, \mu^{2})$ is not small,
but is of the same order as $\Delta q(x, \mu^{2})$.

\subsection{The twist-3 distributions $h_{L}(x, \mu^{2})$ 
and $e(x, \mu^{2})$}

We now proceed to a systematic study of the chiral-odd 
twist-3 distribution functions
$h_{L}(x, \mu^{2})$ and $e(x, \mu^{2})$
of (\ref{eq201}) and (\ref{eq:scalar}).
Our first task is to demonstrate that
these distribution functions contain information
regarding the quark-gluon correlations in the nucleon,
and thus that they are quantities beyond the parton model.
This can be visualized by
re-expressing $h_{L}(x, \mu^{2})$ and $e(x, \mu^{2})$
in terms of the quark-gluon correlation functions
introduced in \S\ref{sec:tp}.
As in the chiral-even case $g_2$ in \S5.3,
the existence of such nontrivial relations reflects the fact that 
the basis of the twist-3 distribution functions
introduced in \S2 is overcomplete:
on account of the QCD equations of motion,
the number of independent dynamical 
degrees of freedom is less than the number of distribution
functions corresponding to independent Lorentz structures.

The standard technique to reveal such constraints is to
derive relations between local operators
which arise by taking the moments of the definitions
(\ref{eq201}), (\ref{eq:scalar}), and (\ref{eq:T3}),
as worked out for $g_{2}(x, Q^{2})$ in \S\S5.1 and 5.3.
For $h_{L}(x, \mu^{2})$, a similar approach requires 
somewhat complicated analysis~\cite{JJ} 
of the trace terms of twist-3, which were subtracted out from the 
local operator (\ref{eq:OT}).
Here we employ another approach, 
which is more convenient for the present case.~\footnote{
A similar technique was applied to the
analysis of the twist-3 light-cone wave functions
of vector mesons.~\cite{BBKT98}}
We directly derive the relations among the relevant distribution functions
in the Bjorken-$x$ space by using exact operator identities
between the nonlocal operators.
The relevant operator identities are (compare with (\ref{eq:identity})) 
\begin{eqnarray}
\lefteqn{
  \frac{\partial}{\partial z_{\mu}}
  \left\{ \bar{\psi}(0) i \gamma_{5} \sigma_{\mu \nu} z^{\nu} [0, z]
  \psi (z) \right\} }\nonumber\\
&=&
  - i \int_{0}^{1}\! dv \left(v-\frac{1}{2}\right)  \bar{\psi} (0) 
  i \gamma_{5} \sigma^{\alpha \beta}z_{\beta}
  [0, vz]gG_{\alpha \nu}(vz)z^{\nu}
  [vz, z] \psi(z)
  \nonumber \\
 && + \, i m_{q} \bar{\psi}(0)
 \!\not\!z \: \gamma_{5}[0, z]\psi(z) \nonumber\\
 & & +\, \frac{i}{2}\left\{ 
  \bar{\psi}(0)(i\lslash{D}-m_{q} )\rlap/{\mkern-1mu z}\gamma_{5}
  [0, z]\psi(z) + \bar{\psi}(0)\rlap/{\mkern-1mu z}\gamma_{5}
  [0, z] (i \lslash{D}-m_{q} ) \psi(z)\right\}\ ,\label{eq:3id1}\\
\lefteqn{\bar{\psi}(0)[0, z]\psi(z) 
      - \bar{\psi}(0) \psi(0) }\nonumber\\
&=&
  \int_{0}^{1} du \left[ \frac{1}{2} \int_{0}^{u}dv
  \bar{\psi}(0)  \sigma^{\alpha \beta}z_{\beta} [0, vz]
  gG_{\alpha \nu} (vz)z^{\nu} [vz, uz] \psi(uz)\right. \nonumber\\
&& - i m_{q} 
 \bar{\psi}(0) \!\not\! z \: [0, uz] \psi(uz) \nonumber\\
& & -\left. \,\frac{i}{2} 
  \left\{ \bar{\psi}(0) (i\lslash{D}-m_{q} )
  \!\not\! z \: [0, uz] \psi(uz) + \bar{\psi}(0) 
\!\not\! z \: [0, uz] (i\lslash{D}-m_{q} )\psi(uz)
\right\}\right]\ .\nonumber\\
\label{eq:3id2}
\end{eqnarray}
These relations are exact up to operators containing total derivatives
which are irrelevant for the parton distribution functions.
It is straightforward to prove the identities
directly for the nonlocal operators~\cite{BBK89,BF90}
(see also Ref.~\citen{BB98}),
or to demonstrate that their formal Taylor expansions
at small quark-antiquark separations
generate towers of identities for the 
corresponding local operators, 
which are derived in Ref.~\citen{JJ} (see also Refs.~\citen{KT} and ~\citen{KN}).

In the light-cone limit $z^{2} \rightarrow 0$,
the nucleon matrix elements of the nonlocal operators 
on both sides of (\ref{eq:3id1}) and (\ref{eq:3id2})
can be expressed in terms of the distribution functions
defined in (\ref{eq:vector})--(\ref{eq:scalar}),
and (\ref{eq:T3}); note that the matrix elements
of the operators involving the equations of motion 
vanish due to (\ref{eq:EOMV}). 
{}For the calculation of the l.h.s. of (\ref{eq:3id1}),
we need the tensor decomposition 
of the matrix element of the nonlocal operator,
where the quark-antiquark separation
is not restricted to being light-like:
\begin{eqnarray}
\lefteqn{\langle PS| \bar{\psi}(0) i \gamma_{5} \sigma_{\mu \nu} z^{\nu}
[0, z]
\psi(z) |PS \rangle}\nonumber\\
&=&
\frac{2}{M}\left\{ \left(S_{\mu}- \frac{S\cdot z}{P\cdot z}P_{\mu}\right)
\left( P \cdot z \right)
\int_{-1}^{1}dx e^{-ixP\cdot z} \left[ \delta q(x) + O(z^{2})\right]
\right.
\nonumber \\
&-& M^{2}\frac{S\cdot z}{P \cdot z}
\left( z_{\mu} - \frac{z^{2}}{P\cdot z}P_{\mu}\right)
\left. \int_{-1}^{1}dx e^{-ix P\cdot z}
\left[ h_{L}(x) - \delta q(x) + O(z^{2})\right]
\right\}\nonumber\ ,\\
\label{eq:decom}
\end{eqnarray}
where the coefficient of each tensor is determined
by matching with the light-cone limit (\ref{eq201}).
The matrix element of (\ref{eq:3id1})
now yields the differential equation
\begin{equation}
-x^{2}\frac{d}{dx}\left(\frac{1}{x}h_{L}(x)\right)
= 2 \delta q(x) - {\rm P}\int_{-1}^{1}dx'\frac{1}{x-x'}
\left( \frac{\partial}{\partial x}- \frac{\partial}{\partial x'}\right)
\widetilde{\Phi} (x, x') - \frac{m_{q}}{M}\frac{d}{dx}\Delta q(x)\ ,
\label{eq:diffeq}
\end{equation}
where ``${\rm P}$'' denotes the principal value, and
we have used the symmetry relation 
$\widetilde{\Phi} (x, x')$ $=$ $ - \widetilde{\Phi} (x', x)$
of (\ref{eq:symm3}).
The solution of this equation with the boundary condition (\ref{eq:region})
reads
\begin{eqnarray}
\lefteqn{h_{L}(x, \mu^{2})}\nonumber\\
 &=&
2x \int_{x}^{\varepsilon(x)}dy \frac{\delta q(y, \mu^{2})}{y^{2}}
- x \int_{x}^{\varepsilon(x)}dy \frac{1}{y^{2}}
{\rm P} \int_{-1}^{1}dy' \frac{1}{y-y'}
\left( \frac{\partial}{\partial y} - \frac{\partial}{\partial y'}
\right) \widetilde{\Phi} (y, y', \mu^{2})
\nonumber \\
&& + \frac{m_{q}(\mu^{2})}{M} \left( \frac{\Delta q(x, \mu^{2})}{x}
-2 x \int_{x}^{\varepsilon(x)}dy \frac{\Delta q(y, \mu^{2})}{y^{3}}
\right)\ ,
\label{eq:solhL}
\end{eqnarray}
where $-1\le x \le 1$ and $\varepsilon(x) = x/|x|$. 
We also obtain from (\ref{eq:3id2}),
by straightforward calculation,~\footnote{
Actually, $e(x, \mu^{2})$ contains the terms
proportional to the delta function $\delta (x)$.
We eliminate these terms by multiplying $e(x, \mu^{2})$ by $x$,
because they are irrelevant to the
partonic interpretation.}
\begin{equation}
xe(x, \mu^{2}) = {\rm P}\int_{-1}^{1}dx'\frac{1}{x-x'}
\Phi(x, x', \mu^{2}) + \frac{m_{q}(\mu^{2})}{M} q(x, \mu^{2})\ ,
\label{eq:sole}
\end{equation}
where $\Phi (x, x') = \Phi (x', x)$
of (\ref{eq:symm3}) has been used.
The results (\ref{eq:solhL}) and (\ref{eq:sole})
were obtained by a 
different method employing the light-like axial gauge in Ref.~\citen{BM97}.
Here we have derived them with manifest
gauge and Lorentz covariance being maintained.

The results (\ref{eq:solhL}) and (\ref{eq:sole})
show that $h_{L}(x, \mu^{2})$ and $e(x, \mu^{2})$ 
can be completely expressed in terms of the other distribution
functions.
According to the various terms on the r.h.s. of 
(\ref{eq:solhL}) and (\ref{eq:sole}),
we decompose the solution in an obvious way as
\begin{eqnarray}
h_{L} (x, \mu^{2}) &=& h_{L}^{WW} (x, \mu^{2})+ h_{L}^{C}(x, \mu^{2})
+ h_{L}^{m}(x, \mu^{2})\ ,
\label{eq:solhL2}\\
e(x, \mu^{2})&=& e^{C}(x, \mu^{2}) + e^{m}(x, \mu^{2})\ .
\label{eq:sole2}
\end{eqnarray}
Here, $h_{L}^{WW}$ denotes the first term of (\ref{eq:solhL}),
which is the contribution from the twist-2 distribution
$\delta q(x, \mu^{2})$ and corresponds to the Wandzura-Wilczek
contribution for the case of $g_{2}(x, Q^{2})$ (see \S5.1).
On the other hand, the other terms stand for 
the effects due to the genuine twist-3 operators:
$h_{L}^{C}$ and $e^{C}$ 
represent the ``dynamical'' twist-3 contributions expressed as
the particular integral of the quark-gluon correlation
functions $\widetilde{\Phi} (x, x')$ and
$\Phi (x, x')$,
whose explicit forms are given 
by appropriate projection of (\ref{eq:T3}) as
\begin{eqnarray}
\lefteqn{\widetilde{\Phi} (x, x')
= \frac{-1}{2M}\int \frac{d\lambda}{2\pi} \frac{d\zeta}{2\pi}
e^{i\lambda x + i \zeta(x' - x)}}\nonumber \\
&\times&
\langle PS_{\parallel}|\bar{\psi}(0)
\gamma_{5}
\sigma^{\alpha \beta}w_{\beta} [0, \zeta w]gG_{\alpha \nu}(\zeta w)
w^{\nu} [\zeta w, \lambda w]\psi(\lambda w) |PS_{\parallel} \rangle\ ,
\label{eq:Phit}
\end{eqnarray}
and similar expression for $\Phi (x, x')$
with $\gamma_{5}\rightarrow 1$ and $S_{\parallel} \rightarrow S$. 
Furthermore,
the quark mass $m_{q}$ generates another type of twist-3 effect,
$h_{L}^{m}$ and $e^{m}$,
which are given by the chiral-even twist-2 operators
multiplied by $m_{q}$.
It is straightforward to show that (\ref{eq:solhL}) and (\ref{eq:sole})
give the following relations~\cite{JJ} in the moment space
(compare with (\ref{eq:momg2f}), (\ref{g2sumrule})):
\begin{eqnarray}
{\cal M}_{n}\left[ h_{L}(\mu^{2})\right]
&=& \frac{2}{n+1}t_{n}(\mu^{2}) + 
{\cal M}_{n}\left[ h^{C}_{L}(\mu^{2})\right]
+\frac{n-1}{n+1}\frac{m_{q}(\mu^{2})}{M}a_{n-1}(\mu^{2})\ ,\nonumber\\
\label{eq:momhL}\\
{\cal M}_{n}\left[ e(\mu^{2})\right]
&=& 
{\cal M}_{n}\left[ e^{C}(\mu^{2})\right]
+\frac{m_{q}(\mu^{2})}{M}v_{n-1}(\mu^{2})\ ,
\label{eq:mome}
\end{eqnarray}
where \,$t_{n}$\,,\, $a_{n-1}$\, and $v_{n-1}$ 
are defined as (\ref{eq:tnq}), (\ref{eq:Djn}) and (\ref{adef}),
respectively, and
\begin{eqnarray}
 {\cal M}_{n}\left[ h_{L}(\mu^{2})\right] &\equiv&
         \int_{-1}^{1}dx x^{n-1} h_{L} (x, \mu^{2})\ ,
   \ {\cal M}_{n}\left[ e(\mu^{2})\right] \equiv
        \int_{-1}^{1}dx x^{n-1} e(x, \mu^{2}) \nonumber \ ,\\ 
{\cal M}_{n}\left[ h^{C}_{L}(\mu^{2})\right]
&\equiv& \int_{-1}^{1} dx x^{n-1} h_{L}^{C}(x, \mu^{2})
           =
\sum_{l=2}^{\kappa_{n}^{-}} \left(1 - \frac{2l}{n+1}\right)
\widetilde{b}_{n,l}(\mu^{2})\ ,
\label{eq:Hn}\\
{\cal M}_{n}\left[ e^{C}(\mu^{2})\right]&\equiv &
\int_{-1}^{1} dx x^{n-1} e^{C}(x, \mu^{2})
= \sum_{l=2}^{\kappa^{+}_{n}}
b_{n,l}(\mu^{2})
- \frac{1 - (-1)^{n}}{4}
b_{n,\kappa^{+}_{n}}(\mu^{2})\ .\nonumber\\
\label{eq:En}
\end{eqnarray}
Here
\begin{equation}
\kappa_{n}^{-} = \left[\frac{n}{2}\right]\ , \;\;\;\;\;\;
\kappa^{+}_{n} = \left[ \frac{n-1}{2}\right] + 1\ ,
\label{eq:kappan}
\end{equation}
and 
\begin{eqnarray}
 \widetilde{b}_{n,l}(\mu^{2}) &=& \int dx x^{n-l-1}x'^{l-2} 
\widetilde{\Phi} (x, x', \mu^{2})\ , \nonumber \\ 
b_{n,l}(\mu^{2}) &=& \int dx x^{n-l-1}x'^{l-2}
\Phi (x, x', \mu^{2}) \ . 
\label{eq:tbb}
\end{eqnarray}
Substituting (\ref{eq:Phit}) and the corresponding expression
for $\Phi(x, x')$ into (\ref{eq:tbb}), $\widetilde{b}_{n,l}$
and $b_{n,l}$ are given by
\begin{eqnarray}
\langle PS| \widetilde{W}_{n,l}^{\mu_{1} \cdots \mu_{n-1}}|PS\rangle
&=& 2 \widetilde{b}_{n,l}
M {\cal S}\left(S^{\mu_{1}}P^{\mu_{2}}\cdots P^{\mu_{n-1}}
\right)\ ,\label{eq:bnl}\\
\langle PS| W_{n,l}^{\mu_{1} \cdots \mu_{n-1}}|PS\rangle
&=& 2 b_{n,l} 
M \left(P^{\mu_{1}}P^{\mu_{2}}\cdots P^{\mu_{n-1}}
\right)\ ,
\label{eq:bnlt}
\end{eqnarray}
with the twist-3 local quark-gluon operators,
\begin{eqnarray}
\widetilde{W}_{n,l}^{\mu_{1}\cdots \mu_{n-1}}
&=& \frac{i^{n-1}}{2} {\cal S}
\bar{\psi}\sigma^{\alpha \mu_{1}}
\gamma_{5} D^{\mu_{2}} \cdots g{G_{\alpha}}^{\mu_{l}}
\cdots D^{\mu_{n-1}} \psi - (l \rightarrow n-l+1)\ ,\nonumber\\
\label{eq:wnl} \\
W_{n,l}^{\mu_{1}\cdots \mu_{n-1}}
&=& \frac{i^{n-1}}{2} {\cal S}
\bar{\psi}\sigma^{\alpha \mu_{1}}
D^{\mu_{2}} \cdots g{G_{\alpha}}^{\mu_{l}}
\cdots D^{\mu_{n-1}} \psi + (l \rightarrow n-l+1)\ .
\label{eq:wnlt}
\end{eqnarray}
On the r.h.s. of (\ref{eq:bnl})--(\ref{eq:wnlt}),
subtraction of the trace terms is implied.
Note that $\widetilde{W}_{n,l}^{\mu_{1}\cdots \mu_{n-1}}$
and $W_{n,l}^{\mu_{1}\cdots \mu_{n-1}}$
are odd and even under the replacement $l \rightarrow n-l+1$,
respectively, corresponding to
the symmetry of $\widetilde{\Phi} (x, x')$ and $\Phi (x, x')$
under the interchange $x \leftrightarrow x'$.
The $\widetilde{W}_{n,l}^{\mu_{1}\cdots \mu_{n-1}}$ 
with $l= 2, \cdots, \kappa^{-}_{n}$
($W_{n,l}^{\mu_{1}\cdots \mu_{n-1}}$
with $l = 2, \cdots, \kappa^{+}_{n}$)
form an independent basis of the twist-3 quark-gluon operators
for $h_{L}(x, \mu^{2})$ ($e(x, \mu^{2})$) in the local operator
approach.

Next, we proceed to the $Q^{2}$-evolution
of $h_{L}(x, Q^{2})$ and $e(x, Q^{2})$
(see (\ref{eq:solhL2}) and (\ref{eq:sole2})).
The $Q^{2}$-dependence of $\delta q(x, Q^{2})$ has been discussed in 
\S6.1, and this completely determines that of 
$h_{L}^{WW}(x, Q^{2})$. Similarly,
$h_{L}^{m}(x, Q^{2})$ and $e^{m}(x, Q^{2})$ are driven by
the $Q^{2}$-evolution of $\Delta q(x, Q^{2})$ and $q(x, Q^{2})$
discussed in \S\S5.2 and 3.1,
combined with the running quark mass $m_{q}(Q^{2})$.
On the other hand, the $Q^{2}$-evolution 
of the quark-gluon correlation 
contributions $h_{L}^{C}(x, Q^{2})$ and $e^{C}(x, Q^{2})$
is complicated: $\widetilde{\Phi} (y, y', \mu^{2})$
and $\Phi (x, x', \mu^{2})$
of (\ref{eq:solhL}) and (\ref{eq:sole})
are respectively governed by a RG equation
similar to (\ref{eq;DGLAP3}) with the corresponding kernel
[up to the mixing of $m_{q}(\mu^{2})\Delta q(x, \mu^{2})$
and $m_{q}(\mu^{2})q(x, \mu^{2})$].

Following the discussion in \S3.1,
let us go over to the moment space to treat
the evolution of the quark-gluon correlations.
Neglecting the quark masses for simplicity,
$\{\widetilde{b}_{n,l}(\mu^{2})\}$ and $\{b_{n,l}(\mu^{2})\}$
of (\ref{eq:bnl}) and (\ref{eq:bnlt})
obey RG equations
similar to (\ref{eq:RG3}) with the corresponding anomalous dimension
matrices. 
In the LO evolution, we denote these anomalous dimension matrices as
$\frac{\alpha_{s}}{2\pi} Y_{n}^{-}$ and 
$\frac{\alpha_{s}}{2\pi} Y^{+}_{n}$, respectively,
where the superscripts $-$ and $+$ refer to the ``parity'' of the relevant
operators (\ref{eq:wnl}) and (\ref{eq:wnlt})
under $l \rightarrow n-l+1$.~\footnote{
The RG equations at LO are not affected by the insertion of  
$\gamma_{5}$ into the relevant operators (\ref{eq:wnl}) 
and (\ref{eq:wnlt}).~\cite{BBKT}
Therefore, we label the mixing matrices by
the ``parity'' of the corresponding operators.}
The RG equations are solved to give (compare with 
(\ref{eq:solfnk}))
\begin{equation}
\widetilde{b}_{n,l}(Q^{2}) = \sum_{k=2}^{\kappa^{-}_{n}}\left[
L^{Y_{n}^{-}/\beta_{0}}
\right]_{lk}\widetilde{b}_{n,k}(Q_{0}^{2})\ , \;\;\;\;\;
b_{n,l}(Q^{2}) = \sum_{k=2}^{\kappa^{+}_{n}}\left[
L^{Y^{+}_{n}/\beta_{0}}
\right]_{lk}b_{n,k}(Q_{0}^{2})\ .
\label{eq:RGsol3}
\end{equation}
The mixing matrices $Y_{n}^{\mp}$ have been obtained
by computing the one-loop corrections to the relevant operators
(\ref{eq:wnl}) and (\ref{eq:wnlt}) in Refs.~\citen{KT} and \citen{KN}
(see also Ref.~\citen{BM97}).~\footnote{
In these works, the mixing of $m_{q} a_{n-1}$ ($m_{q} v_{n-1}$)
with $\widetilde{b}_{n,k}$ ($b_{n,k}$) is also calculated.}
In accord with the general argument in \S4,
the EOM operators, which are generated by Taylor expanding 
the nonlocal EOM operators appearing in (\ref{eq:3id1}) and (\ref{eq:3id2}),
play roles in the course of renormalization.~\footnote{
The BRST noninvariant EOM operators also come into play
and can be treated~\cite{KT,KN}
similarly to the case of \S5.3.}
On the other hand, in contrast to the chiral-even twist-3 case of \S5.3,
one does not encounter the 
mixing of the pure gluonic operators nor the BRST invariant 
alien operators in the singlet channel,
due to the chiral-odd property.

Substituting (\ref{eq:RGsol3}) into the r.h.s. of (\ref{eq:Hn}) 
and (\ref{eq:En}), ${\cal M}_{n}\left[ h_{L}^{C}(Q^{2})\right]$
and ${\cal M}_{n}\left[ e^{C}(Q^{2})\right]$
exhibit a complicated mixing pattern characteristic of
the higher twist operators.
They involve $\kappa^{-}_{n}-1$ 
and $\kappa^{+}_{n}-1$ independent terms
corresponding to the eigenvalues of the mixing matrices $Y_{n}^{-}$
and $Y^{+}_{n}$, respectively. 
More and more terms with different anomalous dimensions contribute to 
${\cal M}_{n}\left[ h_{L}^{C}(Q^{2})\right]$
$\left({\cal M}_{n}\left[ e^{C}(Q^{2})\right]\right)$
for larger $n$,
and ${\cal M}_{n}\left[ h_{L}^{C}(Q^{2})\right]$ for $n \ge 6$
$\left({\cal M}_{n}\left[ e^{C}(Q^{2})\right]\ {\rm for}\ 
n \ge 5\right)$ are not directly related to 
${\cal M}_{n}\left[ h_{L}^{C}(Q_{0}^{2})\right]$
$\left({\cal M}_{n}\left[ e^{C}(Q_{0}^{2})\right]\right)$
by the evolution, in contrast to the twist-2 case of \S6.1.
It would not be possible to distinguish experimentally
between terms with different anomalous dimensions in 
${\cal M}_{n}\left[ h_{L}^{C}(Q^{2})\right]$
and ${\cal M}_{n}\left[ e^{C}(Q^{2})\right]$.

There exist, however, two important limits,
$N_{c} \rightarrow \infty$ and $n\rightarrow \infty$,
where ${\cal M}_{n}\left[ h_{L}^{C}(Q^{2})\right]$
and ${\cal M}_{n}\left[ e^{C}(Q^{2})\right]$
obey simple $Q^{2}$-evolution.
The mixing matrices $Y_{n}^{\mp}$ 
of (\ref{eq:RGsol3}) involve two Casimir operators 
$C_{2}(R) = \frac{N_{c}^{2}-1}{2N_{c}}$ and $C_{2}(A) = N_{c}$.
In the large $N_{c}$ limit 
with the replacement $C_{2}(R) \rightarrow \frac{N_{c}}{2}$,
the exact relations
\begin{eqnarray}
\sum_{l=2}^{\kappa^{-}_{n}}
\left( 1 - \frac{2l}{n+1} \right) \left[ Y_{n}^{-}\right]_{lk}
&=& \left( 1 - \frac{2k}{n+1}\right)\lambda_{n}^{-}\ ,
\label{eq:ERhl} \\
\sum_{l=2}^{\kappa^{+}_{n}} \left[ Y^{+}_{n}\right]_{lk}
- \frac{1 - (-1)^{n}}{4}\left[ Y_{n}^{+}
\right]_{\kappa^{+}_{n} k}
&=& \left( 1 - \frac{1 - (-1)^{n}}{4}\delta_{k, \kappa^{+}_{n}}
\right) \lambda_{n}^{+}
\label{eq:ERe}
\end{eqnarray}
have been derived,~\cite{KN,BBKT} where $\lambda_{n}^{\mp}$ 
are the smallest eigenvalue of the mixing matrices $Y_{n}^{\mp}$ 
in this limit, and are given by
\begin{equation}
\lambda_{n}^{-} = 2N_{c} \left(\sum_{j=1}^{n-1} \frac{1}{j}
- \frac{1}{4} + \frac{3}{2n} \right)\ ,
\;\;\;\;\;\;\;
\lambda^{+}_{n} = 2 N_{c}
\left( \sum_{j=1}^{n-1}\frac{1}{j} - \frac{1}{4}
- \frac{1}{2n} \right)\ .
\label{eq:tln}
\end{equation}
As a consequence of these relations, we obtain (compare with (\ref{eq:g2LNsol}))
\begin{equation}
{\cal M}_{n}\left[ h_{L}^{C}(Q^{2})\right]
= L^{\lambda_{n}^{-}/\beta_{0}}
{\cal M}_{n}\left[ h_{L}^{C}(Q_{0}^{2})\right]\ , 
\;\;\; {\cal M}_{n}\left[ e^{C}(Q^{2})\right]
= L^{\lambda_{n}^{+}/\beta_{0}}
{\cal M}_{n}\left[ e^{C}(Q_{0}^{2})\right]\ ,
\label{eq:Qevo}
\end{equation}
which indicate that 
${\cal M}_{n}\left[ h_{L}^{C}(Q^{2})\right]$
and ${\cal M}_{n}\left[ e^{C}(Q^{2})\right]$ obey simple 
$Q^{2}$-evolution without any complicated operator mixing.
They are governed by the anomalous dimensions 
given in analytic form (\ref{eq:tln}).
The inverse Mellin transformations of (\ref{eq:Qevo})
show that $h_{L}^{C}(x, Q^{2})$ and $e^{C}(x, Q^{2})$
obey DGLAP-type homogeneous evolution equations,
as in the twist-2 nonsinglet case.

The mathematical reason for this simplification
is the same as that for the similar simplification
applied for the chiral-even (nonsinglet)
structure function $g_{2}(x, Q^{2})$
in (\ref{eq:g2LN})--(\ref{eq:g2LNsol}):
The coefficients of the relevant quark-gluon 
contributions $\widetilde{b}_{n,l}$ and $b_{n,l}$ in 
(\ref{eq:Hn}) and (\ref{eq:En}) give the {\it left} eigenvector
of the mixing matrices $Y_{n}^{-}$ and $Y_{n}^{+}$ 
corresponding to the lowest eigenvalue
$\lambda_{n}^{-}$ and $\lambda_{n}^{+}$
(see (\ref{eq:ERhl}) and (\ref{eq:ERe})).
These phenomena can be referred to as the decoupling of the particular combination
of the quark-gluon operators,
which give $h_{L}^{C}(x, \mu^{2})$ and $e^{C}(x, \mu^{2})$
as (\ref{eq:Hn}) and (\ref{eq:En}),
from the evolution (\ref{eq:RGsol3}).
The same decoupling is observed at large $n$ for arbitrary values of 
$N_{c}$.~\cite{ABH,BBKT}
In this case, we obtain (\ref{eq:g2LN}), 
(\ref{eq:ERhl}) and (\ref{eq:ERe}) with the lowest anomalous dimensions
(\ref{eq:varpi}) and (\ref{eq:tln}) shifted according to 
($\varrho_{n} = \varpi_{n}, \lambda_{n}^{-}, \lambda_{n}^{+}$)
\begin{equation}
\varrho_{n}
\rightarrow 
\varrho_{n}
+ (4C_{2}(R) - 2 N_{c}) \left( \sum_{j=1}^{n-1} \frac{1}{j}
- \frac{3}{4}\right)\ ,
\label{eq:shift}
\end{equation}
to ${\cal O}\left(\ln (n)/n\right)$ accuracy.
With these modifications of the anomalous dimensions,
the results (\ref{eq:g2LNsol}) and (\ref{eq:Qevo}) are valid to 
${\cal O}\left( 1/N_{c}^{2}\cdot \ln (n)/n\right)$ accuracy.

We recognize that the simplifications of the $Q^{2}$-evolution in the two limits,
$N_{c}\rightarrow  \infty$ and $n\rightarrow \infty$,
are universal phenomena for all twist-3 distribution
functions $g_{2}(x, Q^{2})$, $h_{L}(x, Q^{2})$ 
and $e(x, Q^{2})$.~\footnote{Similar simplifications are
observed in twist-3 nonsinglet fragmentation functions~\cite{BK97}
and in twist-3 meson wave functions~\cite{BBKT98} 
(see also Ref.~\citen{BDM98}).}
These results provide a powerful framework 
both in confronting experimental data and for model building,
since the results are valid up to corrections $\sim\,1/N_{c}^{2}$,
which are numerically small. 
From a general point of view,
they are interesting in providing us with an example of an interacting 
three-particle system in which one can find the exact energy of the 
lowest state.~\cite{BDM98}
For phenomenology,
the description of each moment of the twist-3 distributions now
requires one single nonperturbative parameter. This is noteworthy
because it means that inclusive measurements of twist-3 distributions
are complete [to ${\cal O}\left( 1/N_{c}^{2}\cdot \ln (n)/n\right)$
accuracy] in the same sense as those of the twist-2 distributions;
i.e., knowledge of the distribution
at one value of $Q_{0}^{2}$ is enough to
predict its value at arbitrary $Q^{2}$
in the spirit of DGLAP evolution equation.
For model building, the results are used~\cite{St93,KK97,KKN98} 
to rescale the model calculations at a low scale to the large values of $Q^{2}$
which correspond to actual experiments.
It is also possible to predict~\cite{rv,CSS} 
the behavior of the twist-3 distributions
in the limits $x \rightarrow 0$ and $x \rightarrow 1$:
The rightmost singularity of the relevant anomalous dimensions
(\ref{eq:varpi}) and (\ref{eq:tln}) are located at $n=0$
on the real axis in the complex $n$ plane, and thus
the $N_{c} \rightarrow  \infty$ DGLAP evolution equations
predict the following behavior for $x \rightarrow 0$ with 
$\left(\alpha_{s}(Q^{2})/\pi\right)\ln(1/ x) \ll 1$:
\begin{equation}
\bar{g}_{2}^{NS}(x, Q^{2}),\ h_{L}^{C}(x, Q^{2}),\
e^{C}(x, Q^{2}) \sim {\rm const.}\ ,
\label{eq:as0}
\end{equation}
where $\bar{g}_{2}^{NS}(x, Q^{2})$ denotes the nonsinglet part
of the genuine twist-3 piece of $g_{2}(x, Q^{2})$,
and ``${\rm const.}$'' denotes a $Q^{2}$-dependent normalization factor
of each distribution.
On the other hand, the $n \rightarrow \infty$ DGLAP evolution 
equations for these distributions
are driven by a common anomalous dimension 
(\ref{eq:shift}), from which we obtain
for $x \rightarrow 1$ with 
$\left(\alpha_{s}(Q^{2})/\pi\right) \ln\left( 1/(1-x)\right)$ $\ll$ $1$
\begin{equation}
\bar{g}_{2}^{NS}(x, Q^{2}),\ h_{L}^{C}(x, Q^{2}),\
e^{C}(x, Q^{2})
\simeq F(\mu^{2}) \frac{e^{(3/4-\gamma_{E})K}}
{\Gamma\left(c(\mu^{2}) +K + 1\right)}L^{N_{c}/\beta_{0}}
(1-x)^{c(\mu^{2}) +K}\ ,
\label{eq:as1}
\end{equation}
where $K \equiv - (4C_{2}(R)/\beta_{0})\ln L$, 
$\Gamma(x)$ is the gamma function, and 
$\gamma_{E}$ is the Euler constant.
$F(\mu^{2})$ and $c(\mu^{2})$ are $\mu^{2}$-dependent constants,
which are specific to each distribution.
The results (\ref{eq:as0}) and (\ref{eq:as1})
give a guide to the expected small $x$ and large $x$ behavior,
which is important for experimental extrapolations.

\section{Summary}

In this article, we have studied and surveyed the polarized 
structure functions in QCD.
Our starting point was the facrotization theorems in QCD,
which provide strong machinery to separate systematically a variety of 
high-energy cross sections into short- and long-distance
contributions.
To describe a ``universal'' long-distance part,
we have defined the parton distribution functions
as the nucleon matrix element of gauge invariant nonlocal 
operators, and discussed complete classification
of quark and gluon distribution functions, as well as those
of twist-3 three-particle correlation functions.
One of the key points of our approach
is that we preserved maximal (gauge (BRST) and Lorentz) symmetries
of the theory at every step of investigation.
This allowed us to derive the properties and relations of the 
distribution functions in an economical way.
Another point is use of the one-to-one correspondence 
between the nonlocal (coordinate or Bjorken-$x$ space)
and the local operator (moment space) approaches.
This again corresponds to the maximum use of the symmetry
to simplify the dynamics, 
because the moment $n$ is invariant under the evolution.
In particular, we discussed in detail the $Q^{2}$-evolution
of the twist-3 distribution functions and the corresponding
three-particle correlation functions in the local operator approach,
where all the relevant steps can be worked out
based on the standard and familiar field theory techniques.   
A characteristic feature 
of higher twist gauge invariant operators was discussed
from a general viewpoint, and it was 
shown how we should deal with the renormalization
of the complete set of twist-3 distribution functions.
We also emphasized the essential role of the QCD equations of motion  
to reveal the interrelation between the different twist-3 distributions
and to derive their $Q^{2}$-evolution.

For all the twist-2 distribution functions,
we now have sufficient theoretical information 
to make a NLO prediction of the QCD evolution.
Although the treatment of twist-3 distribution functions
was very complicated already at LO,
the corresponding $Q^{2}$-evolution has been completely worked out 
giving the mixing among an independent set of nonperturbative 
matrix elements.
These results will provide us with a powerful framework 
to analyze {\it new} experimental data with sufficient accuracy.
As explained in the text, the nature of the parton distribution function
depends on the scheme employed.
Therefore, we need the short-distance parts (hard scattering coefficients)
in the same scheme for each independent hard process
to perform a consistent QCD analysis.
We need a wide variety of processes to extract a consistent set of
information and to get an unified picture of the parton
distributions. 
The polarized Drell-Yan and other processes utilizing hadron colliders
will give us opportunities to fully understand the
behavior of quarks and gluons inside the nucleon.

The transversity distribution 
is the ``new'' and final twist-2 quark distribution and 
is expected to play a distinguished role in spin physics,
complementing the information regarding the nucleon spin structure from 
the conventional helicity distribution.
The ``measurable'' twist-3 distribution functions are 
also of special interest, because they contain information
concerning quark-gluon correlations inside the nucleon and are free from
renormalon ambiguities.   
However, the phenomenology of twist-3 distributions is not as straightforward
as that of twist-2 distributions due to the fact that 
the twist-3 distributions are essentially
three-particle correlation functions: 
We need to determine many parameters to make
precise predictions for the observable structure functions,
so that it is generally
very hard to obtain full information of the twist-3 distributions.
In this respect, drastic and universal
simplification of all twist-3 distributions in the 
$N_{c}\rightarrow \infty$ (or $n\rightarrow \infty$) limit
is crucial, and solves the above problem for all practical purposes. 
Now the $N_{c}\rightarrow \infty$ and $n\rightarrow \infty$
evolution equations allow for the development of phenomenology and model building
of the twist-3 distributions in analogy to 
the twist-2 case to ${\cal O}(1/N_{c}^{2})$ accuracy.

We mention that
there still remain many subtleties and controversial aspects
which were not covered in this article. 
Even in the twist-2 case, we are sometimes forced to 
make several assumptions on 
various unknown corrections in using QCD predictions
to describe experimental data.
Among those problems are: the small $x$ (and large $x$)
behavior of the structure
functions,~\cite{kkh} theoretical treatments of the \lq\lq wee\rq\rq\ 
partons, including the role of massive quarks, 
the flavor separation of the structure functions, etc.
Also, the power corrections in $1/Q^2$ could produce
rather large effects in realistic experiments, and cause some ambiguities.
These power corrections come from
the higher twist terms as well as target mass effects.
The treatment of the latter effects is well established,~\cite{ku}
but, as for the former, it is known that there appear ambiguities 
in the twist $\ge 4$ terms due to the renormalon 
singularities.~\cite{m} 
All these problems require careful theoretical investigations.
Forthcoming precision measurements will also provide us with
information and clues to answer these questions.

Finally, we hope that various kinds of new experiments
and more detailed theoretical investigations will be able to
clarify not only perturbative but also nonperturbative
aspects of QCD related to hadron spin physics.

\section*{Acknowledgements}

The authors would like to thank I. I. Balitsky, V. M. Braun, H. Kawamura,
Y. Koike, S. Matsuda, T. Nasuno, K. Sasaki, H. Tochimura,
Y. Yasui and T. Uematsu
for collaboration on the subject discussed in this work.
The work of J. K. was supported in part by the Monbusho Grant-in-Aid
for Scientific Research No.C(2)09640364.
The work of K. T. was supported in part by the
Grant-in-Aid for Encouragement of Young Scientists No. 09740215
from the Ministry of Education, Science, Sports and Culture.

%%%%%%%%%%%%%%%%%%%%%%%%%%%%%%%%%%%%%%%%%

\end{document}